\documentclass[a4paper,11pt]{article}
\pdfoutput=1

\usepackage{hyperref}

\usepackage{
graphicx,
hyperref,
amsmath,
amssymb,
charter,
xcolor,
ifluatex,
multirow}
        
\definecolor{Gray}{gray}{0.95}
\definecolor{RGray}{gray}{0.85}
\definecolor{CGray}{gray}{0.92}

\definecolor{tit}{rgb}{0.1,0.2,0.4}
\definecolor{blus}{cmyk}{1,1,0,0.6}
\definecolor{verde}{cmyk}{0.92,0,0.59,0.25}

\usepackage{tipa}
\usepackage{amsmath,amssymb,amsfonts, bm}
\usepackage{epsfig}
\usepackage{graphicx}
\usepackage{slashed}
\usepackage{color}

\usepackage{caption}
\usepackage{subcaption}
\usepackage{slashed} 
\usepackage{float}

\newcommand{\eps}{\epsilon}

\newcommand{\Heff}{{\cal H}_\text{NP}}

\newcommand{\meg}{\mu \to e \gamma}

\newcommand{\D}{{\cal D}}

\newcommand{\M}{{\cal M}}

\usepackage{calc}

\newcommand{\be}{\begin{equation}}
\newcommand{\ee}{\end{equation}}

\newcommand{\bea}{\begin{eqnarray}}
\newcommand{\eea}{\end{eqnarray}}

\newcommand{\bfig}{\begin{figure}}
\newcommand{\efig}{\end{figure}}

\newcommand{\meee}{\mu \to e \bar{e} e}
\newcommand{\lag}{\ensuremath{\mathcal{L}}}

\newcommand{\qqquad}{\qquad \qquad}

\newcommand{\br}{\text{BR}}

\makeatletter
\newcommand*{\rom}[1]{\expandafter\@slowromancap\romannumeral #1@}
\makeatother

\begin{document}
\allowdisplaybreaks
\vspace*{-2.5cm}

\vspace{2cm}

\begin{center}
{\LARGE \bf \color{tit} Flavour bounds on the flavon of  a minimal and a non-minimal   $\mathcal{Z}_2 \times \mathcal{Z}_N$  symmetry }\\[1cm]

{\large\bf Gauhar Abbas$^{a}$\footnote{email: gauhar.phy@iitbhu.ac.in} \\ 
Vartika   Singh$^{a}$\footnote{email:vartikasingh.rs.phy19@itbhu.ac.in  }   \\
Neelam  Singh$^{a}$\footnote{email: neelamsingh.rs.phy19@itbhu.ac.in }  \\
Ria Sain$^{b}$\footnote{email:  riasain@rnd.iitg.ac.in } }
\\[7mm]
{\it $^a$ } {\em Department of Physics, Indian Institute of Technology (BHU), Varanasi 221005, India}\\[3mm]
{\it $^b$  Department of Physics, Indian Institute of Technology,  Guwahati  781039,  India } {\em }\\[3mm]

\vspace{1cm}
{\large\bf\color{blus} Abstract}
\begin{quote}
We investigate  flavour bounds on the $\mathcal{Z}_2 \times \mathcal{Z}_5$ and  $\mathcal{Z}_2 \times \mathcal{Z}_9$ flavour symmetries.   These flavour symmetries  are   a minimal  and a non-minimal  forms  of the  $\mathcal{Z}_2 \times \mathcal{Z}_N$ flavour symmetry,  that can provide a simple set-up for the Froggatt-Nielsen mechanism.  The $\mathcal{Z}_2 \times \mathcal{Z}_5$ and  $\mathcal{Z}_2 \times \mathcal{Z}_9$ flavour symmetries are capable of explaining the fermionic masses and mixing pattern of the standard model including that of the neutrinos.  The bounds on the parameter space of the flavon field of   the $\mathcal{Z}_2 \times \mathcal{Z}_5$ and  $\mathcal{Z}_2 \times \mathcal{Z}_9$ flavour symmetries are derived using  the current quark and lepton flavour physics data  and future projected sensitivities of quark and lepton flavour effects.  The strongest bounds on the flavon of the $\mathcal{Z}_2 \times \mathcal{Z}_5$ symmetry come from the $D^0 - \bar D^0$ mixing.   The bounds on the $\mathcal{Z}_2 \times \mathcal{Z}_9$ flavour symmetry are stronger than that of the minimal $\mathcal{Z}_2 \times \mathcal{Z}_5$ symmetry.    The ratio $R_{\mu \mu}$ provides rather robust bounds on the flavon parameters  in the   future phase-\rom{1} and phase-\rom{2}  of the LHCb by leaving only a very small region in the allowed parameter space of the models.
\end{quote}

\thispagestyle{empty}
\end{center}

\begin{quote}
{\large\noindent\color{blus} 
}

\end{quote}

\newpage
\setcounter{footnote}{0}
\section{Introduction}
The $\mathcal{Z}_2 \times \mathcal{Z}_N$ flavour symmetry\cite{Abbas:2018lga} provides a new framework for the celebrated Froggatt-Nielsen (FN) mechanism that eventually furnishes an elegant solution to the flavour problem of the standard model (SM)\cite{Froggatt:1978nt}.   The flavour problem of the SM comprises a set of fundamental questions,  including the origin of the mass pattern of fermions of the SM,  an explanation for the observed quark-mixing,   and the source of neutrino masses and oscillations.  There are various approaches to address this  problem in literature.  For instance, it can have a solution through the hierarchy of vacuum-expectation-values (VEVs) in a technicolour-framework where VEVs are sequential chiral condensates of an extended dark-technicolour sector\cite{Abbas:2017vws,Abbas:2019fza,Abbas:2020frs}.  A possible explanation can be obtained  using Abelian flavor symmetries~\cite{Froggatt:1978nt,flavor_symm1,flavor_symm2,flavor_symm3,Davidson:1983fy,Davidson:1987tr}, creating loop-suppressed couplings to the Higgs~\cite{higgs_coup}, through wave-function localization~\cite{wf_local} or via compositeness~\cite{partial_comp}.

The central idea of the FN mechanism  mechanism is based on an Abelian flavour symmetry $U(1)_F$,  which can distinguish different flavours of fermions among and within the fermionic generations in the SM.  This is achieved by introducing a flavon field $\chi$ in such a way that only top quark gets mass from a renormalized SM interaction, and masses for other fermions are obtained from appropriate non-renormalized higher dimensional operators, which are constructed using  the  flavon field $\chi$.  For instance, if under  the $U(1)_F$ symmetry the fermions $\psi_i^c$ and $\psi_j$ have charges  $\theta_i$ and $\theta_j$ respectively and the charge of the  the SM Higgs field is zero then the Yukawa Lagrangian of the SM is forbidden by the $U(1)_F$ symmetry.  In this scenario, the masses of the SM fermions can be recovered by the non-renormalizable operators of the form,
\begin{eqnarray}
\label{flavon_ops}
\mathcal{O}  &=&  y (\dfrac{\chi}{\Lambda})^{(\theta_i + \theta_j)} \bar{\psi} \varphi \psi, \\ \nonumber
&=&  y \epsilon^{(\theta_i + \theta_j)} \bar{\psi} \varphi \psi = Y \bar{\psi} \varphi \psi
\end{eqnarray}
where $y$ is the dimensionless  coupling constant,  $\Lambda$ is the scale at which these operators are renormalized,  $\epsilon = \dfrac{\langle \chi \rangle}{\Lambda}$, $Y= y \epsilon^{(\theta_i + \theta_j)}$ is the effective Yukawa coupling,  and  the gauge singlet flavon scalar field  $\chi$  transforms under the $SU(3)_c \times SU(2)_L \times U(1)_Y$ symmetry of the SM as,
\begin{eqnarray}
\chi :(1,1,0).
 \end{eqnarray} 

The $U(1)_F$ flavour symmetry is broken spontaneously when the flavon field $\chi$ acquires a VEV.  The scale $\Lambda$ is not provided by the theory, and it can be anywhere between the weak and the Planck scale. We only require that the flavour symmetry should be broken weakly which means the ratio $ \dfrac{\langle \chi \rangle} { \Lambda}  $ should be less than unity. The flavon exchange effects in the SM phenomenology will be highly suppressed if  the scale of new physics  $\Lambda$ is much larger than the weak scale.  However,  if the flavour symmetry is broken close to the weak scale, we can hope to see observable effects on the direct or indirect experimentally measured SM observables such as  mixing and $CP$-violation in mesons.  Therefore, we need to ask how low the flavour scale could be such that it respects the bounds on flavour-changing and CP-violating processes.  Moreover, the nature of the flavour symmetry also plays an important role in  the investigation of the flavour scale.  For example, if the flavour symmetry is a continuous $U(1)_F$,  then we should ask whether it is a gauged or a global symmetry.  In a gauged $U(1)_F$ scenario,  the phenomenology of the flavon field will be affected by the  exchange of the corresponding gauge boson.  If the continuous $U(1)_F$ is global, then a massless Goldstone boson must exist.

The $\mathcal{Z}_2 \times \mathcal{Z}_N$ flavour symmetry,  unlike the conventional continuous  $ U (1)$ flavour symmetry that is employed to achieve the  FN mechanism,   is a  product of two discrete symmetries which can implement the FN mechanism in a unique way such that the flavour structure of the SM including neutrino masses and mixing parameters can be parametrized in terms of a small parameter which is the ratio of the  VEV of the flavon field  and the flavour scale $\Lambda$\cite{Abbas:2018lga}.   We notice  that the origin of the $\mathcal{Z}_2 \times \mathcal{Z}_N$ flavour symmetry  may be traced to an underlying Abelian or non-Abelian continuous symmetry or their products.  For instance,  the $\mathcal{Z}_2 \times \mathcal{Z}_N$ symmetry may be a by-product  of  a   spontaneous breaking of $U (1) \times U (1)$ continuous product symmetry.  

We note  that the discrete $\mathcal{Z}_2$ symmetry is extensively used  in studying the different versions of the two-Higgs-doublet model (2HDM) and the minimal supersymmetric SM (MSSM).  In particular,  in the $\mathcal{Z}_2 \times \mathcal{Z}_N$ flavour symmetry,  the  discrete $\mathcal{Z}_2$ symmetry exactly behaves like the one used in the type-\rom{2} 2HDM\cite{Branco:2011iw}.  Therefore,   the  $\mathcal{Z}_2 \times \mathcal{Z}_N$ flavour symmetry may also be used to implement the FN mechanism in the type-\rom{2} 2HDM and the MSSM.   Moreover,  the discrete $\mathcal{Z}_2$ symmetry  is also found to be useful  in model building,  for instance,  see references \cite{Abbas:2017vws,Abbas:2020frs,Abbas:2017vle,Abbas:2017hzw,Abbas:2016qqc,Abbas:2016xgj}.

In this work, we investigate flavour bounds on the dynamics of the flavon field of a minimal and a non-minimal  form of the   $\mathcal{Z}_2 \times \mathcal{Z}_N$ flavour symmetry that provides a simple  set-up for the FN mechanism.  We do not consider any ultraviolet completion of the $\mathcal{Z}_2 \times \mathcal{Z}_N$ based FN mechanism,  and present our results in a model-independent manner.  The phenomenological investigations of the flavon field of the FN mechanism in the framework of a  continuous $ U (1)$ symmetry and its extensions are dedicatedly performed in literature,  for instance,  flavour bounds are investigated in reference\cite{Dorsner:2002wi},  the LHC phenomenology is explored in  references\cite{Tsumura:2009yf,Berger:2014gga,Huitu:2016pwk,Diaz-Cruz:2014pla,Arroyo-Urena:2018mvl,Arroyo-Urena:2019fyd,Higaki:2019ojq,Arroyo-Urena:2022oft},  a low flavour breaking scale is studied in reference \cite{Bauer:2015kzy},  a study for a future high energy collider is presented in reference\cite{Bauer:2016rxs},  flavon exchange effects in the dark matter interactions are studied in reference \cite{Calibbi:2015sfa},  the texture based investigation of the  FN mechanism can be found in reference \cite{Fedele:2020fvh}.

We shall present our phenomenological analysis  along the following line: In section \ref{sec2},  we investigate a minimal form of the $\mathcal{Z}_2 \times \mathcal{Z}_N$ flavour symmetry that can implement the FN mechanism. An explanation to  neutrino masses and mixing parameters is discussed in section \ref{neutrino_masses}. A non-minimal form of the $\mathcal{Z}_2 \times \mathcal{Z}_N$ flavour symmetry that implements the FN mechanism is discussed in section \ref{sec3}. The scalar potential of our model is discussed in section \ref{potential}.  Phenomenological bounds based on the quark flavour physics on the parameter space of a minimal and a non-minimal form of the $\mathcal{Z}_2 \times \mathcal{Z}_N$ flavour symmetry are derived in section \ref{quark_flavour}. Leptonic flavour constraints are investigated in section \ref{lepton_flavour}. A summary of the work is presented in section \ref{summ}.  
 \section{A minimal  $\mathcal{Z}_2 \times \mathcal{Z}_N$  flavour symmetry} 
  \label{sec2}
We now discuss the question   of  a minimal form of the $\mathcal{Z}_2 \times \mathcal{Z}_N$ flavour symmetry that can provide a simple set-up of the FN mechanism.  Our guiding principle for this purpose is the observation that a minimal suppression of the effective Yukawa couplings will require a minimal form of the  $\mathcal{Z}_2 \times \mathcal{Z}_N$ flavour symmetry.  For instance,  we assume that the mass of the top quark originates from the tree level SM Yukawa operator,  then,  following the principle of minimum suppression (PMS),  the mass of the bottom quark is obtained from the operator having the suppression of the order $y \epsilon$,  the mass of the  charm quark  from the operator having the suppression of the order  $y \epsilon^2$,  the mass of the  strange quark  from the operator having the suppression of the order $y \epsilon^3$, and  the mass of the  up  and down quarks  from the operators having at least the suppression of the order $y \epsilon^4$.

Additionally,  we need to count the number of  hierarchical energy scales  needed to account for the fermionic mass hierarchy in the SM.  For instance,  for the quark sector,  we need three energy scales to explain the mass hierarchy among the three fermionic  families.  We note that only the second and the third quark families have  intra-generational mass hierarchies,  which require only two hierarchical energy scales to achieve an explanation for the mass hierarchy within the second and  third quark families.  These hierarchical energy scales are created by the different non-renormalizable operators of the flavon fields as given in equation \ref{flavon_ops}.   Since the mass of the top quark is generated by the renormalized SM Yuakwa operators,  at least four  energy scales are required to be created through the operators of the form given in equation \ref{flavon_ops} for providing an explanation for the hierarchical quark mass pattern.

The symmetry $ \mathcal{Z}_N$ in the $\mathcal{Z}_2 \times \mathcal{Z}_N$ flavour symmetry is  responsible for providing such operators.  The symmetry $ \mathcal{Z}_2$ distinguishes between the up-type and the down-type quarks,  which makes sure that identical non-renormalizable operators of the flavon fields,  as given in equation \ref{flavon_ops},  do not appear in the up  and down-type quark mass matrices.   For creating four energy scales, the required $ \mathcal{Z}_N$ symmetry,  therefore,  should have at least four non-trivial charges.  Therefore,  the size of a minimal symmetry will be determined by this requirement and through the application of the PMS.

Finally,  we must note that a minimal form of the  $\mathcal{Z}_2 \times \mathcal{Z}_N$   flavour symmetry should not only produce correct pattern of the charged fermion masses,  it should be capable of explaining the quark mixing pattern,  neutrino masses,  and  more importantly,   it should predict correct pattern of  the neutrino mixing angles.   

After taking into account above considerations,   $\mathcal{Z}_2 \times \mathcal{Z}_{N}$ flavour symmetry  allows us to write the following generic Lagrangian which provides masses to the charged fermions of the SM,
\bea
\label{mass1}
-{\mathcal{L}}_{\rm Yukawa} &=&    \left[  \dfrac{ \chi(\chi^{\dagger})}{\Lambda} \right]^{n_{ij}^u}     y_{ij}^u \bar{ \psi}_{L_i}^q  \tilde{\varphi} \psi_{R_j}^{u}  +   \left[  \dfrac{ \chi(\chi^{\dagger})}{\Lambda} \right]^{n_{ij}^d}      y_{ij}^d \bar{ \psi}_{L_i}^q  \varphi \psi_{R_j}^{d}  \nonumber \\
&+&    \left[  \dfrac{ \chi(\chi^{\dagger})}{\Lambda} \right]^{n_{ij}^\ell}        y_{ij}^\ell \bar{ \psi}_{L_i}^\ell  \varphi \psi_{R_j}^{\ell} 
+  {\rm H.c.}, \\ \nonumber
&=&  Y^u_{ij} \bar{ \psi}_{L_i}^q  \tilde{\varphi} \psi_{R_j}^{u}
+ Y^d_{ij} \bar{ \psi}_{L_i}^q  \varphi \psi_{R_j}^{d}
+ Y^\ell_{ij} \bar{ \psi}_{L_i}^\ell  \varphi \psi_{R_j}^{\ell}   + \text{H.c.}, 
\eea
where $\chi$ or $\chi^\dagger$ may appear in the numerator of the term inside the square brackets.  We note that in the above Lagrangian  $i$ and $j$   are family indices,  $ \psi_{L}^q,  \psi_{L}^\ell  $ are quark and leptonic doublets,  $ \psi_{R}^u,  \psi_{R}^d, \psi_{R}^\ell$ are right-handed up, down type singlet quarks and  leptons,  $\varphi$ and $ \tilde{\varphi}= -i \sigma_2 \varphi^* $  are the SM Higgs field and its conjugate and $\sigma_2$ is the second Pauli matrix.  The effective Yukawa couplings $Y_{ij}$ are defined in terms of the expansion parameter  $   \dfrac{\langle \chi \rangle} { \Lambda} = \dfrac{f}{\sqrt{2} \Lambda}= \epsilon <<1$ such that $Y_{ij} = y_{ij} \epsilon^{n_{ij}}$.

\subsection{$\mathcal{Z}_2 \times \mathcal{Z}_2$ flavour symmetry  } 
The simplest choice is the $\mathcal{Z}_2 \times \mathcal{Z}_2$ flavour symmetry, which turns out to be a trivial selection since the  only charges of the $ \mathcal{Z}_2$ symmetry are $\pm1$,  which are too trivial to provide four energy scales or equivalently non-trivial operators of the form given in equation \ref{flavon_ops}.  Hence,  we conclude that this symmetry cannot be used to create a simple FN mechanism.
\subsection{$\mathcal{Z}_2 \times \mathcal{Z}_3$ flavour symmetry  } 
The first non-trivial form of the $\mathcal{Z}_2 \times \mathcal{Z}_N$ flavour symmetry,  which may provide an implementation of the FN mechanism,  is $\mathcal{Z}_2 \times \mathcal{Z}_3$.  The symmetry $ \mathcal{Z}_3$ has two non-trivial charges characterized by $\omega$ and $\omega^2$,  where $\omega$ is the cube root of unity.   In the first scenario,  following the PMS,  we assign the charges to the SM  and flavon fields as given in table \ref{tab_z3_1}.

 \begin{table}[H]
\begin{center}
\begin{tabular}{|c|c|c|}
  \hline
  Fields             &        $\mathcal{Z}_2$                    & $\mathcal{Z}_3$        \\
  \hline
  $u_{R}, c_{R}, t_{R}$                 &   +  & $ \omega$                             \\
   $d_{R},  s_{R}, b_{R}, e_R, \mu_R, \tau_R$                 &   -  &     $ 1$                              \\
   $\psi_{L_1}^q$               &   +  &    $\omega^2 $                          \\
    $\psi_{L_2}^q$                  &   +  &     $ 1 $                         \\
     $\psi_{L_3}^q$               &   +  &      $ \omega $                         \\
       $\psi_{L_1}^\ell$                 &   +  &    $\omega^2 $                          \\
     $\psi_{L_2}^\ell$               &   +  &     $ 1 $                         \\
     $\psi_{L_3}^\ell$               &   +  &      $ \omega $                         \\
    $\chi$                        & -  &       $ \omega$                                        \\
    $\varphi$              &   +        &     1 \\
  \hline
     \end{tabular}
\end{center}
\caption{The charges of left and  right-handed fermions of three families of the SM,  Higgs  and the flavon field under the $\mathcal{Z}_2$ and $\mathcal{Z}_3$  product symmetry,  where $\omega$ is the cube root of unity. }
 \label{tab_z3_1}
\end{table}

We observe that the masses of $s$ and $b$ quarks can be recovered from this charge assignment,  for instance,  mass of the $s$ quark is of the order  $ \epsilon^3$,  and that of the $b$ quark is of the order  $ \epsilon$.  However,  the mass of the $u$  and $d$ quarks  are  produced by the operators  $y (\dfrac{\chi^\dagger}{\Lambda})^{2} \bar{\psi} \varphi u_R$  and $y (\dfrac{\chi^\dagger}{\Lambda}) \bar{\psi} \varphi d_R$ instead of the operators  $y (\dfrac{\chi}{\Lambda})^{4} \bar{\psi} \varphi u_R$ and $y (\dfrac{\chi}{\Lambda})^{5} \bar{\psi} \varphi d_R$.  Any other charge assignment also does not  reproduce  masses of  every quark.

As an additional check,  we may assume that exactly identical diagonal operators for the  $u$ and $d$ quarks in their mass matrices as given in table \ref{tab_z3_2}. This charge assignment is against the original sprite of the $\mathcal{Z}_2 \times \mathcal{Z}_N$ flavour symmetry,  where the $\mathcal{Z}_2$  is exactly like the symmetry used in the type \rom{2} 2HDM.  It turns out that even in this case,  the mass of the $u$ quark  is of the order $ \epsilon$.   This conclusion does not change even if we provide different non-trivial charge assignments to the fermions and flavon fields  under the $\mathcal{Z}_2 \times \mathcal{Z}_3$ flavour symmetry.

 \begin{table}[H]
\begin{center}
\begin{tabular}{|c|c|c|}
  \hline
  Fields             &        $\mathcal{Z}_2$                    & $\mathcal{Z}_3$        \\
  \hline
  $c_{R}, t_{R}$                 &   +  & $ \omega$                             \\
   $u_{R},  d_{R},  s_{R}, b_{R}, e_R, \mu_R, \tau_R$                 &   -  &     $ 1$                              \\
   $\psi_{L_1}^q$              &   +  &    $\omega^2 $                          \\
  $\psi_{L_2}^q$               &   +  &     $ 1 $                         \\
     $\psi_{L_3}^q$                     &   +  &      $ \omega $                         \\
     $\psi_{L_1}^\ell$                &   +  &    $\omega^2 $                          \\
   $\psi_{L_2}^\ell$       &   +  &     $ 1 $                         \\
    $\psi_{L_3}^\ell$                           &   +  &      $ \omega $                         \\
    $\chi$                        & -  &       $ \omega$                                        \\
    $\varphi$              &   +        &     1 \\
  \hline
     \end{tabular}
\end{center}
\caption{The charges of left and  right-handed fermions of three families of the SM,  Higgs and the flavon fields under the $\mathcal{Z}_2$ and $\mathcal{Z}_3$  product symmetry,  where $\omega$ is the cube root of unity. }
 \label{tab_z3_2}
\end{table} 

\subsection{$\mathcal{Z}_2 \times \mathcal{Z}_4$ flavour symmetry  } 
The charges of the  symmetry $\mathcal{Z}_2 \times \mathcal{Z}_4$  are characterized by the fourth roots of unity,  which are $1,  \omega, \omega^2 $ and $\omega^3$ where $\omega = i$, $\omega^3 = \omega^*$ and $\omega^2 =-1$.  We particularly note that minimally suppressed diagonal operator of the form  $y (\dfrac{\chi}{\Lambda})^{4} \bar{\psi}_L \varphi u_R$ is not the dominant operator no matter what charge we assign to the flavon and fermionic   fields. This is because the tree-level SM Yukawa operator $y  \bar{\psi}_L \varphi u_R$ is allowed for any charge assignment for the $y (\dfrac{\chi}{\Lambda})^{4} \bar{\psi}_L \varphi u_R$  operator  under the  $ \mathcal{Z}_2 \times \mathcal{Z}_4$ flavour symmetry.

 \begin{table}[H]
\begin{center}
\begin{tabular}{|c|c|c|}
  \hline
  Fields             &        $\mathcal{Z}_2$                    & $\mathcal{Z}_4$        \\
  \hline
  $c_{R}, t_{R}$                 &   +  & $ \omega^2$                             \\
   $u_{R},  d_{R},  s_{R}, b_{R}, e_R, \mu_R, \tau_R$                 &   -  &     $ \omega$                              \\
   $\psi_{L_1}^q$                   &   +  &    $\omega^2 $                          \\
    $\psi_{L_2}^q$                  &   +  &     $ 1 $                         \\
    $\psi_{L_3}^q$                        &   +  &      $ \omega^2 $                         \\
    $\psi_{L_1}^\ell$                     &   +  &    $\omega^2 $                          \\
    $\psi_{L_2}^\ell$                 &   +  &     $ 1 $                         \\
    $\psi_{L_3}^\ell$                     &   +  &      $ \omega^2 $                         \\
    $\chi$                        & -  &       $ \omega$                                        \\
    $\varphi$              &   +        &     1 \\
  \hline
     \end{tabular}
\end{center}
\caption{The charges of left and  right-handed fermions of three families of the SM,  Higgs and the flavon fields under the $\mathcal{Z}_2$ and $\mathcal{Z}_4$  product symmetry,  where $\omega$ is the fourth root of unity.  }
 \label{tab_z4_1}
\end{table}  

Therefore,  to produce the mass of the $u$-quark,  we either choose a non-trivial transformation of the $u_R$-quark or  the first family of the quarks under the $ \mathcal{Z}_2$ symmetry, in addition to the  next to the minimal suppressed operator of the order $\epsilon^5$.  One such charge assignment is given in table \ref{tab_z4_1}.   In this case,  the masses of the down-type quarks are produced correctly through the operators with minimal suppression.  However,  in the case of up-type quarks,  still non-diagonal tree-level SM operators dominate the mass of the $u$-quark.  Other alternative charge assignments  also do not work  for creating an FN mechanism through  the $\mathcal{Z}_2 \times \mathcal{Z}_4$ symmetry.

\subsection{ $\mathcal{Z}_2 \times \mathcal{Z}_5$ flavour symmetry  } 
We now impose the next flavour symmetry,  that is,  the $\mathcal{Z}_2 \times \mathcal{Z}_5$  symmetry on the SM in a way that the various fields of the SM transform under this symmetry as given in table \ref{tab_z5}\cite{Abbas:2018lga}.   As  discussed  earlier,    we need to create at least four hierarchical energy scales as an origin of the quark mass spectrum.   This means,  for creating these energy scales,   a non-trivial and a minimal $ \mathcal{Z}_N$ symmetry should have at least four non-trivial charges.  Thus, the symmetry  $ \mathcal{Z}_5$  could be such a symmetry.   Moreover,  we note that the transformation of fields under  the $\mathcal{Z}_2 \times \mathcal{Z}_5$  symmetry is chosen such that the symmetry  $\mathcal{Z}_2$    exactly acts like the way used in the type-\rom{2} 2HDM \footnote{Adding an additional Higgs doublet to this model such that it is odd under the  $\mathcal{Z}_2$   symmetry will result in a type-\rom{1} like 2HDM.}.

 \begin{table}[H]
\begin{center}
\begin{tabular}{|c|c|c|}
  \hline
  Fields             &        $\mathcal{Z}_2$                    & $\mathcal{Z}_5$        \\
  \hline
  $u_{R}, c_{R}, t_{R}$                 &   +  & $ \omega^2$                             \\
   $d_{R},  s_{R}, b_{R}, e_R, \mu_R, \tau_R$                 &   -  &     $\omega $                              \\
   $ \nu_{e_R},   \nu_{\mu_R}, \nu_{\tau_R} $                 &   -  &     $\omega^3 $                              \\
   $\psi_{L_1}^q$                 &   +  &    $\omega $                          \\
    $\psi_{L_2}^q$                 &   +  &     $\omega^4 $                         \\
     $\psi_{L_3}^q$                 &   +  &      $ \omega^2 $                         \\
  $\psi_{L_1}^\ell$                 &   +  &    $\omega $                          \\
    $\psi_{L_2}^\ell$                 &   +  &     $\omega^4 $                         \\
     $\psi_{L_3}^\ell$                 &   +  &      $ \omega^2 $                         \\     
    $\chi$                        & -  &       $ \omega$                                        \\
    $\varphi$              &   +        &     1 \\
  \hline
     \end{tabular}
\end{center}
\caption{The charges of left and  right-handed fermions of three families of the SM,  right-handed neutrinos,  Higgs,  and singlet scalar fields under $\mathcal{Z}_2$ and $\mathcal{Z}_5$  symmetries,  where $\omega$ is the fifth root of unity. }
 \label{tab_z5}
\end{table}

The $\mathcal{Z}_2 \times \mathcal{Z}_5$ flavour symmetry allows us to write the following Lagrangian which provides masses to the charged fermions of the SM,
\bea
\label{massz5}
-{\mathcal{L}}_{\rm Yukawa} &=&    \left(  \dfrac{ \chi}{\Lambda} \right)^{4}  y_{11}^u \bar{ \psi}_{L_1}^q \tilde{\varphi} u_{R}+  \left(  \dfrac{ \chi}{\Lambda} \right)^{4}  y_{12}^u \bar{ \psi}_{L_1}^q \tilde{\varphi} c_{R} +  \left(  \dfrac{ \chi}{\Lambda} \right)^{4}  y_{13}^u \bar{ \psi}_{L_1}^q \tilde{\varphi}  t_{R} +  \left(  \dfrac{ \chi}{\Lambda} \right)^{2}  y_{21}^u \bar{ \psi}_{L_2}^q \tilde{\varphi} u_{R}\nonumber \\
&+& \left(  \dfrac{ \chi}{\Lambda} \right)^{2}  y_{22}^u \bar{ \psi}_{L_2}^q \tilde{\varphi} c_{R} + \left(  \dfrac{ \chi}{\Lambda} \right)^{2}  y_{23}^u \bar{ \psi}_{L_2}^q \tilde{\varphi} t_{R}+   y_{31}^u \bar{ \psi}_{L_3}^q \tilde{\varphi} u_{R} +   y_{32}^u \bar{ \psi}_{L_3}^q \tilde{\varphi} c_{R} +   y_{33}^u \bar{ \psi}_{L_3}^q \tilde{\varphi} t_{R} \nonumber \\
&+&  
 \left(  \dfrac{ \chi}{\Lambda} \right)^{5} y_{11}^d \bar{ \psi}_{L_1}^q  \varphi d_{R} + \left(  \dfrac{ \chi}{\Lambda} \right)^{5} y_{12}^d \bar{ \psi}_{L_1}^q  \varphi s_{R} + \left(  \dfrac{ \chi}{\Lambda} \right)^{5} y_{13}^d \bar{ \psi}_{L_1}^q  \varphi b_{R} + \left(  \dfrac{ \chi}{\Lambda} \right)^{3} y_{21}^d \bar{ \psi}_{L_2}^q  \varphi d_{R} \nonumber \\
&+& \left(  \dfrac{ \chi}{\Lambda} \right)^{3} y_{22}^d \bar{ \psi}_{L_2}^q  \varphi s_{R} + \left(  \dfrac{ \chi}{\Lambda} \right)^{3} y_{23}^d \bar{ \psi}_{L_2}^q  \varphi b_{R} + \left(  \dfrac{ \chi}{\Lambda} \right) y_{31}^d \bar{ \psi}_{L_3}^q  \varphi d_{R}+ \left(  \dfrac{ \chi}{\Lambda} \right) y_{32}^d \bar{ \psi}_{L_3}^q  \varphi s_{R}   \nonumber \\ 
&+& \left(  \dfrac{ \chi}{\Lambda} \right) y_{33}^d \bar{ \psi}_{L_3}^q  \varphi b_{R} + 
\left(  \dfrac{ \chi}{\Lambda} \right)^{5} y_{11}^\ell \bar{ \psi}_{L_1}^\ell  \varphi e_{R} + \left(  \dfrac{ \chi}{\Lambda} \right)^{5} y_{12}^\ell \bar{ \psi}_{L_1}^\ell  \varphi \mu_{R} + \left(  \dfrac{ \chi}{\Lambda} \right)^{5} y_{13}^\ell \bar{ \psi}_{L_1}^\ell  \varphi \tau_{R}  \nonumber \\
&+& \left(  \dfrac{ \chi}{\Lambda} \right)^{3} y_{21}^\ell \bar{ \psi}_{L_2}^\ell  \varphi e_{R} + \left(  \dfrac{ \chi}{\Lambda} \right)^{3} y_{22}^\ell \bar{ \psi}_{L_2}^\ell  \varphi \mu_{R} + \left(  \dfrac{ \chi}{\Lambda} \right)^{3} y_{23}^\ell \bar{ \psi}_{L_2}^\ell  \varphi \tau_{R} +  \left(  \dfrac{ \chi}{\Lambda} \right) y_{31}^\ell \bar{ \psi}_{L_3}^\ell  \varphi e_{R} \nonumber \\
&+& \left(  \dfrac{ \chi}{\Lambda} \right) y_{32}^\ell \bar{ \psi}_{L_3}^\ell  \varphi \mu_{R} \nonumber + \left(  \dfrac{ \chi}{\Lambda} \right) y_{33}^\ell \bar{ \psi}_{L_3}^\ell  \varphi \tau_{R} 
 + \text{H.c.}
\eea

The mass matrices for  up- and down-type quarks and charged leptons  can be written now in terms of the expansion parameter $\epsilon$,
\begin{equation}
\M_u = \dfrac{v}{\sqrt{2}}
\begin{pmatrix}
y_{11}^u  \epsilon^4 &  y_{12}^u \epsilon^4  & y_{13}^u \epsilon^4    \\
y_{21}^u  \epsilon^2    & y_{22}^u \epsilon^2  &  y_{23}^u \epsilon^2    \\
y_{31}^u     &  y_{32}^u      &  y_{33}^u
\end{pmatrix}, 
\M_d = \dfrac{v}{\sqrt{2}}
\begin{pmatrix}
y_{11}^d  \epsilon^5 &  y_{12}^d \epsilon^5 & y_{13}^d \epsilon^5   \\
y_{21}^d  \epsilon^3  & y_{22}^d \epsilon^3 &  y_{23}^d \epsilon^3  \\
 y_{31}^d \epsilon &  y_{32}^d \epsilon   &  y_{33}^d \epsilon
\end{pmatrix},
\M_\ell =  \dfrac{v}{\sqrt{2}}
\begin{pmatrix}
y_{11}^\ell  \epsilon^5 &  y_{12}^\ell \epsilon^5  & y_{13}^\ell \epsilon^5   \\
y_{21}^\ell  \epsilon^3  & y_{22}^\ell \epsilon^3  &  y_{23}^\ell \epsilon^3  \\
 y_{31}^\ell \epsilon   &  y_{32}^\ell \epsilon   &  y_{33}^\ell \epsilon
\end{pmatrix}.
\end{equation}

The masses of quarks and charged leptons approximately are\cite{Rasin:1998je},
\begin{align}
\label{eqn5}
\{m_t, m_c, m_u\} &\simeq \{|y_{33}^u| , ~ \left |y_{22}^u- \frac {y_{23}^u y_{32}^u} {|y_{33}^u|} \right| \epsilon^2,\\&
~ \left |y_{11}^u- \frac {y_{12}^u y_{21}^u}{|y_{22}^u-y_{23}^u y_{32}^u/y_{33}^u|}- \frac{y_{13}^u |y_{31}^u y_{22}^u-y_{21}^u y_{32}^u|-y_{31}^u y_{12}^u y_{23}^u}{|y_{22}^u- y_{23}^u y_{32}^u/y_{33}^u| |y_{33}^u|} \right| \epsilon^4\}v/\sqrt{2} ,\nonumber &\\ 
\label{eqn6}
\{m_b, m_s, m_d\} & \simeq \{|y_{33}^d| \epsilon, ~ \left |y_{22}^d- \frac {y_{23}^d y_{32}^d} {|y_{33}^d|} \right| \epsilon^3,\\&
~  \left |y_{11}^d- \frac {y_{12}^d y_{21}^d}{|y_{22}^d-y_{23}^d y_{32}^d/y_{33}^d|}- \frac{y_{13}^d |y_{31}^d y_{22}^d-y_{21}^d y_{32}^d|-y_{31}^d y_{12}^d y_{23}^d}{|y_{22}^d- y_{23}^d y_{32}^d/y_{33}^d| |y_{33}^d|} \right| \epsilon^5\}v/\sqrt{2} ,\nonumber &\\
\{m_\tau, m_\mu, m_e\} & \simeq \{|y_{33}^l| \epsilon, ~ \left|y_{22}^l- \frac {y_{23}^l y_{32}^l} {|y_{33}^l|} \right| \epsilon^3,\\& ~  \left |y_{11}^l- \frac {y_{12}^l y_{21}^l}{|y_{22}^l-y_{23}^l y_{32}^l/y_{33}^l|}- \frac{y_{13}^l |y_{31}^l y_{22}^l-y_{21}^l y_{32}^l|-y_{31}^l y_{12}^l y_{23}^l}{|y_{22}^l- y_{23}^l y_{32}^l/y_{33}^l| |y_{33}^l|} \right| \epsilon^5\}v/\sqrt{2} ,\nonumber &\\
\end{align}

The mixing angles of quarks are found to be\cite{Rasin:1998je},
\begin{eqnarray}
\sin \theta_{12}  \simeq |V_{us}| &\simeq& \left|{y_{12}^d \over y_{22}^d}  -{y_{12}^u \over y_{22}^u}  \right| \epsilon^2, 
\sin \theta_{23}  \simeq |V_{cb}| \simeq  \left|{y_{23}^d \over y_{33}^d}  -{y_{23}^u \over y_{33}^u}  \right|  \epsilon^2,\nonumber \\
\sin \theta_{13}  \simeq |V_{ub}| &\simeq& \left|{y_{13}^d \over y_{33}^d}  -{y_{12}^u y_{23}^d \over y_{22}^u y_{33}^d} 
- {y_{13}^u \over y_{33}^u} \right|  \epsilon^4.
\end{eqnarray}

We notice that the $\sin \theta_{12}$ and $\sin \theta_{23}$ have the same order.  The similar result is also reported in reference \cite{flavor_symm2}.

We present a fit  of  the experimental data to the masses of fermions in appendix.  It turns out  that some of the couplings are not order one.    We discuss a theoretical scenario for such couplings in the appendix.  This kind of not order one couplings are also reported in references \cite{flavor_symm2,Dorsner:2002wi}.  

\subsubsection{Neutrino masses and mixing}
\label{neutrino_masses}
The neutrino masses are obtained by adding  three right-handed neutrinos   as  shown in table \ref{tab_z5}.    The Lagrangian for the tree-level Majorana mass is,
\bea
\mathcal{L}_{\rm M_R}  &=& c_{ij} \left[  \frac{\chi^\dagger}{\Lambda}  \right]^5 \chi^\dagger \bar{\nu^c}_{i,R} \nu_{j,R},
\eea
where $i,j$ are flavour indices.

The Majorana mass matrices $\M_{R}$ is,
\begin{equation}
\label{NM}
\M_{R} =  M
\begin{pmatrix}
c_{11}& c_{12}    & c_{13} \\
  c_{12}  & c_{22} & c_{23} \\
c_{13}   & c_{23}   & c_{33}
\end{pmatrix},
\end{equation}
where $ M=  \langle \chi \rangle   \left[  \frac{  \langle \chi \rangle}{\Lambda} \right]^5 = \frac{f}{\sqrt{2}} \epsilon^5 $.

The Dirac mass Lagranginan for neutrinos can be written as, 

\begin{eqnarray}
\label{mass1}
-{\mathcal{L}}_{\rm Yukawa}^{\nu} &=&       y_{11}^\nu \bar{ \psi}_{L_1}^\ell  H \nu_{e_R}  \left[  \dfrac{ \chi}{\Lambda} \right]^{3} + y_{12}^\nu \bar{ \psi}_{L_1}^\ell  H \nu_{\mu_R}  \left[  \dfrac{ \chi}{\Lambda} \right]^{3} + y_{13}^\nu \bar{ \psi}_{L_1}^\ell  H   \nu_{\tau_R}   \left[  \dfrac{ \chi}{\Lambda} \right]^{3} + y_{21}^\nu \bar{ \psi}_{L_2}^\ell  H \nu_{e_R} \left[  \dfrac{ \chi}{\Lambda} \right] \\ \nonumber
&+& y_{22}^\nu \bar{ \psi}_{L_2}^\ell  H \nu_{\mu_R} \left[  \dfrac{ \chi}{\Lambda} \right] + y_{23}^\nu \bar{ \psi}_{L_2}^\ell   H \nu_{\tau_R}   \left[  \dfrac{ \chi}{\Lambda} \right]
 + y_{31}^\nu \bar{ \psi}_{L_3}^\ell  H \nu_{e_R} \left[  \dfrac{ \chi^\dagger }{\Lambda} \right] + y_{32}^\nu \bar{ \psi}_{L_3}^\ell H \nu_{\mu_R} \left[  \dfrac{ \chi^\dagger}{\Lambda} \right] \\ \nonumber
&+& y_{33}^\nu \bar{ \psi}_{L_3}^\ell  H \nu_{\tau_R}  \left[  \dfrac{ \chi^\dagger}{\Lambda} \right]
+  {\rm H.c.}.
\end{eqnarray}

The Dirac mass matrix  is given by,
\begin{equation}
\label{NM}
\M_{\D} = \dfrac{v}{\sqrt{2}}
\begin{pmatrix}
y_{11}^\nu  \epsilon^3 &  y_{12}^\nu \epsilon^3 & y_{13}^\nu   \epsilon^{3}  \\
y_{21}^\nu  \epsilon  & y_{22}^\nu \epsilon &  y_{23}^\nu   \epsilon \\
y_{31}^\nu \epsilon   &  y_{32}^\nu  \epsilon   &  y_{33}^\nu \epsilon
\end{pmatrix}.
\end{equation}

The  mass matrix  of neutrinos after including the Majorana mass terms can be written as,
\begin{equation}
\label{NM}
\M = 
\begin{pmatrix}
\M_{L}   &  \M_{\D} \\
\M_{\D}^T & \M_{R}   \\
\end{pmatrix}.
\end{equation}

Since  $v<< f$,  we ignore the contribution of the mass matrix $\M_{L}$ to the neutrino masses.\footnote{Alternatively,  we can assume that it is forbidden by some discrete symmetry.   For instance,   if three left-handed fermionic doublets of quarks and leptons,  and the Higgs doublet have a charge $\omega$  under a  $\mathcal{Z}_3$ symmetry,  the mass matrix $\M_{L}$ is forbidden.}   Now, we can use the type-\rom{1} seesaw mechanism to determine the neutrino masses by assuming  $ \M_{\D} << \M_{R}$\cite{seesaw}.  Thus the  light neutrino mass matrix is,
\begin{eqnarray}
{ \M}~ & \approx & ~ -  \M_{\D}  \M_{R}^{-1}  \M_{\D}^T,  \\ \nonumber
& \approx & \frac{v}{\sqrt{2} } \epsilon^\prime
\left(
\begin{array}{ccc}
 -\frac{\epsilon^4 \left(c_{22} c_{33} y_{11}^{\nu 2}-2
  c_{22} y_{11}^\nu +c_{22} -2c_{33}
  y_{11}^\nu +c_{33}-y_{11}^{\nu 2}+4
  y_{11}^\nu -2\right)}{(c_{22}-1) (c_{33}-1)} &
   -\epsilon^2 y_{11}^\nu & -\frac{\epsilon^2 ( c_{33}
  y_{11}^\nu- y_{11}^\nu
  y_{33}^\nu + y_{33}^\nu -1)}{c_{33}-1} \\
 -\epsilon^2 y_{11}^\nu & -1 & -1 \\
 -\frac{\epsilon^2 (c_{33} y_{11}^\nu - y_{11}^\nu
  y_{33}^\nu + y_{33}^\nu -1)}{c_{33}-1} & -1 &
   -\frac{ (c_{33}+(y_{33}^\nu-2)
  y_{33}^\nu )}{c_{33} -1} \\
\end{array}
\right),
\end{eqnarray}

where $\epsilon^{\prime } = \frac{v}{ f \epsilon^3} $,  and  we have assumed each and every coupling exactly one except those appearing in above equation.

We obtain two degenerate  neutrino masses.   The masses  approximately are  given by,
\bea
m_1 &\approx &  \frac{ \left(-y_{11}^{\nu 2}+2
  y_{11}^\nu -1\right)}{c_{22} -1}  \epsilon^{4} \epsilon^{\prime } v/\sqrt{2}, \\ \nonumber
  m_2 & \approx&    \frac{ \left(-\sqrt{4 c_{33}^2-8
   c_{33}+y_{33}^{\nu 4}-4 y_{33}^{\nu 3}+6 y_{33}^{\nu 2}-4
   y_{33}^\nu +5}-2 y_{33}^\nu -y_{33}^{\nu 2}+2
   y_{33}^\nu +1\right)}{2 (c_{33}-1)}  ~ \epsilon^{\prime }  v/\sqrt{2}, \\ \nonumber
  m_3 & \approx&    \frac{ \left(\sqrt{4 c_{33}^2-8
   c_{33}+y_{33}^{\nu 4}-4 y_{33}^{\nu 3}+6 y_{33}^{\nu 2}-4
   y_{33}^\nu +5}-2 c_{33}- y_{33}^{\nu 2} +2
   y_{33}^\nu +1\right)}{2 (c_{33}-1)}  ~ \epsilon^{\prime }  v/\sqrt{2}.
\eea 
 This kind of approximate degenerate neutrino masses are well studied in literature,  for instance, see references \cite{Abbas:2016qbl,Abbas:2015vba,Abbas:2014ala,Abbas:2013uqh}.

The leptonic mixing angles can be written as,
\begin{eqnarray}
\sin \theta_{12}  &\simeq& \left|{y_{12}^\ell \over y_{22}^\ell}  -y_{11}^\nu \right| \epsilon^2, \\ \nonumber
\sin \theta_{23} &\simeq & \left|{1-c_{33} \over  c_{33}+(y_{33}^\nu -2)y_{33}^\nu}  \right|, \\ \nonumber
\sin \theta_{13} &\simeq& \left|  \frac{ ( c_{33}
 y_{11}^\nu-y_{11}^\nu
  y_{33}^\nu + y_{33}^\nu  -1)}{ c_{33}+(y_{33}^\nu -2 )y_{33}^\nu}  \right|  \epsilon^2.
\end{eqnarray}

The remarkable observation is the pattern of the neutrino mixing angles.  The mixing angle $ \theta_{12}$ and $ \theta_{13}$ are of the same order of magnitude,  where $ \theta_{13}$ is closer to the Cabibbo angle,  and the  mixing angle $ \theta_{23}$ is completely unsuppressed.

\section{A non-minimal $\mathcal{Z}_2 \times \mathcal{Z}_9$ flavour symmetry  } 
\label{sec3}
We note from the previous section that  some of the Yukawa couplings for the minimal model based on the  $\mathcal{Z}_2 \times \mathcal{Z}_5$ flavour symmetry are not order one,  which is a preferred choice in literature.  However,  so far purpose  has been to introduce the $\mathcal{Z}_2 \times \mathcal{Z}_N$ flavour paradigm.  In this section,  we show a non-minimal model based on the $\mathcal{Z}_2 \times \mathcal{Z}_N$ flavour paradigm where the Yukawa couplings turn out to be order one,  and are given in the appendix.  We adopt a non-minimal  $\mathcal{Z}_2 \times \mathcal{Z}_9$ flavour symmetry,  and assign the charges to different fields as shown in table \ref{tab_z9}.

 \begin{table}
\begin{center}
\begin{tabular}{|c|c|c|}
  \hline
  Fields             &        $\mathcal{Z}_2$                    & $\mathcal{Z}_9$        \\
  \hline
  $u_{R}, t_{R}$                 &   +  & $ 1$                             \\
  $c_{R}$                 &   +  & $ \omega^4$                             \\
   $d_{R},  s_{R},  b_{R}, e_R, \mu_R, \tau_R $                 &   -  &     $\omega^3 $                              \\
   $ \nu_{e_R},   \nu_{\mu_R},  \nu_{\tau_R}  $                 &   -  &     $\omega^7 $                              \\
   $\psi_{L_1}^q$                 &   +  &    $\omega $                          \\
    $\psi_{L_2}^q$                 &   +  &     $\omega^8 $                         \\
     $\psi_{L_3}^q$                 &   +  &      $ 1 $                         \\
     $\psi_{L_1}^\ell$                 &   +  &    $\omega $                          \\
    $\psi_{L_2}^\ell$                 &   +  &     $\omega^8 $                         \\
     $\psi_{L_3}^\ell$                 &   +  &      $ \omega^6 $                         \\     
    $\chi$                        & -  &       $ \omega$                                        \\
    $\varphi$              &   +        &     1 \\
  \hline
     \end{tabular}
\end{center}
\caption{The charges of left and  right-handed fermions of three families of the SM,  right-handed neutrinos,  Higgs,  and singlet scalar field under $\mathcal{Z}_2$ and $\mathcal{Z}_9$  symmetries,  where $\omega$ is the ninth root of unity. }
 \label{tab_z9}
\end{table} 
The $\mathcal{Z}_2 \times \mathcal{Z}_9$ flavour symmetry allows us to write the following Lagrangian which provides masses to the charged fermions of the SM,
\bea
\label{massz9}
-{\mathcal{L}}_{\rm Yukawa} &=&    \left(  \dfrac{ \chi^\dag}{\Lambda} \right)^{8}  y_{11}^u \bar{ \psi}_{L_1}^q \tilde{\varphi} u_{R}+  \left(  \dfrac{ \chi}{\Lambda} \right)^{6}  y_{12}^u \bar{ \psi}_{L_1}^q \tilde{\varphi} c_{R} +  \left(  \dfrac{ \chi^\dag}{\Lambda} \right)^{8}  y_{13}^u \bar{ \psi}_{L_1}^q \tilde{\varphi}  t_{R} +  \left(  \dfrac{ \chi}{\Lambda} \right)^{8}  y_{21}^u \bar{ \psi}_{L_2}^q \tilde{\varphi} u_{R}\nonumber \\
&+& \left(  \dfrac{ \chi}{\Lambda} \right)^{4}  y_{22}^u \bar{ \psi}_{L_2}^q \tilde{\varphi} c_{R} + \left(  \dfrac{ \chi}{\Lambda} \right)^{8}  y_{23}^u \bar{ \psi}_{L_2}^q \tilde{\varphi} t_{R}+   y_{31}^u \bar{ \psi}_{L_3}^q \tilde{\varphi} u_{R} + \left(  \dfrac{ \chi^\dag}{\Lambda} \right)^{4}  y_{32}^u \bar{ \psi}_{L_3}^q \tilde{\varphi} c_{R} +   y_{33}^u \bar{ \psi}_{L_3}^q \tilde{\varphi} t_{R} \nonumber \\
&+&  
 \left(  \dfrac{ \chi}{\Lambda} \right)^{7} y_{11}^d \bar{ \psi}_{L_1}^q  \varphi d_{R} + \left(  \dfrac{ \chi}{\Lambda} \right)^{7} y_{12}^d \bar{ \psi}_{L_1}^q  \varphi s_{R} + \left(  \dfrac{ \chi}{\Lambda} \right)^{7} y_{13}^d \bar{ \psi}_{L_1}^q  \varphi b_{R} + \left(  \dfrac{ \chi}{\Lambda} \right)^{5} y_{21}^d \bar{ \psi}_{L_2}^q  \varphi d_{R} \nonumber \\
&+& \left(  \dfrac{ \chi}{\Lambda} \right)^{5} y_{22}^d \bar{ \psi}_{L_2}^q  \varphi s_{R} + \left(  \dfrac{ \chi}{\Lambda} \right)^{5} y_{23}^d \bar{ \psi}_{L_2}^q  \varphi b_{R} + \left(  \dfrac{ \chi^\dag}{\Lambda} \right)^{3} y_{31}^d \bar{ \psi}_{L_3}^q  \varphi d_{R}+ \left(  \dfrac{ \chi^\dag}{\Lambda} \right)^{3} y_{32}^d \bar{ \psi}_{L_3}^q  \varphi s_{R}   \nonumber \\ 
&+& \left(  \dfrac{ \chi^\dag}{\Lambda} \right)^{3} y_{33}^d \bar{ \psi}_{L_3}^q  \varphi b_{R} + 
\left(  \dfrac{ \chi}{\Lambda} \right)^{7} y_{11}^\ell \bar{ \psi}_{L_1}^\ell  \varphi e_{R} + \left(  \dfrac{ \chi}{\Lambda} \right)^{7} y_{12}^\ell \bar{ \psi}_{L_1}^\ell  \varphi \mu_{R} + \left(  \dfrac{ \chi}{\Lambda} \right)^{7} y_{13}^\ell \bar{ \psi}_{L_1}^\ell  \varphi \tau_{R}  \nonumber \\
&+& \left(  \dfrac{ \chi}{\Lambda} \right)^{5} y_{21}^\ell \bar{ \psi}_{L_2}^\ell  \varphi e_{R} + \left(  \dfrac{ \chi}{\Lambda} \right)^{5} y_{22}^\ell \bar{ \psi}_{L_2}^\ell  \varphi \mu_{R} + \left(  \dfrac{ \chi}{\Lambda} \right)^{5} y_{23}^\ell \bar{ \psi}_{L_2}^\ell  \varphi \tau_{R} +  \left(  \dfrac{ \chi}{\Lambda} \right)^{3} y_{31}^\ell \bar{ \psi}_{L_3}^\ell  \varphi e_{R} \nonumber \\
&+& \left(  \dfrac{ \chi}{\Lambda} \right)^{3} y_{32}^\ell \bar{ \psi}_{L_3}^\ell  \varphi \mu_{R} \nonumber + \left(  \dfrac{ \chi}{\Lambda} \right)^{3} y_{33}^\ell \bar{ \psi}_{L_3}^\ell  \varphi \tau_{R} 
 + \text{H.c.}
\eea

The mass matrices for  up and down-type quarks and charged leptons  turn out to be,
\begin{equation}
\M_u = \dfrac{v}{\sqrt{2}}
\begin{pmatrix}
y_{11}^u  \epsilon^8 &  y_{12}^u \epsilon^{6}  & y_{13}^u \epsilon^{8}    \\
y_{21}^u  \epsilon^8    & y_{22}^u \epsilon^4  &  y_{23}^u \epsilon^8   \\
y_{31}^u      &  y_{32}^u  \epsilon^4     &  y_{33}^u  
\end{pmatrix}, 
\M_d = \dfrac{v}{\sqrt{2}}
\begin{pmatrix}
y_{11}^d  \epsilon^7 &  y_{12}^d \epsilon^7 & y_{13}^d \epsilon^7   \\
y_{21}^d  \epsilon^5  & y_{22}^d \epsilon^5 &  y_{23}^d \epsilon^5  \\
 y_{31}^d \epsilon^3 &  y_{32}^d \epsilon^3   &  y_{33}^d \epsilon^3
\end{pmatrix},
\M_\ell =  \dfrac{v}{\sqrt{2}}
\begin{pmatrix}
y_{11}^\ell  \epsilon^7 &  y_{12}^\ell \epsilon^7  & y_{13}^\ell \epsilon^7   \\
y_{21}^\ell  \epsilon^5  & y_{22}^\ell \epsilon^5  &  y_{23}^\ell \epsilon^5  \\
 y_{31}^\ell \epsilon^3   &  y_{32}^\ell \epsilon^3   &  y_{33}^\ell \epsilon^3
\end{pmatrix}.
\end{equation}

The masses of charged fermions are approximately  given by\cite{Rasin:1998je},
\begin{align}
\label{eqn5}
\{m_t, m_c, m_u\} &\simeq \{|y_{33}^u| , ~ \left |y_{22}^u   \epsilon^4 - \frac {y_{23}^u y_{32}^u} {|y_{33}^u|   }  \epsilon^{12} \right|,\\&
~ \left |y_{11}^u \epsilon^8 - \frac {y_{12}^u y_{21}^u}{|y_{22}^u|} \epsilon^{10}- \frac{y_{13}^u |y_{31}^u y_{22}^u-y_{21}^u y_{32}^u|}{|y_{22}^u| |y_{33}^u|} \epsilon^8 \right| \}v/\sqrt{2} ,\nonumber &\\ 
\label{eqn6}
\{m_b, m_s, m_d\} & \simeq \{|y_{33}^d| \epsilon^3, ~ \left |y_{22}^d- \frac {y_{23}^d y_{32}^d} {|y_{33}^d|} \right| \epsilon^5,\\&
~  \left |y_{11}^d- \frac {y_{12}^d y_{21}^d}{|y_{22}^d-y_{23}^d y_{32}^d/y_{33}^d|}- \frac{y_{13}^d |y_{31}^d y_{22}^d-y_{21}^d y_{32}^d|-y_{31}^d y_{12}^d y_{23}^d}{|y_{22}^d- y_{23}^d y_{32}^d/y_{33}^d| |y_{33}^d|} \right| \epsilon^7\}v/\sqrt{2} ,\nonumber &\\
\{m_\tau, m_\mu, m_e\} & \simeq \{|y_{33}^l| \epsilon^3, ~ \left|y_{22}^l- \frac {y_{23}^l y_{32}^l} {|y_{33}^l|} \right| \epsilon^5,\\& ~  \left |y_{11}^l- \frac {y_{12}^l y_{21}^l}{|y_{22}^l-y_{23}^l y_{32}^l/y_{33}^l|}- \frac{y_{13}^l |y_{31}^l y_{22}^l-y_{21}^l y_{32}^l|-y_{31}^l y_{12}^l y_{23}^l}{|y_{22}^l- y_{23}^l y_{32}^l/y_{33}^l| |y_{33}^l|} \right| \epsilon^7\}v/\sqrt{2} ,\nonumber &\\
\end{align}

Similarly,  the mixing angles of quarks read\cite{Rasin:1998je},
\begin{eqnarray}
\sin \theta_{12}  \simeq |V_{us}| &\simeq& \left|{y_{12}^d \over y_{22}^d}  -{y_{12}^u \over y_{22}^u}  \right| \epsilon^2, 
\sin \theta_{23}  \simeq |V_{cb}| \simeq  \left|{y_{23}^d \over y_{33}^d}   \epsilon^2 -{y_{23}^u \over y_{33}^u} \epsilon^8  \right| ,\nonumber \\
\sin \theta_{13}  \simeq |V_{ub}| &\simeq& \left|{y_{13}^d \over y_{33}^d}  \epsilon^4  -{y_{12}^u y_{23}^d \over y_{22}^u y_{33}^d}       \epsilon^4
- {y_{13}^u \over y_{33}^u}  \epsilon^8 \right|  .
\end{eqnarray}

\subsection{Neutrino masses and mixing}
\label{neutrino_masses2}
The masses and mixing of neutrinos in the non-minimal model is identical to that of the minimal model.  Thus, we write the Majorana Lagrangian for right-handed neutrinos as,
\bea
\mathcal{L}_{\rm M_R}  &=& c_{ij} \left[  \frac{\chi}{\Lambda}  \right]^3 \chi \bar{\nu^c}_{i,R} \nu_{j,R},
\eea
where $i,j$ are flavour indices.

The Majorana mass matrices $\M_{R}$ can be written as,

\begin{equation}
\label{NM}
\M_{R} =  M
\begin{pmatrix}
c_{11}& c_{12}    & c_{13} \\
  c_{12}  & c_{22} & c_{23} \\
c_{13}   & c_{23}   & c_{33}
\end{pmatrix},
\end{equation}

where $ M=  \langle \chi \rangle   \left[  \frac{  \langle \chi \rangle}{\Lambda} \right]^3 = \frac{f}{\sqrt{2}} \epsilon^3 $.

The Dirac mass Lagranginan for neutrinos is, 

\begin{eqnarray}
\label{mass1}
-{\mathcal{L}}_{\rm Yukawa}^{\nu} &=&       y_{11}^\nu \bar{ \psi}_{L_1}^\ell  H \nu_{e_R}  \left[  \dfrac{ \chi}{\Lambda} \right]^{3} + y_{12}^\nu \bar{ \psi}_{L_1}^\ell  H \nu_{\mu_R}  \left[  \dfrac{ \chi}{\Lambda} \right]^{3} + y_{13}^\nu \bar{ \psi}_{L_1}^\ell  H   \nu_{\tau_R}   \left[  \dfrac{ \chi}{\Lambda} \right]^{3} + y_{21}^\nu \bar{ \psi}_{L_2}^\ell  H \nu_{e_R} \left[  \dfrac{ \chi}{\Lambda} \right] \\ \nonumber
&+& y_{22}^\nu \bar{ \psi}_{L_2}^\ell  H \nu_{\mu_R} \left[  \dfrac{ \chi}{\Lambda} \right] + y_{23}^\nu \bar{ \psi}_{L_2}^\ell   H \nu_{\tau_R}   \left[  \dfrac{ \chi}{\Lambda} \right]
 + y_{31}^\nu \bar{ \psi}_{L_3}^\ell  H \nu_{e_R} \left[  \dfrac{ \chi^\dagger }{\Lambda} \right] + y_{32}^\nu \bar{ \psi}_{L_3}^\ell H \nu_{\mu_R} \left[  \dfrac{ \chi^\dagger}{\Lambda} \right] \\ \nonumber
&+& y_{33}^\nu \bar{ \psi}_{L_3}^\ell  H \nu_{\tau_R}  \left[  \dfrac{ \chi^\dagger}{\Lambda} \right]
+  {\rm H.c.}.
\end{eqnarray}

The Dirac mass matrix for neutrinos now reads,

\begin{equation}
\label{NM}
\M_{\D} = \dfrac{v}{\sqrt{2}}
\begin{pmatrix}
y_{11}^\nu  \epsilon^3 &  y_{12}^\nu \epsilon^3 & y_{13}^\nu   \epsilon^{3}  \\
y_{21}^\nu  \epsilon  & y_{22}^\nu \epsilon &  y_{23}^\nu   \epsilon \\
y_{31}^\nu \epsilon   &  y_{32}^\nu  \epsilon   &  y_{33}^\nu \epsilon
\end{pmatrix}.
\end{equation}

The  mass matrix  of neutrinos after including the Majorana mass term is,
\begin{equation}
\label{NM}
\M = 
\begin{pmatrix}
\M_{L}   &  \M_{\D} \\
\M_{\D}^T & \M_{R}   \\
\end{pmatrix}.
\end{equation}

The  light neutrino mass matrix is,
\begin{eqnarray}
{ \M}~ & \approx & ~ -  \M_{\D}  \M_{R}^{-1}  \M_{\D}^T,  \\ \nonumber
& \approx & \frac{v}{\sqrt{2} } \epsilon^\prime
\left(
\begin{array}{ccc}
 -\frac{\epsilon^4 \left(c_{22} c_{33} y_{11}^{\nu 2}-2
  c_{22} y_{11}^\nu +c_{22} -2c_{33}
  y_{11}^\nu +c_{33}-y_{11}^{\nu 2}+4
  y_{11}^\nu -2\right)}{(c_{22}-1) (c_{33}-1)} &
   -\epsilon^2 y_{11}^\nu & -\frac{\epsilon^2 ( c_{33}
  y_{11}^\nu- y_{11}^\nu
  y_{33}^\nu + y_{33}^\nu -1)}{c_{33}-1} \\
 -\epsilon^2 y_{11}^\nu & -1 & -1 \\
 -\frac{\epsilon^2 (c_{33} y_{11}^\nu - y_{11}^\nu
  y_{33}^\nu + y_{33}^\nu -1)}{c_{33}-1} & -1 &
   -\frac{ (c_{33}+(y_{33}^\nu-2)
  y_{33}^\nu )}{c_{33} -1} \\
\end{array}
\right),
\end{eqnarray}

where $\epsilon^{\prime } = \frac{v}{ f \epsilon} $,  and  we have again assumed each and every coupling exactly one except those appearing in above equation.
 
The neutrino  masses  approximately are,
\bea
m_1 &\approx &  \frac{ \left(-y_{11}^{\nu 2}+2
  y_{11}^\nu -1\right)}{c_{22} -1}  \epsilon^{4} \epsilon^{\prime } v/\sqrt{2}, \\ \nonumber
  m_2 & \approx&    \frac{ \left(-\sqrt{4 c_{33}^2-8
   c_{33}+y_{33}^{\nu 4}-4 y_{33}^{\nu 3}+6 y_{33}^{\nu 2}-4
   y_{33}^\nu +5}-2 y_{33}^\nu -y_{33}^{\nu 2}+2
   y_{33}^\nu +1\right)}{2 (c_{33}-1)}  ~ \epsilon^{\prime }  v/\sqrt{2}, \\ \nonumber
  m_3 & \approx&    \frac{ \left(\sqrt{4 c_{33}^2-8
   c_{33}+y_{33}^{\nu 4}-4 y_{33}^{\nu 3}+6 y_{33}^{\nu 2}-4
   y_{33}^\nu +5}-2 c_{33}- y_{33}^{\nu 2} +2
   y_{33}^\nu +1\right)}{2 (c_{33}-1)}  ~ \epsilon^{\prime }  v/\sqrt{2}.
\eea

The neutrino mixing angles are,
\begin{eqnarray}
\sin \theta_{12}  &\simeq& \left|{y_{12}^\ell \over y_{22}^\ell}  -y_{11}^\nu \right| \epsilon^2, \\ \nonumber
\sin \theta_{23} &\simeq & \left|{1-c_{33} \over  c_{33}+(y_{33}^\nu -2)y_{33}^\nu}  \right|, \\ \nonumber
\sin \theta_{13} &\simeq& \left|  \frac{ ( c_{33}
 y_{11}^\nu-y_{11}^\nu
  y_{33}^\nu + y_{33}^\nu  -1)}{ c_{33}+(y_{33}^\nu -2 )y_{33}^\nu}  \right|  \epsilon^2.
\end{eqnarray}

\section{The scalar potential}
\label{potential}
The scalar potential of the model can be written in the following form,
\begin{align}
- \lag_\text{potential}
=- \mu^2 \varphi^\dagger \varphi +\lambda (\varphi^\dagger \varphi)^2 - \mu_\chi^2\, \chi^*  \chi
  + \lambda_\chi\, (\chi^* \chi)^2 
  +  \, (\rho  \ \chi^2 + \rm H.c. ) 
  + \lambda_{\varphi  \chi}  (\chi^* \chi)  (\varphi^\dagger \varphi),
\label{eq:potential}
\end{align}
where we have introduced a soft breaking of the $\mathcal{Z}_5$ symmetry in the fifth term.  We are assuming  $ \lambda_{\varphi  \chi}   =0$, i.e., no Higgs-flavon mixing\cite{Bauer:2016rxs}.  If this term is non-zero, the phenomenology of the flavon field will be different,  for instance,  see reference\cite{Berger:2014gga}. The only parameter which can have a phase in the scalar potential  is $\rho$.  However, this phase can be removed by a phase rotation of the flavon field $\chi$ leading to a real value of the VEV of the field $\chi$.

We can parametrize the flavon field  by excitations around its VEV,
\begin{align}
 \chi(x)=\frac{f + s(x) +i\, a(x)}{\sqrt{2}}.
\label{chi}
\end{align}

In a similar manner, the Higgs field can be written as,
\begin{align}
 \varphi (x)=\frac{v+ h(x) }{\sqrt{2}}.
\label{phi}
\end{align}

The minimization conditions can be written in terms of the scalar and pseudo-scalar  components having the following masses:
\begin{align}
m_s =\sqrt{\mu_\chi- 2 \rho} = \sqrt{\lambda_\chi} f 
\qqquad \text{and} \qqquad
m_a= \sqrt{-2 \rho}.
\label{eq:masses}
\end{align}

We observe that the mass of the pseudoscalar component of the flavon field depends on the soft-breaking parameter $\rho$.  Therefore, it is a free parameter of the model.   Now using equation \ref{chi},  we can write 
\begin{equation}
\label{eps}
\frac{\chi}{\Lambda} = \epsilon  [ 1 +  \frac{s + i a}{f} ].
\end{equation}

The couplings of the scalar and pseudoscalar components of the flavon field are obtained  from equation \ref{mass1} by writing the effective Yukawa couplings in the following form:
\begin{equation}
\label{coupling}
Y_{ij}^f \varphi = y_{ij}^f  \left(\frac{\chi}{\Lambda} \right)^{n^{f}_{ij}} \left(\frac{v+ h }{\sqrt{2}}.\right) 
\cong  y_{ij}^f  \epsilon^{n^{f}_{ij}}  \frac{v}{\sqrt{2}} \left[1 + \frac{n^f_{ij} (s +  i  a )}{f} + \frac{h}{v}\right] = \M_f \left[1 + \frac{n^f_{ij} (s +  i  a )}{f} + \frac{h}{v}\right],
\end{equation}
where $f= u,d,\ell$, and $n^{f}_{ij}$ is the power of the parameter $\epsilon$ appearing in the mass matrices $\M_f$.

We note that for our phenomenological investigation, we have only retained the terms linear in the flavon field components $s$ and $a$ in equation \ref{coupling}.  The terms which are higher than the linear terms are not interesting in the present work.   The couplings of the Higgs boson field $h$ to the charged fermions are real and diagonal since the mass matrices $\M_f$ can be diagonalized resulting in real and positive masses of the charged fermions.  However, the couplings of the scalar and pseudoscalar components $s$ and $a$  of the flavon field are given by $n^f_{ij} \M_f$.  This product cannot be diagonalized exactly and as a consequence, the couplings of $s$ and $a$ cannot be made real and diagonal.  This in turn gives rise to the flavour-changing and $CP$-violating interactions of the flavon field.  

The couplings of $a$ field with fermions for minimal $\mathcal{Z}_2 \times \mathcal{Z}_5$ symmetry are now given by,
\begin{eqnarray}
\label{fup}
y_{af_{iL} f_{jR}}^{u} &\equiv & y_{aij}^u = \frac{v}{\sqrt{2}f} 
\begin{pmatrix}
4 y_{11}^u  \epsilon^4 & 4 y_{12}^u \epsilon^4  &  4 y_{13}^u \epsilon^4    \\
2 y_{21}^u  \epsilon^2    & 2 y_{22}^u \epsilon^2  &  2 y_{23}^u \epsilon^2     \\
0   & 0   & 0
\end{pmatrix},
y_{aij}^{d} = \frac{v}{\sqrt{2}f} 
\begin{pmatrix}
5 y_{11}^d  \epsilon^5 & 5 y_{12}^d \epsilon^5 & 5 y_{13}^d \epsilon^5   \\
3 y_{21}^d  \epsilon^3  & 3 y_{22}^d \epsilon^3 & 3 y_{23}^d \epsilon^3  \\
  y_{31}^d \epsilon &   y_{32}^d \epsilon   &  y_{33}^d \epsilon
\end{pmatrix}, \\ \nonumber
y_{aij}^{\ell} &=& \frac{v}{\sqrt{2}f} 
\begin{pmatrix}
5 y_{11}^\ell  \epsilon^5 & 5 y_{12}^\ell \epsilon^5 & 5 y_{13}^\ell \epsilon^5   \\
3 y_{21}^\ell  \epsilon^3  & 3 y_{22}^\ell \epsilon^3 & 3 y_{23}^\ell \epsilon^3  \\
  y_{31}^\ell \epsilon &   y_{32}^\ell \epsilon   &  y_{33}^\ell \epsilon
\end{pmatrix}.
\end{eqnarray}
In the similar way, the couplings of $a$ field with fermions for non-minimal $\mathcal{Z}_2 \times \mathcal{Z}_9$ symmetry are given by\begin{eqnarray}
\label{fup_z2z9}
y_{af_{iL} f_{jR}}^{u} &\equiv & y_{aij}^u = \frac{v}{\sqrt{2}f} 
\begin{pmatrix}
8 y_{11}^u  \epsilon^8 & 6 y_{12}^u \epsilon^6  &  8 y_{13}^u \epsilon^8    \\
8 y_{21}^u  \epsilon^8    & 4 y_{22}^u \epsilon^4  &  8 y_{23}^u \epsilon^8  \\
0   & 4 y_{32}^u \epsilon^4   & 0
\end{pmatrix},
y_{aij}^{d} = \frac{v}{\sqrt{2}f} 
\begin{pmatrix}
7 y_{11}^d  \epsilon^7 & 7 y_{12}^d \epsilon^7 & 7 y_{13}^d \epsilon^7   \\
5 y_{21}^d  \epsilon^5  & 5 y_{22}^d \epsilon^5 & 5 y_{23}^d \epsilon^5  \\
 3 y_{31}^d \epsilon^3 &  3 y_{32}^d \epsilon^3   &  3 y_{33}^d \epsilon^3
\end{pmatrix}, \\ \nonumber
y_{aij}^{\ell} &=& \frac{v}{\sqrt{2}f} 
\begin{pmatrix}
7 y_{11}^\ell  \epsilon^7 & 7 y_{12}^\ell \epsilon^7 & 7 y_{13}^\ell \epsilon^7   \\
5 y_{21}^\ell  \epsilon^5  & 5 y_{22}^\ell \epsilon^5 & 5 y_{23}^\ell \epsilon^5  \\
  3 y_{31}^\ell \epsilon^3 &  3 y_{32}^\ell \epsilon^3   & 3 y_{33}^\ell \epsilon^3
\end{pmatrix}.
\end{eqnarray}

For the pseudoscalar component of flavon field, the following notation is used:
\begin{equation}
y_{ij}= y_{sf_{iL} f_{iR}} = - i  y_{af_{iL} f_{iR}}. 
\end{equation}
The couplings of $a$ to fermions are identical to that of $s$ except with a relative phase factor $i$. This factor becomes trivial  in the squared amplitude of a Feynman diagram mediated by $a$.  Thus, it does not give rise to $CP$-violating interactions to the order investigated in this work.

\section{Quark flavour physics in the minimal and non-minimal $\mathcal{Z}_2 \times \mathcal{Z}_N$ flavour symmetry}
\label{quark_flavour}
The quark flavour physics places stronger bounds on the parameter space of our model,  which is parametrized by the VEV of the flavon field $f$,    the mass of the pseudoscalar flavon $m_a$, and the quartic coupling $\lambda_\chi$.  In particular,  measurement of  the loop-induced processes in the SM such as neutral meson mixing and rare mesonic decays constrain the parameter space of the model.   The numerical inputs used in this work are given in table \ref{tab5}.

\subsection{Neutral meson mixing}
The non-diagonal  couplings of the flavon to fermions introduce the FCNC interactions at tree-level.  Therefore,  they are expected to be highly suppressed from neutral meson-antimeson mixing.  These interactions of the flavon can be parametrized by writing the   $\Delta F =2$  effective Hamiltonian as follows,
\begin{align}
\Heff^{\Delta F=2}&=C_1^{ij} \,( \bar q^i_L\,\gamma_\mu \, q^j_L)^2+\widetilde C_1^{ij} \,( \bar q^i_R\,\gamma_\mu \, 
q^j_R)^2 +C_2^{ij} \,( \bar q^i_R \, q^j_L)^2+\widetilde C_2^{ij} \,( \bar q^i_L \, q^j_R)^2\notag\\
&+ C_4^{ij}\, ( \bar q^i_R \, q^j_L)\, ( \bar q^i_L \, q^j_R)\,+C_5^{ij}\, ( \bar q^i_L \,\gamma_\mu\, q^j_L)\, ( \bar q^i_R \,
\gamma^\mu q^j_R)\,+ \text{H.c.}, 
\label{eq:heffdf2}
\end{align}
where $q_{R,L} = \frac{1\pm \gamma_5}{2} q$ and the colour indices are omitted for simplicity.    

\begin{table}[H]
\begin{center}
\begin{tabular}{|c|c || c|c|}\hline
$G_{F}$ & $1.166 \times 10^{-5}$ \,$~{\rm GeV}$ \cite{Zyla:2021} &$v$  & 246.22  \,GeV \cite{Zyla:2021}  \\  \cline{1-4}
     
     $\alpha_{s}[M_{Z}]$ & $0.1184$ \cite{Y.Aoki:2021} & $m_{u}$  & $ (2.16^{+0.49}_{-0.26} )\times 10^{-3}$~{\rm GeV} \cite{Zyla:2021} \\  \cline{1-2}
     
$M_{W}$  & $80.387 \pm 0.016$ \,GeV \cite{Zyla:2021} & $m_{d}$ & $ (4.67^{+0.48}_{-0.17}) \times 10^{-3}$~{\rm GeV} \cite{Zyla:2021} \\   \cline{1-2}     
      $f_{K}$  &  $159.8$ \,MeV \cite{Ciuchini:1998ix} & $m_{c}$ & $ 1.27 \pm 0.02 $~{\rm GeV} \cite{Zyla:2021}  \\  
     
      $m_{K}$  &  $497.611 \pm 0.013$ \,MeV \cite{Zyla:2021}  & $m_{s}$ & $ 93.4^{+8.6}_{-3.4}$~{\rm GeV}  \cite{Zyla:2021} \\ 
    $\hat{B_{K}}$  & $0.7625$  \cite{Y.Aoki:2021}  & $m_{t}$ & $ 172.69 \pm 0.30 $~{\rm GeV} \cite{Zyla:2021} \\ 
   $B_{1}^K$  & $0.60 (6)$  \cite{Ciuchini:1998ix}  & $m_{b}$ & $ 4.18^{+0.03}_{-0.02}$~{\rm GeV} \cite{Zyla:2021} \\ 
     
    $B_{2}^K$ & $0.66 (4)$ \cite{Ciuchini:1998ix} & $m_{c}(m_{c})$ & $ 1.275$~{\rm GeV}  \\ 
    $B_{3}^K$ & $1.05 (12)$ \cite{Ciuchini:1998ix}  &  $m_{b}(m_{b})$ & $ 4.18$~{\rm GeV} \\  
     $B_{4}^K$ & $ 1.03 (6)$ \cite{Ciuchini:1998ix}  &  $m_{t}(m_{t})$ & $ 162.883$~{\rm GeV}  \\ \cline{3-4}
     $B_{5}^K$ & $0.73 (10)$ \cite{Ciuchini:1998ix}  &$\alpha$ & $ 1/137.035$ \cite{Zyla:2021} \\ 
    
    $\eta_{1}$ & $1.87 \pm 0.76$ \cite{Brod:2012}& $e$ & $0.302862$~{\rm GeV} \\  
   $\eta_{2}$ & $0.574$ \cite{Buchalla:1996}& $m_{e}$ & $ 0.51099 $~{\rm MeV} \cite{Zyla:2021} \\
   $\eta_{3}$ & $0.496 \pm 0.047$ \cite{Brod:2010}& $m_{\mu}$ & $ 105.65837$~{\rm MeV} \cite{Zyla:2021} \\ \cline{1-2}
    
 $f_{B_{s}}$  &  $230.3 $ \,MeV \cite{Y.Aoki:2021} & $m_{\tau}$ & $ 1776.86 \pm 0.12$~{\rm MeV} \cite{Zyla:2021}  \\ 
 
$m_{B_{s}}$  &  $5366.88$ \,MeV \cite{Zyla:2021} & $\tau_{\mu}$ & $ 2.196811 \times 10^{-6}$~{\rm sec} \cite{Zyla:2021}\\ 

$\hat{B_{B_{s}}}$  & $1.232$\cite{Y.Aoki:2021}  & $\tau_{\tau}$ & $ (290.3 \pm 0.5) \times 10^{-15}$~{\rm sec} \cite{Zyla:2021}  \\

$B_{1}^{B_{s}}$ & $0.86 (2) (^{+5} _{-4})$ \cite{Becirevic:2001xt} & $m_{p}$ & $ 938.272$~{\rm MeV} \cite{Zyla:2021} \\
$B_{2}^{B_{s}}$ & $0.83 (2) (4)$ \cite{Becirevic:2001xt}& $m_{n}$ & $ 939.565$~{\rm MeV} \cite{Zyla:2021} \\ \cline{3-4}
$B_{3}^{B_{s}}$ & $1.03 (4) (9)$ \cite{Becirevic:2001xt}& $m_{D}$ & $ 1864.83$~{\rm MeV} \cite{Zyla:2021}\\
$B_{4}^{B_{s}}$ & $1.17 (2) (^{+5} _{-7})$ \cite{Becirevic:2001xt}& $f_{D}$ & $ 212$~{\rm MeV}\cite{Y.Aoki:2021}\\ 
$B_{5}^{B_{s}}$ & $1.94 (3) (^{+23} _{-7})$ \cite{Becirevic:2001xt}& $B_{1}^{D}$ & $ 0.861$ \cite{Bona:2007vi} \\ \cline{1-2}
$\eta_{2B}$ & $0.551$ \cite{Buchalla:1996}& $B_{2}^{D}$ & $ 0.82$ \cite{Bona:2007vi}\\ \cline{1-2}

$f_{B_{d}}$  &  $190.0 $ \,MeV \cite{Y.Aoki:2021} & $B_{3}^{D}$ & $ 1.07$ \cite{Bona:2007vi}\\ 
 
$m_{B_{d}}$  &  $5279.65$ \,MeV \cite{Zyla:2021}&  $B_{4}^{D}$ & $ 1.08$ \cite{Bona:2007vi}\\ 
$\hat{B_{B_{d}}}$  & $1.222$\cite{Y.Aoki:2021}& $B_{5}^{D}$ & $ 1.455$ \cite{Bona:2007vi}\\ \cline{3-4}
$B_{1}^{B_{d}}$ & $0.87 (4) (^{+5} _{-4})$ \cite{Becirevic:2001xt}& $\tau_{B_{d}}$ & $ (1.520 \pm 0.004)\times 10^{-12}$ ~{\rm sec} \cite{HFLAV:2016hnz} \\
$B_{2}^{B_{d}}$ & $0.82 (3) (4)$ \cite{Becirevic:2001xt}& $\tau_{B_{s}}$ & $ (1.505 \pm 0.005) \times 10^{-12}$ ~{\rm sec} \cite{HFLAV:2016hnz}\\
$B_{3}^{B_{d}}$ &$1.02 (6) (9)$ \cite{Becirevic:2001xt}& $\tau_{K_{L}}$ & $ (5.116 \pm 0.021) \times 10^{-8}$~{\rm sec} \cite{Zyla:2021}\\ 

$B_{4}^{B_{d}}$ & $1.16 (3) (^{+5} _{-7})$ \cite{Becirevic:2001xt} & $\tau_{D}$ & $ (410.1 \pm 1.5) \times 10^{-15}$~{\rm sec} \cite{Zyla:2021} \\
 
$B_{5}^{B_{d}}$ & $1.91 (4) (^{+22} _{-7})$ \cite{Becirevic:2001xt} &  &   \\   
\hline  
\end{tabular} \\
\end{center}

\caption{Values of the experimental and theoretical quantities used as input parameters.}
   \label{tab5}
   \end{table}

The tree-level contribution to neutral meson mixing due to the flavon exchange  gives rise to the following  Wilson coefficients~\cite{Buras:2013rqa,Crivellin:2013wna},
\begin{align}
C_2^{ij} &= -(y_{ji}^*)^2\left(\frac{1}{m_s^2}-\frac{1}{m_a^2}\right)\notag \\
\tilde C_2^{ij} &= -y_{ij}^2\left(\frac{1}{m_s^2}-\frac{1}{m_a^2}\right)\notag \\
C_4^{ij} &= -\frac{y_{ij}y_{ji}}{2}\left(\frac{1}{m_s^2}+\frac{1}{m_a^2}\right),
\label{eq:wilsons}
\end{align}
where $m_{s}$ and $m_{a}$ are the masses of scalar and pseudoscalar component of flavon field,  respectively.

The Wilson coefficients $C_i$ are computed at a scale $\Lambda$,   where heavier new degrees of freedom are integrated out.   They need to be evolved down to the   hadronic scales $4.6$~GeV for bottom mesons,  $2.8$~GeV for charmed mesons,  and $2$~GeV for kaons.  These particular scales are used in the lattice computations of the corresponding matrix elements
\cite{Ciuchini:1998ix, Becirevic:2001xt, Bona:2007vi}.   In this work,  renormalization group running of the matrix elements is implemented  as discussed in reference~\cite{Bona:2007vi} and matrix elements are taken from reference~\cite{Ciuchini:1998ix, Becirevic:2001xt}.   Thus,  the new physics  contribution to the  $B_q -\bar B_q$ mixing amplitudes due to the Wilson coefficients $C_i$ at a scale $\Lambda$ can be written as~\cite{Bona:2007vi},
\begin{equation}
\label{eq:magicbb}
\langle \bar B_q \vert {\cal H}_{\rm eff}^{\Delta B=2} \vert B_q \rangle_i = 
\sum_{j=1}^5 \sum_{r=1}^5
             \left(b^{(r,i)}_j + \eta \,c^{(r,i)}_j\right)
             \eta^{a_j} \,C_i(\Lambda)\, \langle \bar B_q \vert Q_r^{bq}
             \vert B_q \rangle\,,
\end{equation}
where $q=d,s$,  $\alpha_s$ is the strong coupling constant,  $\eta=\alpha_s(\Lambda)/\alpha_s(m_t)$,  and   $a_j$, $b^{(r,i)}_j$,  $c^{(r,i)}_j$  are the so-called the magic numbers which are taken from reference~\cite{Becirevic:2001jj}.   We can write a similar formula for  $D^0 - \bar D^0$ mixing with magic numbers given in reference~\cite{Bona:2007vi}.  For  $K^0-\bar K^0$ mixing, the formula becomes~\cite{Bona:2007vi},
\begin{equation}
  \label{eq:kkbsm}
  \langle \bar K^0 \vert {\cal H}_{\rm eff}^{\Delta S=2} \vert K^0 \rangle_i = 
  \sum_{j=1}^5 \sum_{r=1}^5
  \left(b^{(r,i)}_j + \eta \,c^{(r,i)}_j\right)
  \eta^{a_j} \,C_i(\Lambda)\, R_r \, \langle \bar K^0 \vert Q_1^{sd}
  \vert K^0 \rangle,
\end{equation}
where $R_r$ are the ratio of the matrix elements of NP operators over that of SM \cite{Donini:1999nn} and their numerical values are  directly taken from reference  \cite{Bona:2007vi} for our analysis.  The magic numbers for $K^0-\bar K^0$ mixing are taken from reference \cite{Ciuchini:1998ix}.

The mixing observables of the  $K^0-\bar K^0$ mixing can be used now to constrain the flavon mass and VEV by employing their experimental measurements.  These are \cite{Bona:2007vi},
\begin{align}
C_{\eps_K}&=\frac{ \text{Im} \langle K^0|\mathcal{H}_\text{eff}^{\Delta F=2}|\bar K^0\rangle}{\text{Im} \langle K^0| \mathcal{H}_\text{SM}^{\Delta F=2} |\bar K^0 \rangle} = 1.12_{-0.25}^{+0.27},
C_{\Delta m_K} =\frac{\text{Re}\langle K^0|\mathcal{H}_\text{eff}^{\Delta F=2}|\bar K^0\rangle}{\text{Re} \langle K^0| \mathcal{H}_\text{SM}^{\Delta F=2} |\bar K^0 \rangle} = 0.93_{-0.42}^{+1.14} ,
\end{align}
where numbers are given at $95\%$ C.L.,  $\mathcal{H}_\text{eff}^{\Delta F=2}$ contains the SM and flavon contributions,  and  $\mathcal{H}_\text{SM}^{\Delta F=2}$ represents only  the SM contribution.

The mixing observables for the  $B_q -\bar B_q$ mixing  are,
\begin{align*}
C_{B_{q}}e^{2i\phi_{B_{q}}}&=\frac{ \text{Im} \langle B_{q}^0|\mathcal{H}^{\Delta F=2}|\bar B_{q}^0\rangle}{\text{Im} \langle B_{q}^0| \mathcal{H}_\text{SM}^{\Delta F=2} |\bar B_{q}^0 \rangle} 
\end{align*}
where $q=s,d$ for $B_{s}$ and $B_{d}$ mixing respectively.   The following  measurements at 95 \% CL limits are used in this work \cite{Bona:2007vi},
\begin{align*}
C_{B_{s}}= 1.110 \pm 0.090 \hspace{0.1cm} [0.942, 1.288] ,\hspace{0.5cm} \phi_{B_{s}}^o= 0.42 \pm 0.89 \hspace{0.1cm} [-1.35, 2.21]\\
C_{B_{d}}= 1.05 \pm 0.11 \hspace{0.1cm} [0.83, 1.29], \hspace{0.8cm} \phi_{B_{d}}^o= -2.0 \pm 1.8 \hspace{0.1cm} [ -6.0, 1.5]
\end{align*}

 The new physics contributions to neutral meson mixing can be written as,
 \be
 M_{12}^{d,s,K} =  (M_{12}^{d,s,K} )_{\rm SM} \left(  1 +  h_{d,s,K} e^{2 i \sigma_{d,s,K}}    \right).
  \ee
  
We assume the minimum flavour violation scenario that corresponds to $\sigma_{d,s,K} =0$.     We adopt the following future sensitivity phases in this work\cite{Charles:2020dfl}:
\begin{enumerate}
\item
Phase \rom{1} which is $50 fb^{-1}$ LHCb and $50 ab^{-1}$ Belle \rom{2} (late 2020s);
\item
Phase \rom{2} which is $300 fb^{-1}$ LHCb and $250 ab^{-1}$ Belle \rom{2} (late 2030s).
\end{enumerate}

The expected sensitivities to $C_{\Delta m_K} $ and $C_{B_{q}}$   in future  phase \rom{1} and \rom{2} of LHCb and Belle \rom{2} can be obtained from  table \ref{hdsk}.
\begin{table}[tb]
\centering
\begin{tabular}{l|ccc}
\text{Observables} & \text{Phase \rom{1}} & \text{Phase \rom{2}} & \text{Ref.}  \\
\hline
$h_d$ & $0-0.04$    &   $0-0.028$    &  \cite{Charles:2020dfl}    \\
$h_s$  & $0-0.036$ &  $0-0.025$& \cite{Charles:2020dfl} \\
$h_K$  & $0-0.3$  & $-$  &    \cite{Charles:2013aka}\\
\hline
\end{tabular}
\caption{Future projected sensitivity of the neutral meson mixing. }
\label{hdsk}
\end{table}

\begin{figure}[H]
	\centering
	\begin{subfigure}[]{0.49\linewidth}
    \includegraphics[width=\linewidth]{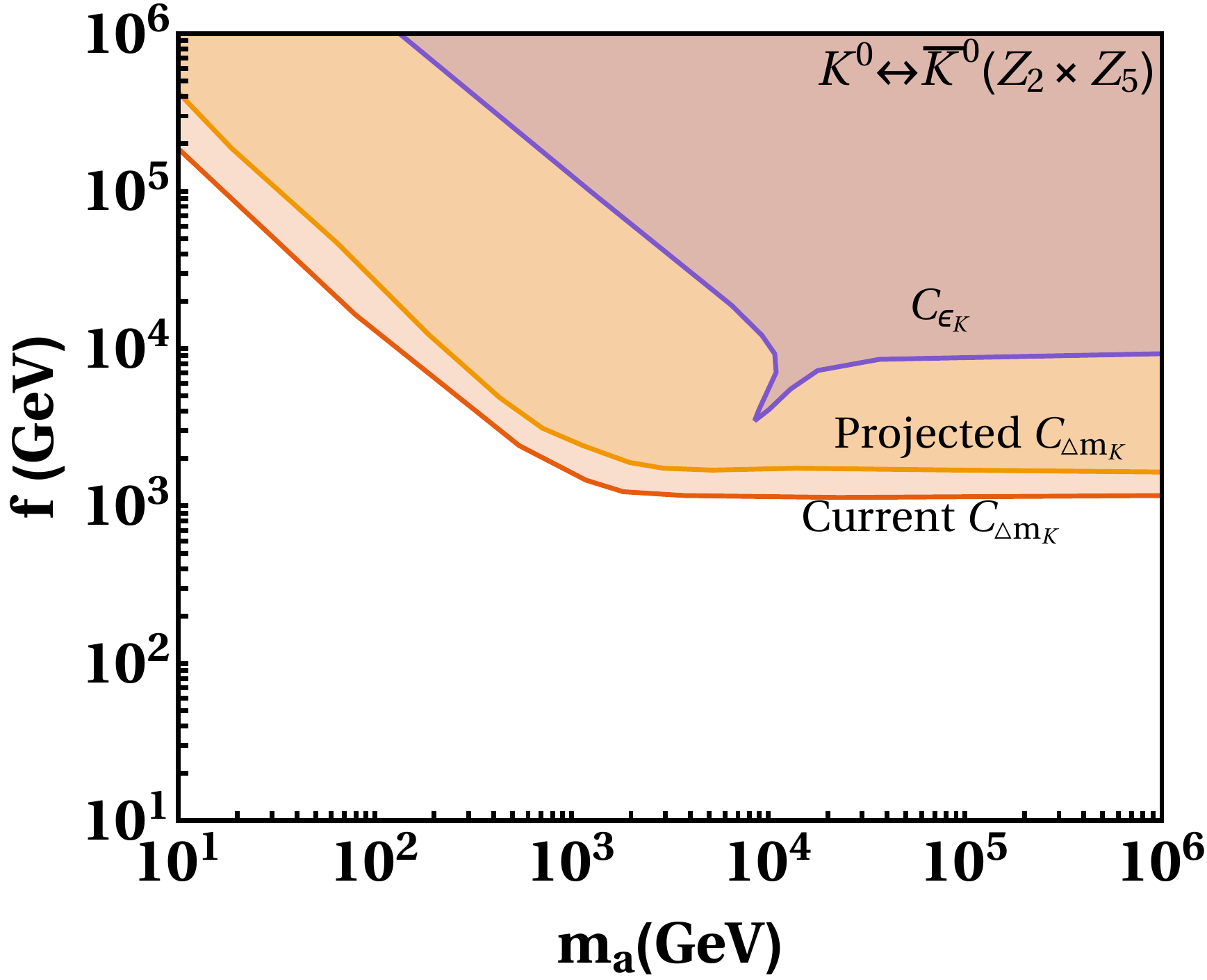}
    \caption{}
         \label{figkk1}	
\end{subfigure}
 \begin{subfigure}[]{0.49\linewidth}
 \includegraphics[width=\linewidth]{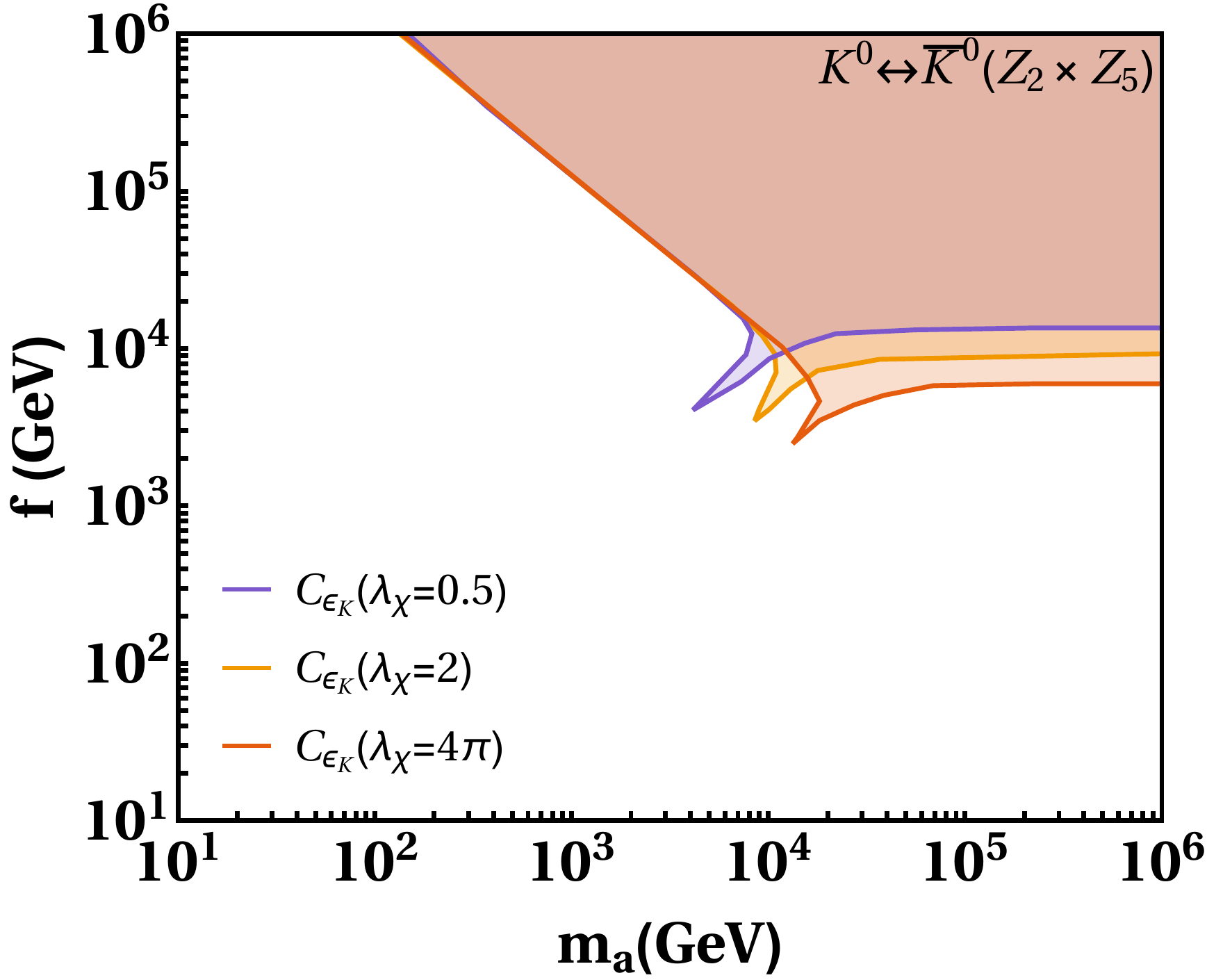}
 \caption{}
         \label{figkk3}
 \end{subfigure} 
 \caption{The allowed parameter space by flavour observables $C_{\epsilon_K}$ and  $C_{\Delta m_K}$  in the $m_a - f$ plane for the minimal  ($\mathcal{Z}_2 \times \mathcal{Z}_5$) flavour symmetry.   On the  left panel in figure \ref{figkk1},  the allowed bounds  for  $\lambda_\chi= 2$ with current limits for $C_{\Delta m_K}$ and $C_{\epsilon_K}$ are shown by red and  violet boundaries, respectively.   Also, the allowed bound with projected limits of $C_{\Delta m_K}$  is shown with  yellow boundary.  The effect of the  variation of the  quartic coupling  $\lambda_\chi$ on the observable $C_{\epsilon_K}$ is shown on the right panel  in figure \ref{figkk3}. }
  \label{figkk_z5}
	\end{figure}

In figure \ref{figkk_z5},  we show the bounds on the VEV of the flavon and the mass of the pseudo-scalar flavon arising due to the neutral kaon mixing observables $C_{\epsilon_K}$ and $C_{\Delta m_K}$ for the minimal model based on the  $\mathcal{Z}_2 \times \mathcal{Z}_5$  flavour symmetry.    On the left,  the allowed region by the observables $C_{\epsilon_K}$ and $C_{\Delta m_K}$  is shown for the quartic coupling   $\lambda_\chi= 2$.  There is a sudden dip in the allowed parameter space given by $C_{\epsilon_K}$  which appears due to a cancellation in the Wilson coefficients $C_2^{ij}$ and $\tilde C_2^{ij} $ when masses of scalar and pseudoscalar flavon become identical. For  $C_{\Delta m_K}$,  this dip  is not visible in this plot, and is excluded  for the quartic coupling   for $\lambda_\chi= 2$.  The region bounded by the yellow curve is the allowed parameter space by the future projected sensitivity as shown in table \ref{hdsk}.  On the right panel,  we show the allowed regions of the parameter space by the observable $C_{\epsilon_K}$ for  $\lambda_\chi= 0.5,~ 2,~ 4\pi$.  It is observed that the allowed region shrinks as $\lambda_\chi$ approaches smaller values.

\begin{figure}[H]
	\centering
	\begin{subfigure}[]{0.49\linewidth}
 \includegraphics[width=\linewidth]{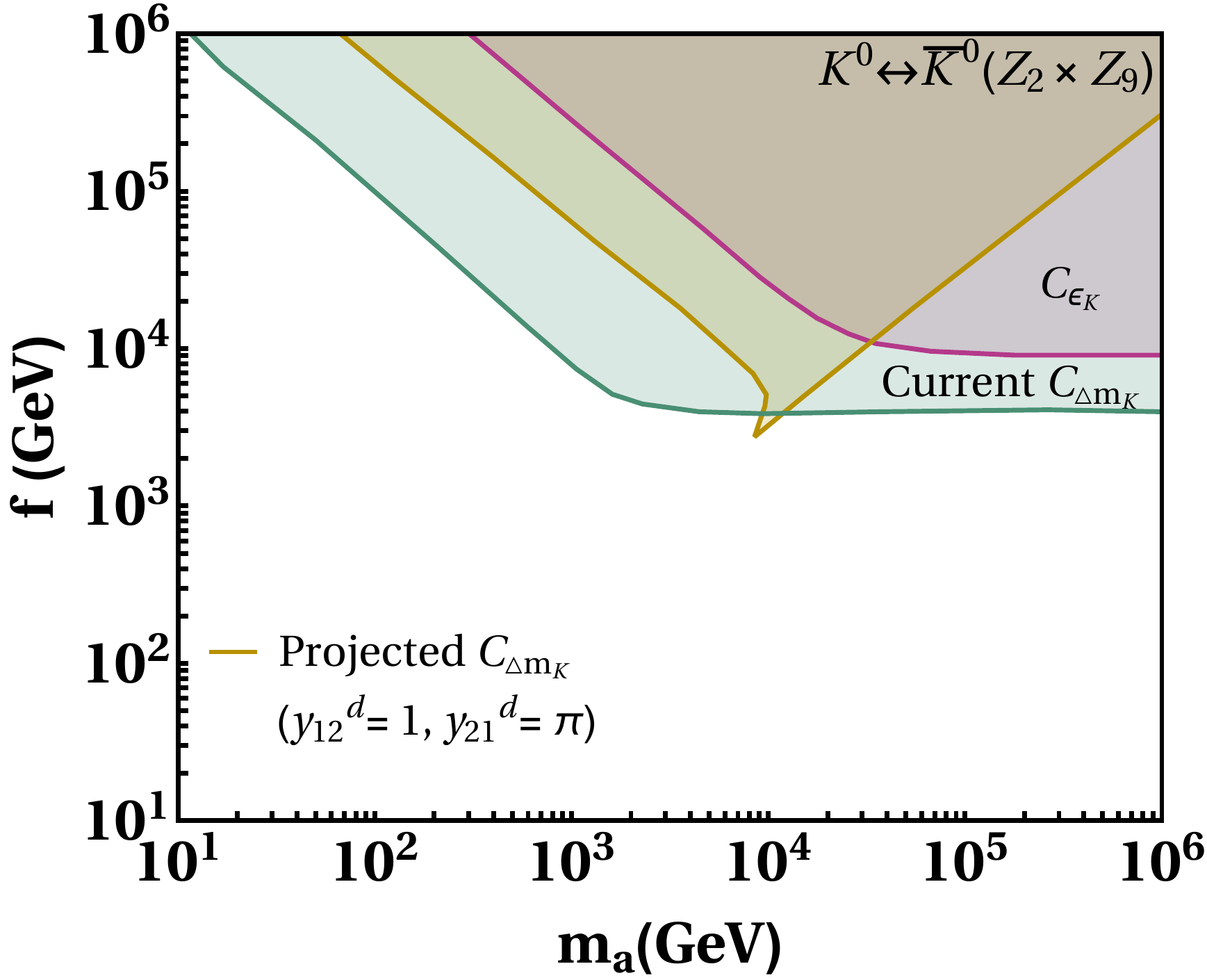}
 \caption{}
         \label{figkk2}
 \end{subfigure} 
 \begin{subfigure}[]{0.49\linewidth}
 \includegraphics[width=\linewidth]{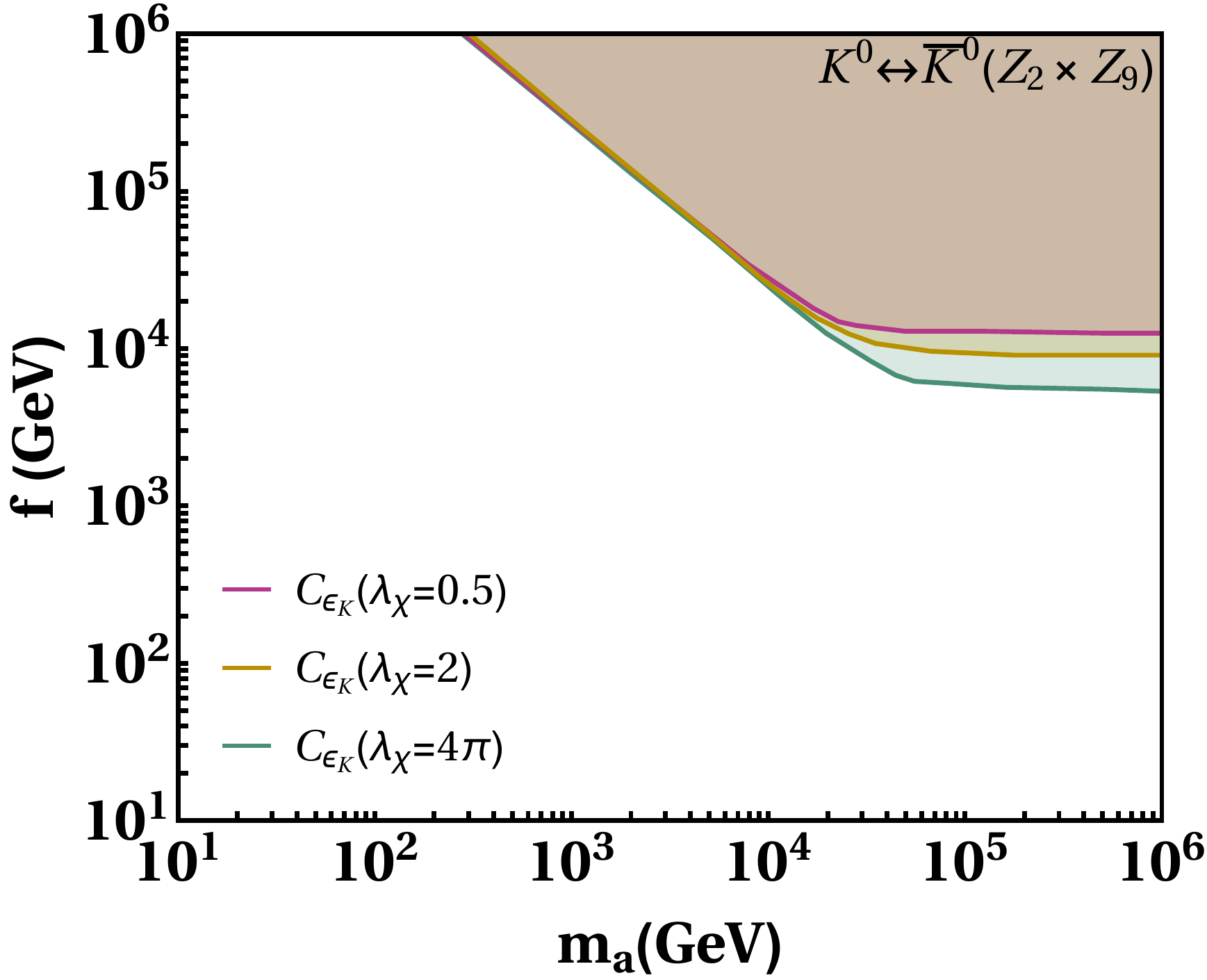}
 \caption{}
         \label{figkk4}
 \end{subfigure}
 \caption{The allowed parameter space by flavour observables $C_{\epsilon_K}$ and  $C_{\Delta m_K}$  in the $m_a - f$ plane for the non-minimal $\mathcal{Z}_2 \times \mathcal{Z}_9$ flavour symmetry.   On the  left panel in figure \ref{figkk2},  the allowed bounds  for  $\lambda_\chi= 2$ with current limits for $C_{\Delta m_K}$ and $C_{\epsilon_K}$ are shown by green and magenta boundaries,  respectively.   Also, the allowed bound with projected limits of $C_{\Delta m_K}$  is shown with  olive coloured boundary.  The effect of the  variation of the  quartic coupling  $\lambda_\chi$ on the observable $C_{\epsilon_K}$ is shown in the right panel  in figure \ref{figkk4}. }
  \label{figkkz9}
	\end{figure}

Similar results for non-minimal model based on the  $\mathcal{Z}_2 \times \mathcal{Z}_9$ flavour symmetry are shown in figure \ref{figkkz9}.  The constraints on the allowed parameter space  by the neutral kaon mixing observables $C_{\epsilon_K}$ and $C_{\Delta m_K}$, in this case, are more stringent in comparison to that of the minimal  ($\mathcal{Z}_2 \times \mathcal{Z}_5$) flavour symmetry.   In particular,   we observe that the region with the sudden dip  is excluded for the non-minimal model.   Moreover,   there is no allowed parameter space for the  future projected sensitivity of the observable $C_{\Delta m_K}$ for the bench-mark values of the couplings given in the appendix.   Therefore,  we show these bounds for different values ( $|y_{12}^d| = 1$ and $|y_{21}^d| = \pi$) of the couplings which are allowed by a more relaxed fit.

\begin{figure}[H]
	\centering
	\begin{subfigure}[]{0.49\linewidth}
	 \includegraphics[width=\linewidth]{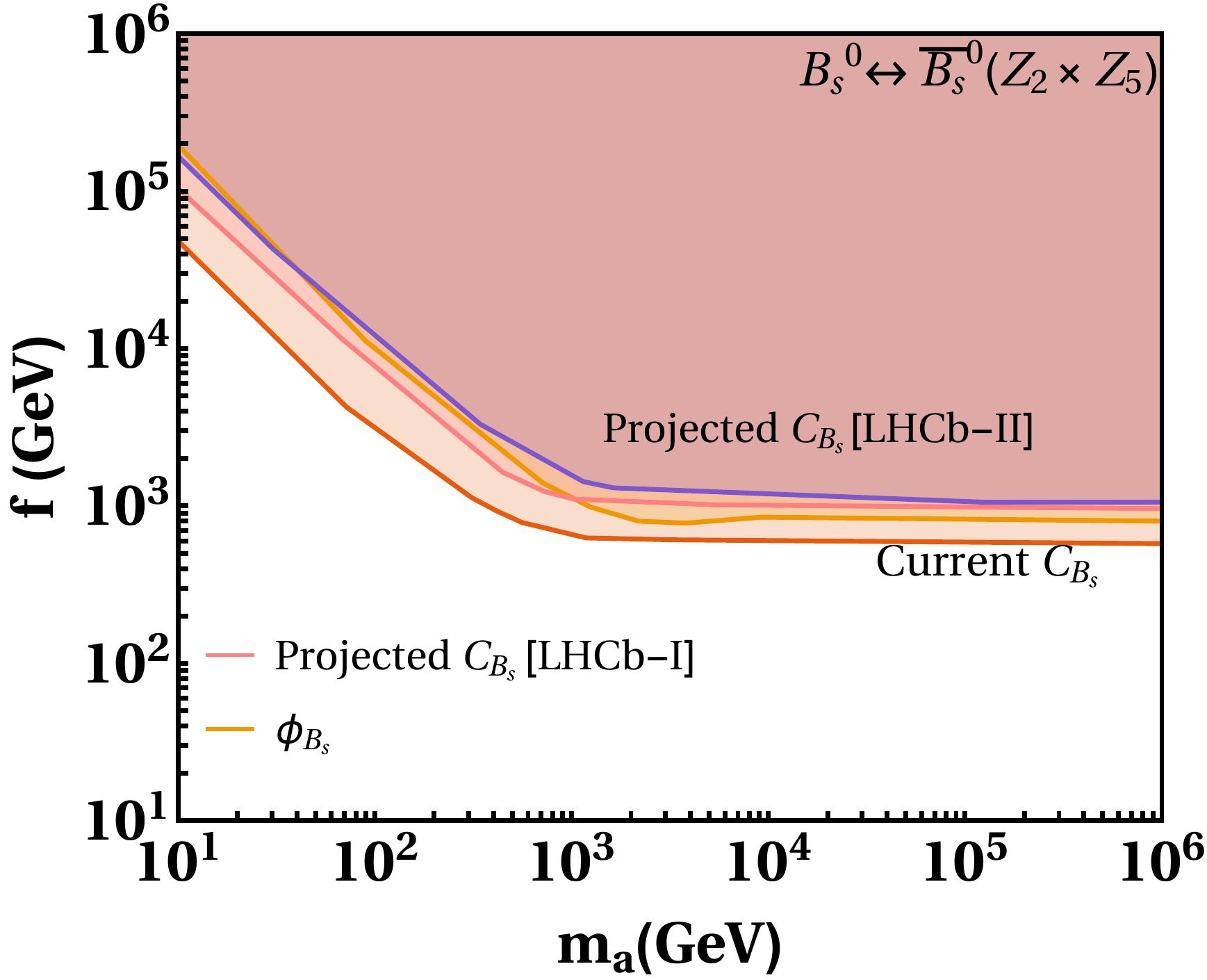}
	 \caption{}
         \label{figbsbs1}	
\end{subfigure} \begin{subfigure}[]{0.49\linewidth}
 \includegraphics[width=\linewidth]{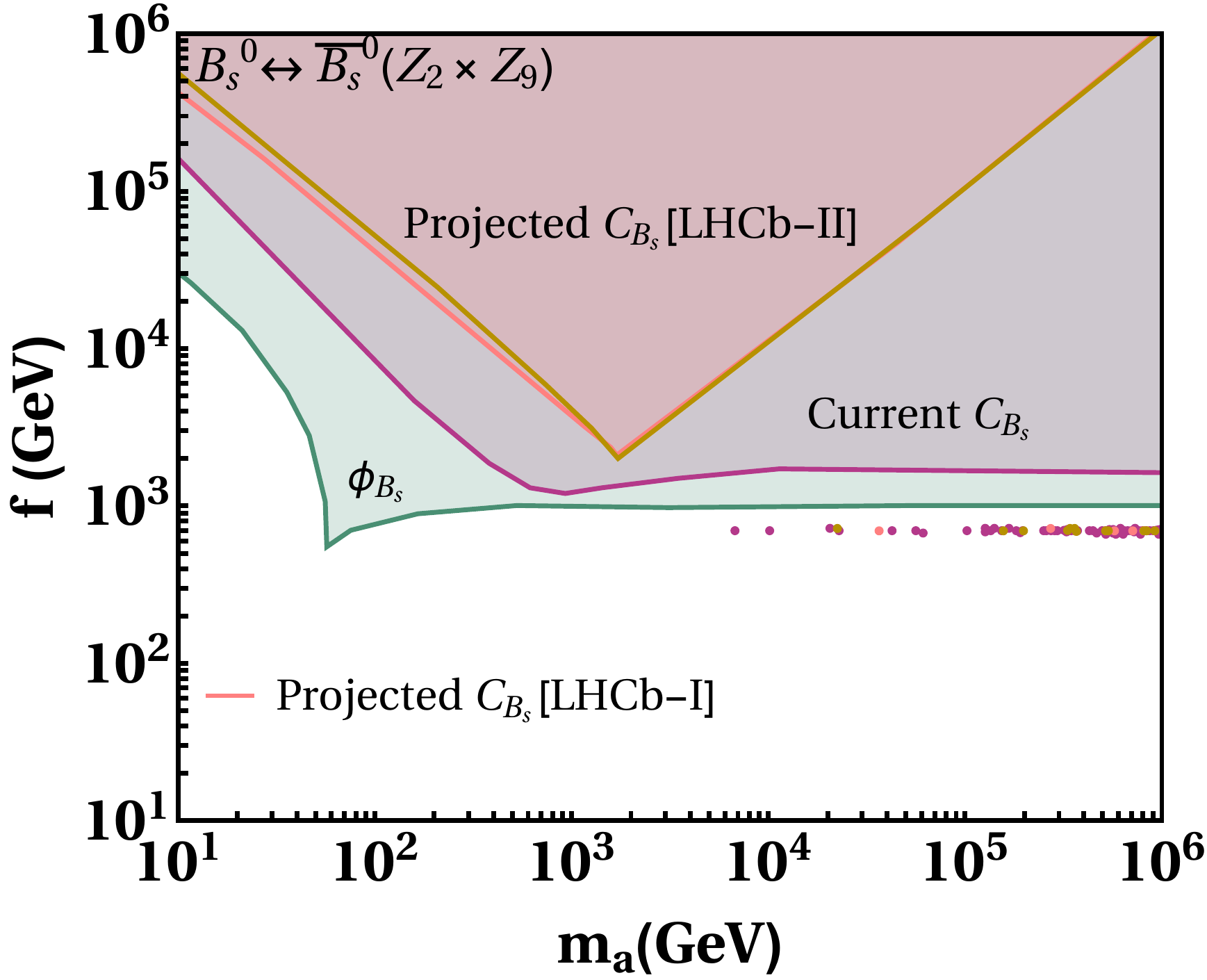}
  \caption{}
         \label{figbsbs2}	
 \end{subfigure} 
\caption{The parameter space allowed by flavour observables $C_{B_{s}}$ and $\phi_{B_{s}}$ for $\lambda_\chi= 2$ in the $m_a - f$ plane for the minimal  ($\mathcal{Z}_2 \times \mathcal{Z}_5$) model in the left panel and for the  non-minimal ($\mathcal{Z}_2 \times \mathcal{Z}_9$) model in the right panel.  }
\label{figbsbs}
\end{figure}

We show the allowed regions of parameter space by the  $B_s -\bar B_s $ mixing  observables $C_{B_{s}}$ and $\phi_{B_{s}}$ for $\lambda_\chi= 2$ in the $m_a - f$ plane for the minimal model based on the  $\mathcal{Z}_2 \times \mathcal{Z}_5$  flavour symmetry and the non-minimal model  based on the $\mathcal{Z}_2 \times \mathcal{Z}_9$  flavour symmetry  in figure \ref{figbsbs}.      In the left panel in figure \ref{figbsbs1},    the red and yellow coloured boundaries  are representing allowed flavon contribution for current values of $C_{B_{s}}$ and $\phi_{B_{s}}$,  respectively in the minimal model, while that for the non-minimal  model are shown in the right panel in figure \ref{figbsbs2},  by magenta and green coloured boundaries,  respectively. We note that the region  in the left panel,  bounded by the blue curve,   represents the allowed bounds for the observable $C_{B_{s}}$  with projected limits of LHCb Phase-\rom{2} for the minimal model,  while the same for the non-minimal model is shown by the region surrounded by the olive coloured curve in the right panel. For the LHCb phase-\rom{1}, the bounds are shown by the pink coloured curves for the minimal as well as for  the non-minimal models,  which are not appreciably different than that of the LHCb phase-\rom{2}.  We also notice an isolated allowed strip of the parameter space for the non-minimal model below the green boundary in the right panel.  The effects of the observables $C_{B_{s}}$ and $\phi_{B_{s}}$ are relatively strong  in the non-minimal model  based on the $\mathcal{Z}_2 \times \mathcal{Z}_9$  flavour symmetry,  which is obvious from the figure itself.

\begin{figure}[H]
	\centering
	\begin{subfigure}[]{0.49\linewidth}
	 \includegraphics[width=\linewidth]{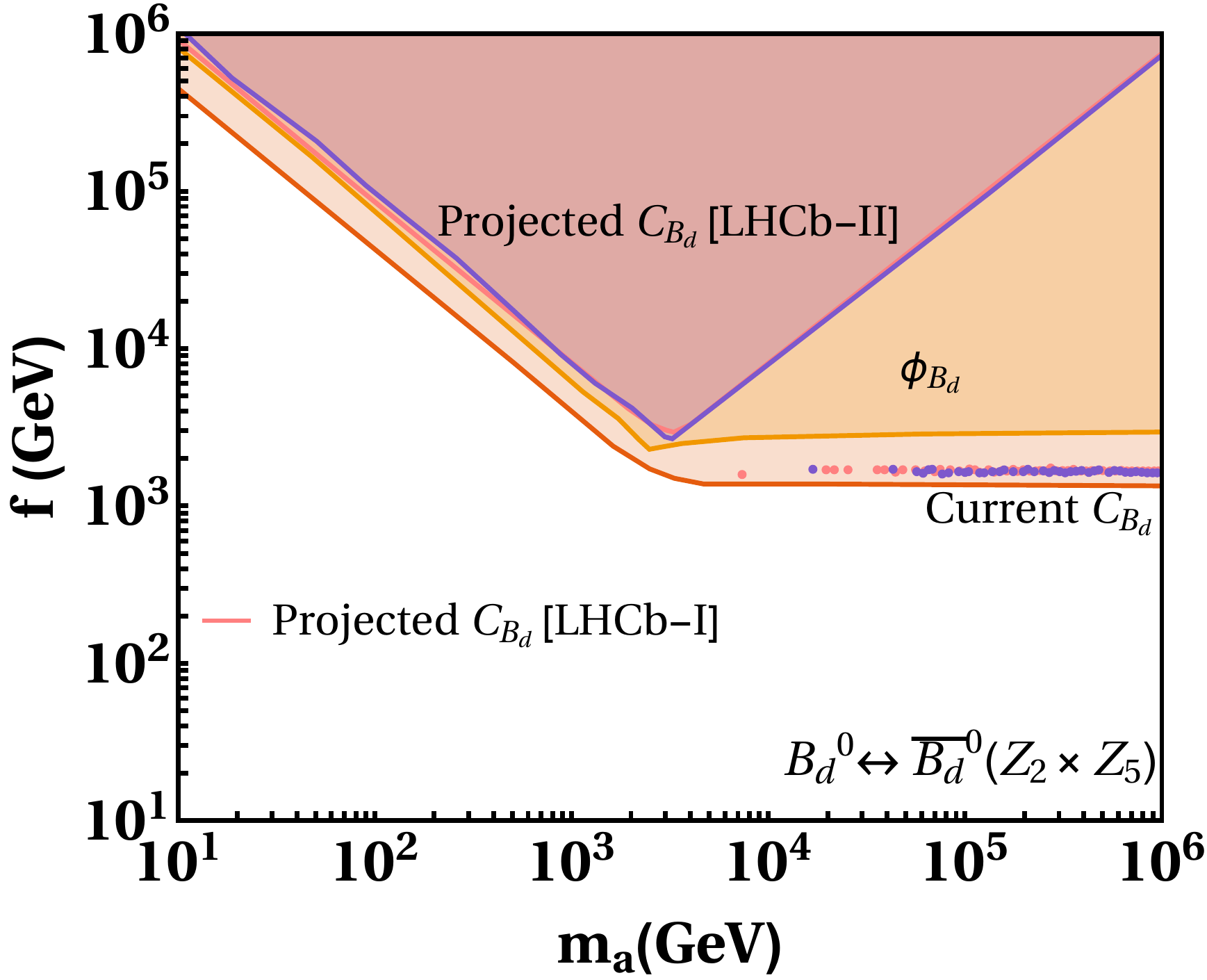}	
	 \caption{}
         \label{bdbd1}	
\end{subfigure}
 \begin{subfigure}[]{0.49\linewidth}
 \includegraphics[width=\linewidth]{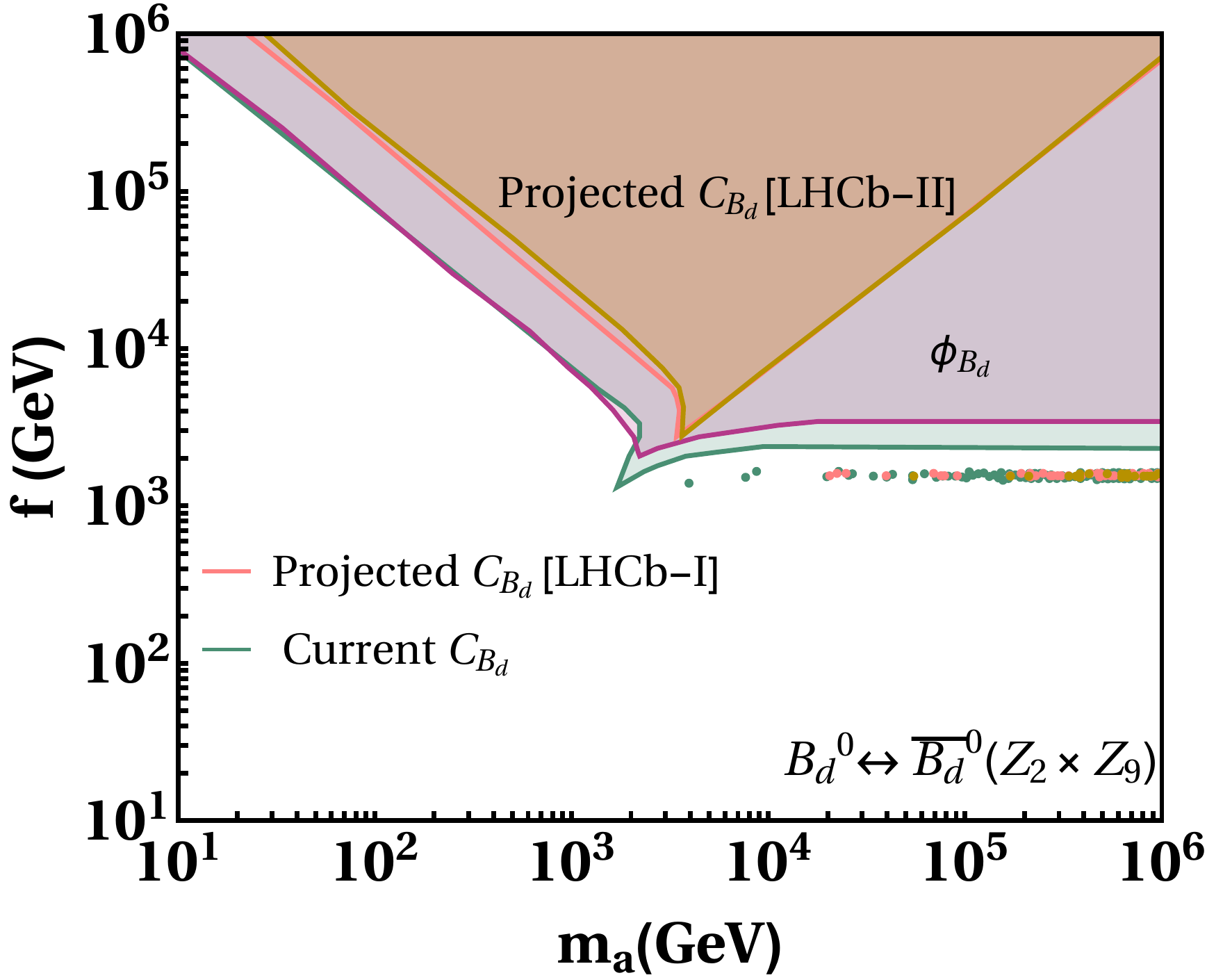}
  \caption{}
         \label{bdbd2}	
 \end{subfigure} 
\caption{The allowed parameter space by flavour observables $C_{B_{d}}$ and $\phi_{B_{d}}$ for $\lambda_\chi= 2$ in the $m_a - f$ plane for the minimal  ($\mathcal{Z}_2 \times \mathcal{Z}_5$) model in the left panel and for the non-minimal ($\mathcal{Z}_2 \times \mathcal{Z}_9$) model in the right panel. }
\label{figbdbd}
\end{figure}

Figure \ref{figbdbd}  shows the allowed parameter space by flavour observables $C_{B_{d}}$ and $\phi_{B_{d}}$ of  the  $B_d -\bar B_d $ mixing for $\lambda_\chi= 2$ in the $m_a - f$ plane.   This is shown for the minimal model on the left in figure \ref{bdbd1} and for the non-minimal model on the right in figure \ref{bdbd2}.   In the left panel  in figure \ref{bdbd1} ,  the red and yellow coloured boundaries are representing allowed flavon contribution for current values of $C_{B_{d}}$ and $\phi_{B_{d}}$, respectively for the minimal  model based on the  $\mathcal{Z}_2 \times \mathcal{Z}_5$ symmetry while that for non-minimal model is shown in the right panel in figure \ref{bdbd2} by green and purple coloured boundaries, respectively.    Moreover,  the region surrounded by the blue curve in the left panel shows the allowed parameter space for  the observable $C_{B_{d}}$  with projected limits of LHCb Phase-\rom{2} for the minimal model while the same for the non-minimal model is shown by olive coloured boundary  in the right panel.   The bounds for the LHCb phase-\rom{1} are shown by the  pink coloured  boundaries for the minimal as well as for  the non-minimal model,  and are very similar to that of the LHCb phase-\rom{2}.

The SM contribution to $D^0 - \bar D^0$ mixing  is  marred by large hadronic uncertainties.   Therefore,  for constraining the parameter space of our model,  we keep only the flavon contribution to  $D^0 - \bar D^0$ mixing such that it always lies within the $2\sigma$  experimental bound \cite{Bona:2017kam}.
\begin{align}
|M_{12}^D|= |\langle D^0|\mathcal{H}^{\Delta F=2}|\bar D^0\rangle | < 7.5 \times 10^{-3}  ps^{-1}
\end{align}

\begin{figure}[H]
	\centering
	\includegraphics[width=0.49\linewidth]{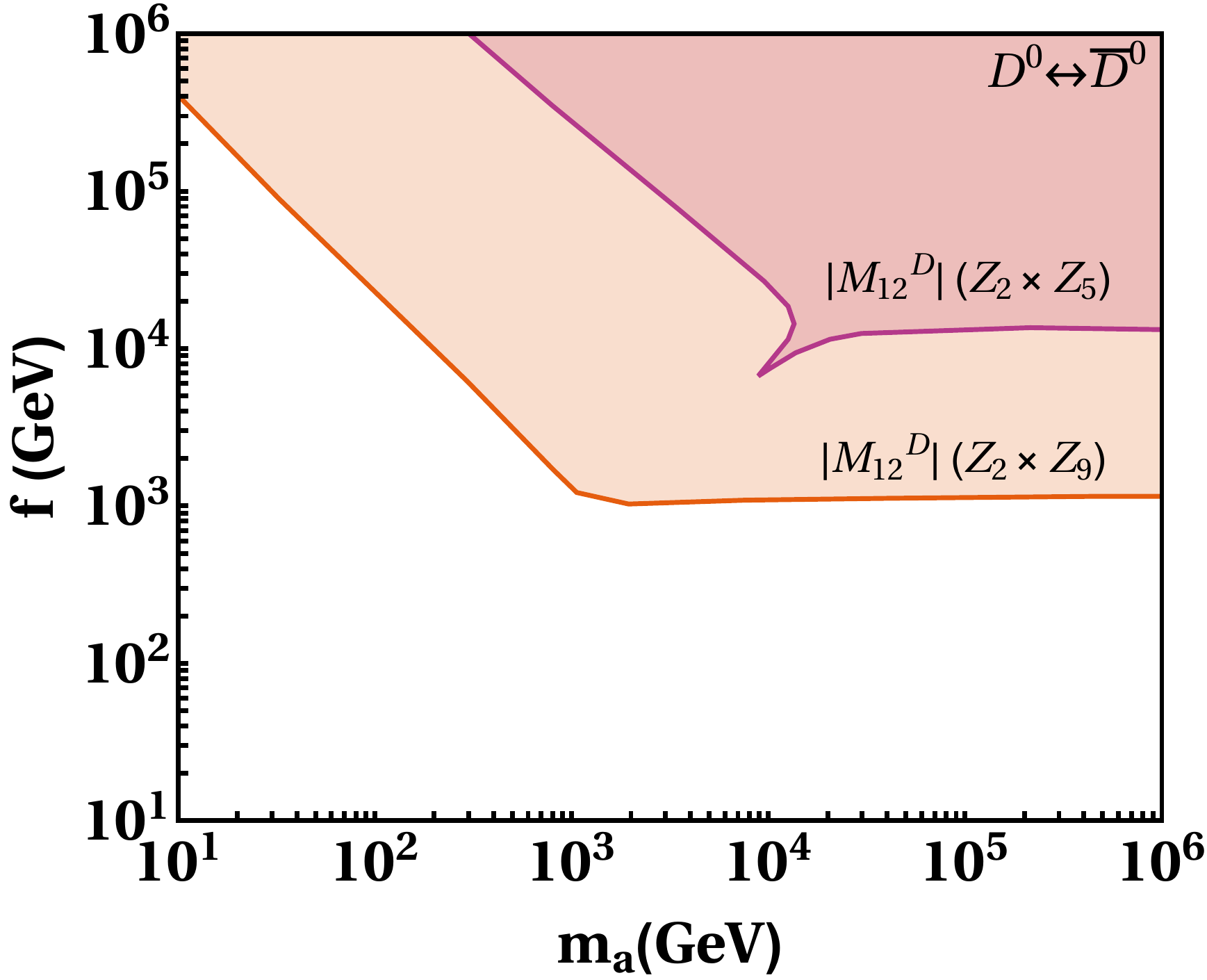}	
  \caption{The allowed parameter space by $|M_{12}^D|$  for $\lambda_\chi= 2$ in the $m_a - f$ plane for the minimal  ($\mathcal{Z}_2 \times \mathcal{Z}_5$) model is shown by the magenta coloured boundary while that for the non-minimal ($\mathcal{Z}_2 \times \mathcal{Z}_9$) model is shown by the red coloured boundary.}
  \label{DDm}
\end{figure}

The bound in the $m_a - f$ panel  arising  from the  $D^0 - \bar D^0$ mixing is shown in figure \ref{DDm} for the minimal model based on the  $\mathcal{Z}_2 \times \mathcal{Z}_5$  flavour symmetry and the non-minimal model  based on the $\mathcal{Z}_2 \times \mathcal{Z}_9$  flavour symmetry.    The first remarkable observation is the allowed parameter space  for the minimal model is much smaller than that of the non-minimal model.  This is because the $D^0 - \bar D^0$ mixing has an enhancement of the order $\epsilon^2$ (see the coupling $2 y_{21}^u \epsilon^2 $ in equation \ref{fup}) in the  minimal model based on the  $\mathcal{Z}_2 \times \mathcal{Z}_5$  flavour symmetry,  which is not present in the case of the non-minimal model.    Therefore,  the bound derived by the   $D^0 - \bar D^0$ mixing is the most stringent bound among the bounds given by the mixing observables  in the minimal model.  However,  this is not the case for the non-minimal model.

The one-loop contribution to this mixing from the box diagram depends on the relatively large $y_{ct}$ and $y_{tc}$ couplings of the flavon to fermions.  In the minimal  model,  $y_{tc}$ is zero.  Thus,  this contribution is proportional to $\epsilon^4 /(4 \pi^2 f^2)$ in the minimal model and proportional to $\epsilon^{16} /(4 \pi^2 f^2)$ in the non-minimal model.  Therefore,   this contribution is highly suppressed with respect to the tree-level contribution used in deriving the bounds in  figure \ref{DDm}.  

\subsection{Leptonic decays of mesons}
The effective Hamiltonian for flavon mediated  decays of neutral mesons into two charged leptons  can be written as,
\begin{align}
\mathcal{H}_\text{eff}=-\frac{G_F^2m_W^2}{\pi^2}\,\left(C_S^{ij}\, (\bar q_iP_L q_j)\bar \ell \ell +\tilde  C_S^{ij}\, (\bar q_iP_R q_j)\bar \ell \ell + C_P^{ij}(\bar q_iP_L q_j)\bar \ell \gamma_5 \ell+\tilde C_P^{ij}(\bar q_iP_R q_j)\bar \ell \gamma_5 \ell \right)+ \text{H.c.}.
\end{align}

The branching ratio of a meson decaying to two charged leptons reads,
\begin{align}
\br(M\rightarrow \ell^+ \ell^- )=&
\frac{G_F^4m_W^4}{8\pi^5} \beta \,m_M f_M^2 m_\ell^2 \tau_M \notag\\
&\left(   \left|\frac{m_M^2\big(C_P^{ij}-\tilde C^{ij}_P\big)}{2m_\ell (m_i+m_j)}-C_A^\text{SM}\right|^2+
\left|\frac{m_M^2\big(C_S^{ij}-\tilde C^{ij}_S\big)}{2m_\ell (m_i+m_j)}\right|^2\beta^2
\right) \; ,
\end{align}
where $\beta(x)=\sqrt{1-4x^2}$ with $x = m_\ell/m_M$.  

The Wilson coefficients having tree-level contribution of the flavon are given as~\cite{Buras:2013rqa,Crivellin:2013wna},
\begin{align}
C_S^{ij}& =\frac{\pi^2}{2G_F^2 m_W^2}\frac{2y_{\ell\ell}y_{ji}}{m_s^2}  \notag\\
\tilde C_S^{ij}&=\frac{\pi^2}{2G_F^2 m_W^2}\frac{2y_{\ell\ell}y_{ij}}{m_s^2} \notag\\
C_P^{ij}&=\frac{\pi^2}{2G_F^2 m_W^2}\frac{2y_{\ell\ell}y_{ji}}{m_a^2}  \notag\\
\tilde C_P^{ij}&=\frac{\pi^2}{2G_F^2 m_W^2}\frac{2y_{\ell\ell}y_{ij}}{m_a^2}\; .
\end{align}

In the SM,  processes of  mesons decaying to two charged leptons  are induced by one-loop contribution,  and for the  $B_s$ meson it is given by\cite{Crivellin:2013wna},
\begin{align}
\label{CA}
C_A^\text{SM}= -V_{tb}^*V_{ts}\,Y\left(\frac{m_t^2}{m_W^2}\right) -V_{cb}^*V_{cs}\,Y\left(\frac{m_c^2}{m_W^2}\right) \; ,
\end{align}
where Inami-Lim function $Y(x)$ is given by\cite{Inami:1980fz},
\begin{align}
Y(x)=\eta_\text{QCD}  \frac{x}{8}\left[\frac{4-x}{1-x}+\frac{3x}{(1-x)^2} \log x \right] ,
\end{align}
where $\eta_\text{QCD}=1.0113$  includes NLO QCD effects~\cite{Buras:2012ru}.    For $B_d$ meson,  the SM predictions are obtained by a simple replacement of indices  in equation \ref{CA}.

The  average of the branching fraction of    $B_{s}   \rightarrow \mu^+\mu^-$  from HFLAV group is \cite{HFLAV:2022pwe}, 

\begin{align}
\label{Hflav}
\br(B_s \rightarrow \mu^+\mu^-) & =(3.45 \pm 0.29 )  \times 10^{-9}.
\end{align}

The latest measurement of the  branching fraction of  $B_{d}   \rightarrow \mu^+\mu^-$  is\cite{LHCb:2021vsc,LHCb:2021awg},
\begin{align}
\br(B_d\rightarrow \mu^+\mu^-)& <  2.6 \times  10^{-10}\,.
\end{align}

As observed in reference \cite{DeBruyn:2012wk},  due to sizeable width difference,  of the $B_s$ meson,  theoretical branching ratio can be converted to  experimental branching ratio by multiplying  $(1-y_s)^{-1}$,  where $y_s=0.088\pm0.014$~\cite{Fleischer:2012bu}.   This correction is  negligible in the case of the $B_d$ meson.

\begin{table}[H]
\centering
\begin{tabular}{l|ccccc}
\text{Observables} & \text{Current} & \text{LHCb-\rom{1}} & \text{LHCb-\rom{2}}  & \text{CMS}  & \text{ATLAS} \\
\hline
$\br(B_s\rightarrow \mu^+\mu^-)   (\times 10^9)$ & $\pm 0.38$    &   $\pm 0.30$    &  $\pm 0.16$   &  $-$      & $\pm 0.50$  \\
$ \mathcal{R}_{\mu\mu} $ & $\sim 70 \%$ &  $\sim 34 \%$& $\sim 10 \%$ &  $\sim 21 \%$   & $-$\\
$\tau_{\mu\mu}$ & $\sim 12 \%$  & $\pm 0.16$ ps  &   $\pm 0.04$ ps  &   $-$     & $-$\\
\hline
\end{tabular}
\caption{The current and expected experimental precision  for rare  $B$ decays observables  where \text{LHCb-\rom{1}} corresponds to $23 fb^{-1}$,   \text{LHCb-\rom{2}} corresponds to $300 fb^{-1}$,   CMS and ATLAS correspond to $3 ab^{-1}$\cite{CMS:2022dbz,Altmannshofer:2022hfs}.}
\label{Bsmmu}
\end{table}

In addition to $\br(B_s\rightarrow \mu^+\mu^-) $ branching ratio,  the LHCb collaboration has also measured the ratio of the $\br(B_d\rightarrow \mu^+\mu^-) $ and $\br(B_s\rightarrow \mu^+\mu^-) $ branching fractions, $ \mathcal{R}_{\mu\mu} $\cite{LHCb:2021vsc,LHCb:2021awg}.  The CMS has measured the effective lifetime,  $\tau_{\mu\mu}$,  of the $B_s\rightarrow \mu^+\mu^- $ decay \cite{CMS:2022dbz}.  We note that the ratio $ \mathcal{R}_{\mu\mu} $ is an excellent observable to probe the minimal flavour violation\cite{Altmannshofer:2022hfs}.   On the other side,  the effective lifetime,  $\tau_{\mu\mu}$,  can be used to discriminate between the contributions due to any possible  new scalar and pseudoscalar mediators \cite{Altmannshofer:2022hfs}.   The measured value of the ratio of branching fractions, $ \mathcal{R}_{\mu\mu} $,  is  \cite{LHCb:2021vsc,LHCb:2021awg},

\bea
 \mathcal{R}_{\mu\mu} &=& \frac{\br(B_d\rightarrow \mu^+\mu^-) }{\br(B_s\rightarrow \mu^+\mu^-)} = 0.039^{+0.030+ 0.006}_{-0.024-0.004}.
\eea

The effective lifetime $\tau_{\mu\mu}$  and the branching fraction of $B_s\rightarrow \mu^+\mu^-$   are also measured by CMS,  and are \cite{CMS:2022dbz},

\bea
 \tau_{\mu\mu} &=& 1.83^{+0.23+ 0.04}_{-0.20-0.04}  ~\rm{ps},
\eea

\begin{align}
\br(B_s\rightarrow \mu^+\mu^-) & =3.83^{+0.38+ 0.19+0.14}_{-0.36-0.16-0.13}   \times 10^{-9}.  
\end{align}

We have also taken HFLAV measurement average into account for effective lifetime $\tau_{\mu\mu}$, which is \cite{HFLAV:2022pwe}, 
\bea
\label{t_hflav}
 \tau_{\mu\mu} &=& 2.00^{+0.27}_{-0.26}  ~\rm{ps},
\eea
with $\br(B_s\rightarrow \mu^+\mu^-)$ given in equation \ref{Hflav}.\\
The current and future sensitivities of these observables are summarized in table \ref{Bsmmu}.   The effective lifetime can be written in the following form \cite{DeBruyn:2012wj},

\be
 \tau_{\mu\mu} =  \tau_{B_s} \frac{(B_s\rightarrow \mu^+\mu^-)^{\rm experiment}}{(B_s\rightarrow \mu^+\mu^-)^{\rm theory}},
\ee

where we have assumed the SM value of the  final state dependent observable $\mathcal{A}^f_{\Delta \Gamma} =1$ \cite{Fleischer:2012bu}.

\begin{figure}[H]
	\centering
	\begin{subfigure}[]{0.49\linewidth}
	 \includegraphics[width=\linewidth]{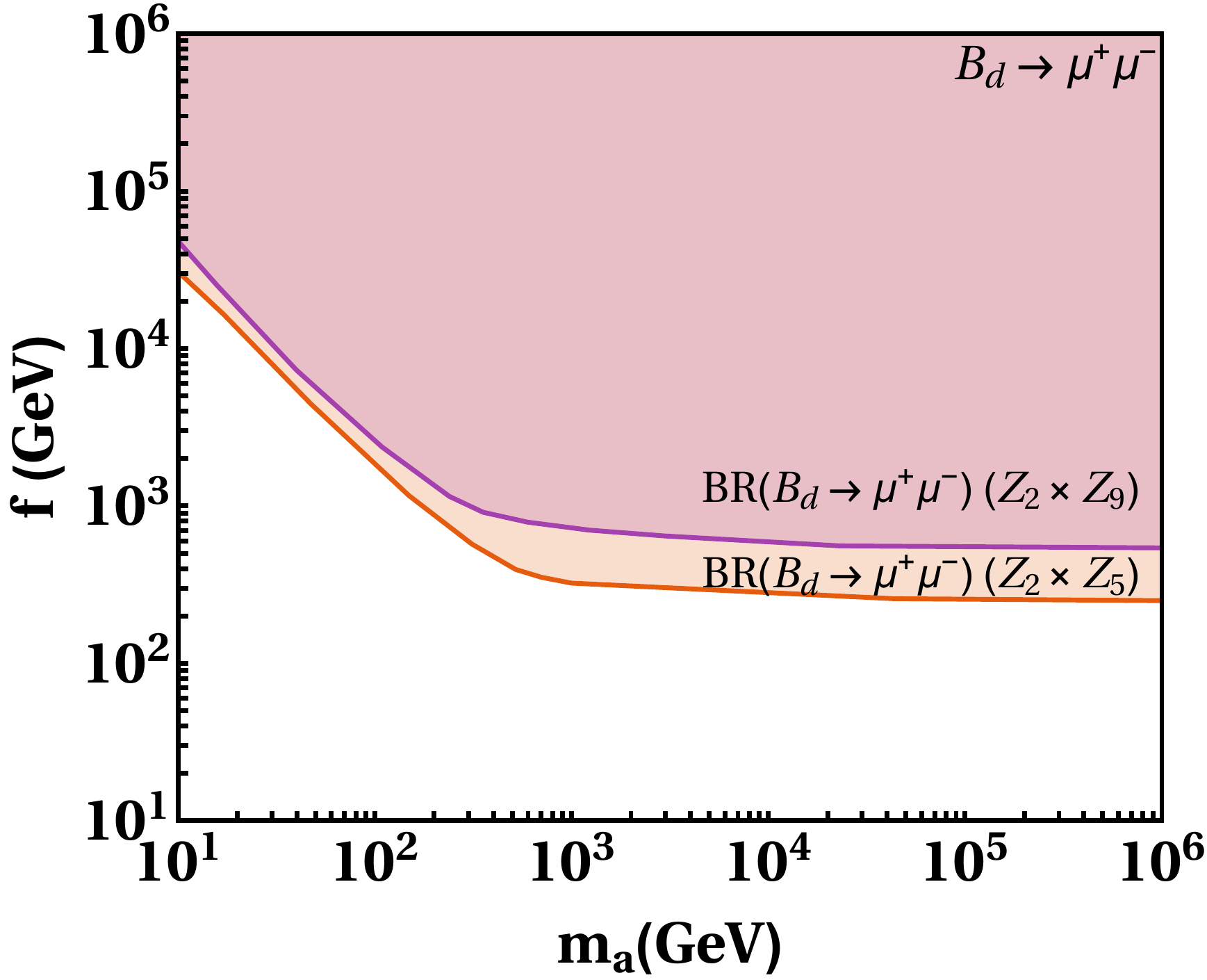}	
	  \caption{}
         \label{bmumu1}	
\end{subfigure} 
	\begin{subfigure}[]{0.49\linewidth}
	 \includegraphics[width=\linewidth]{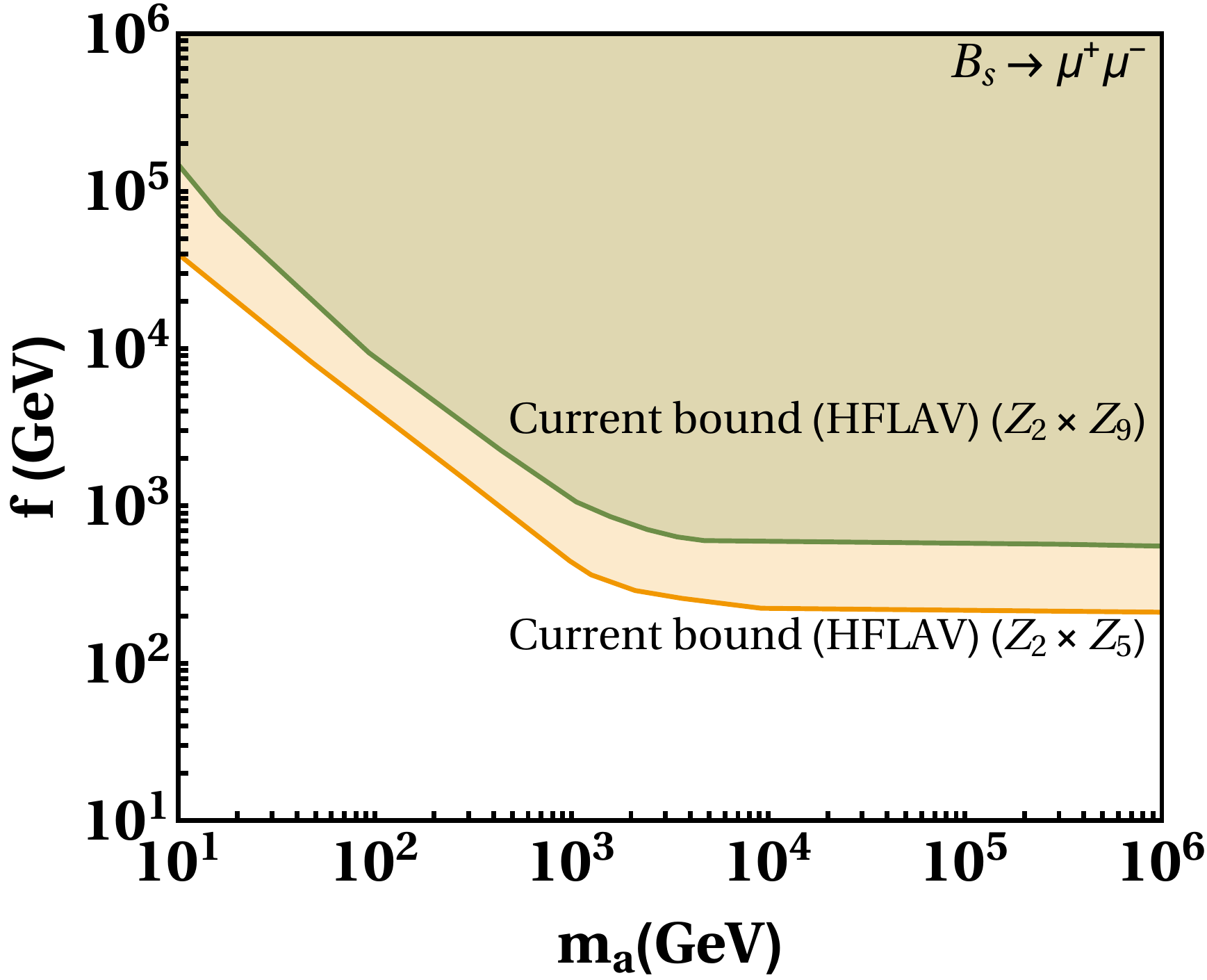}	
	   \caption{}
         \label{bmumu2}	
\end{subfigure} 
\caption{The allowed parameter space by $\br(B_d\rightarrow \mu^+\mu^-)$ in the left panel  and that for the  $\br(B_s\rightarrow \mu^+\mu^-)$  in the right panel for $\lambda_\chi= 2$  with current experimental bounds for the minimal and the non-minimal models.}
\label{bdbs}
\end{figure}

In figure  \ref{bdbs},  the bounds coming from the  branching ratios $\br(B_d\rightarrow \mu^+\mu^-)$ and  the $\br(B_s\rightarrow \mu^+\mu^-)$  for $\lambda_\chi= 2$ in the $m_a - f$ plane are shown.   The  branching ratios $\br(B_d\rightarrow \mu^+\mu^-)$ and  $\br(B_s\rightarrow \mu^+\mu^-)$ place weaker constraints on the parameter space of the minimal model.   However,  for the non-minimal model,  the bounds,  particularly from the branching ratios $\br(B_s\rightarrow \mu^+\mu^-)$,  are  quite stronger.  For the  projected sensitivities of the  LHCb  Phase-\rom{1} and \rom{2} and   of the ATLAS for the $\br(B_s\rightarrow \mu^+\mu^-)$,   we do not obtain any appreciable improvements in our bounds.  Therefore,  we do not show them in figure  \ref{bdbs}.

\begin{figure}[H]
	\centering
	\begin{subfigure}[]{0.49\linewidth}
	 \includegraphics[width=\linewidth]{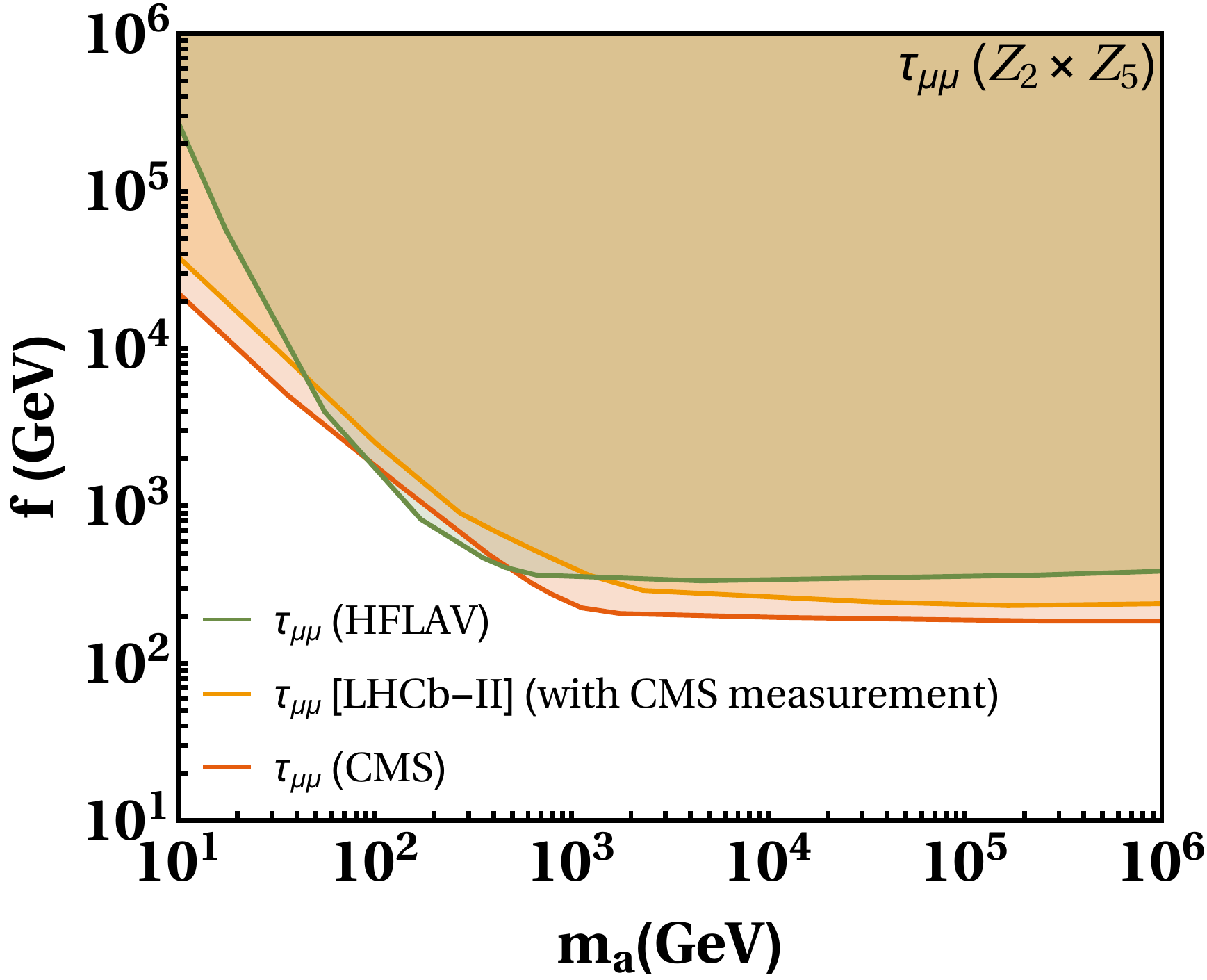}	
	  \caption{}
         \label{LT1}	
\end{subfigure} 
	\begin{subfigure}[]{0.49\linewidth}
	 \includegraphics[width=\linewidth]{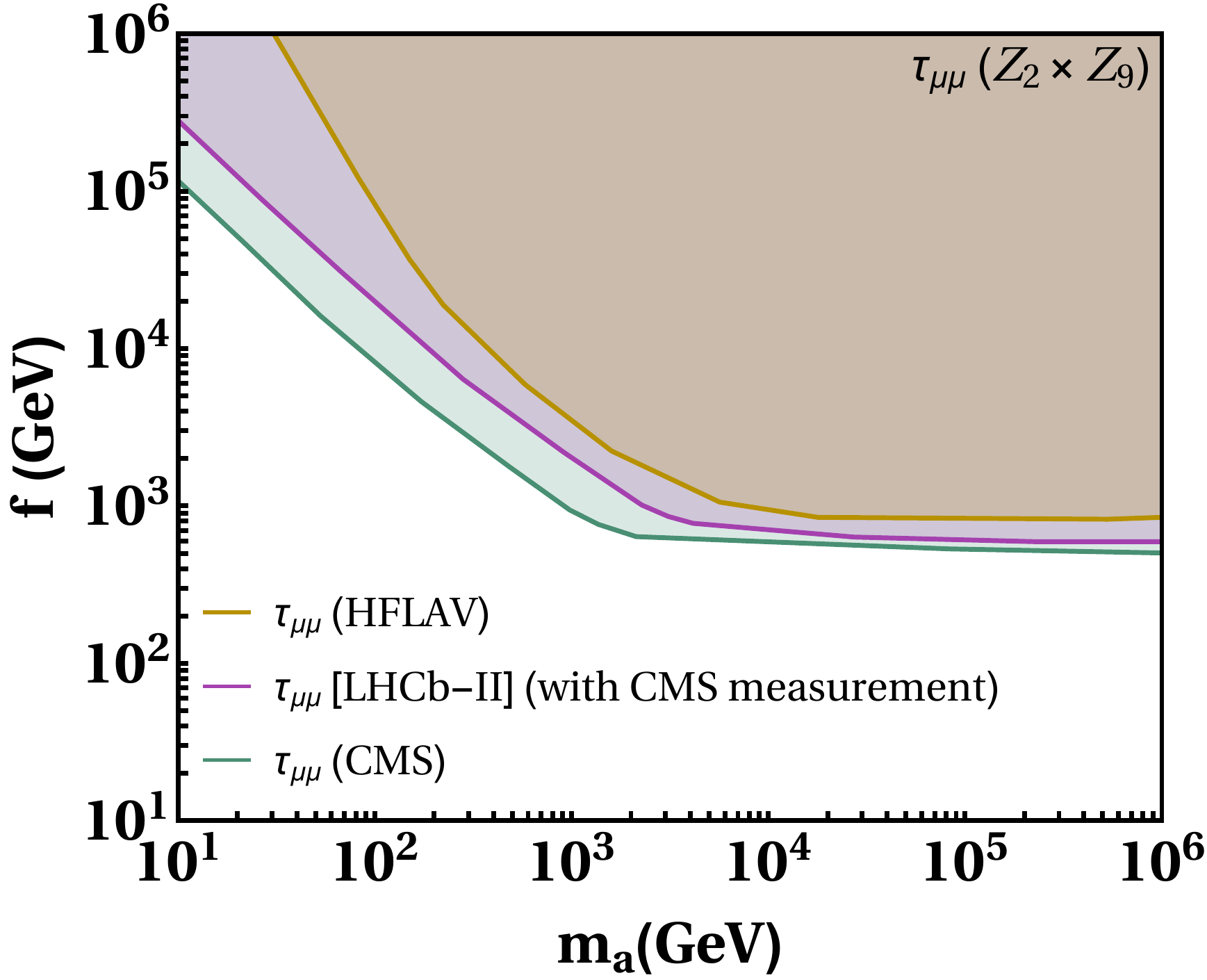}	
	   \caption{}
         \label{LT2}	
\end{subfigure} 
\caption{The allowed parameter space by  $ \tau_{\mu\mu} $ for $\lambda_\chi= 2$ in the $m_a - f$ plane for the recent measurement and the future projected sensitivity of the  LHCb  for the minimal  ($\mathcal{Z}_2 \times \mathcal{Z}_5$) model on the left,  and for the non-minimal  ($\mathcal{Z}_2 \times \mathcal{Z}_9$) model  on the right.}
\label{LT}
\end{figure}

Figure \ref{LT} shows the allowed bounds derived from the HFLAV average of the effective lifetime  $ \tau_{\mu\mu} $  and the branching-ratio  $\br(B_s\rightarrow \mu^+\mu^-)$ by the green coloured curve for the minimal model in figure \ref{LT1} and through the olive coloured boundary for the non-minimal model in figure \ref{LT2}. The bounds from the current measurement by CMS,  and  from the future projected sensitivity of the LHCb phase-\rom{2} for  the minimal model based on $\mathcal{Z}_2 \times \mathcal{Z}_5$, are shown by red and yellow coloured boundaries respectively, in figure \ref{LT1}. The same bounds for the non-minimal model  based on the $\mathcal{Z}_2 \times \mathcal{Z}_9$  flavour symmetry are shown in figure \ref{LT2} by green and magenta coloured curves respectively.  We observe that the bounds are stronger for the non-minimal model in comparison to that of the minimal model.  For the projected sensitivity of the  LHCb  Phase-\rom{1},  we do not find any appreciable  improvement  in the allowed bounds over that from the current measurement.  Therefore, we do not show it in figure  \ref{LT}.

\begin{figure}[H]
	\centering
	\begin{subfigure}[]{0.49\linewidth}
	 \includegraphics[width=\linewidth]{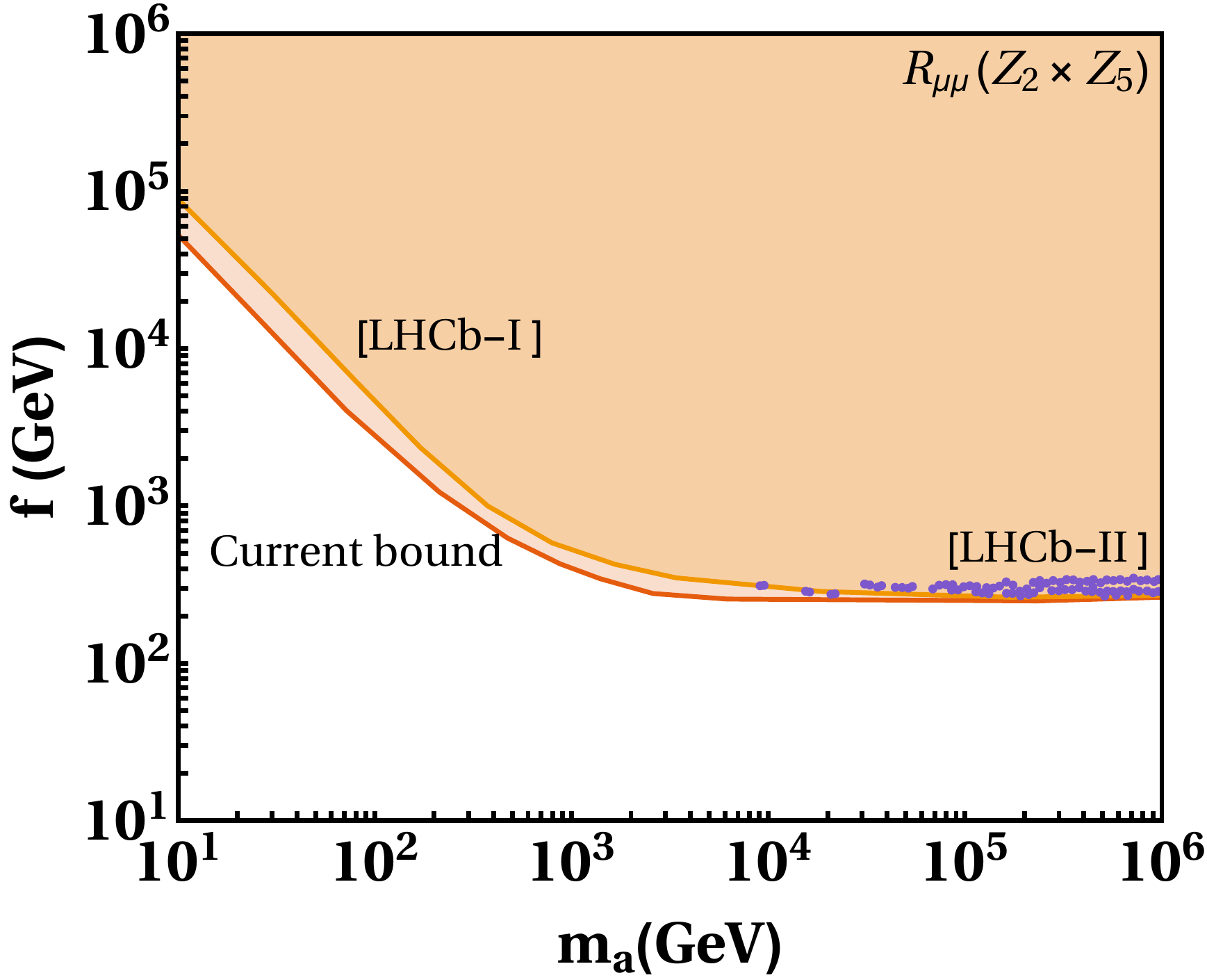}	
	   \caption{}
         \label{Rmm1}	
\end{subfigure}
 \begin{subfigure}[]{0.49\linewidth}
 \includegraphics[width=\linewidth]{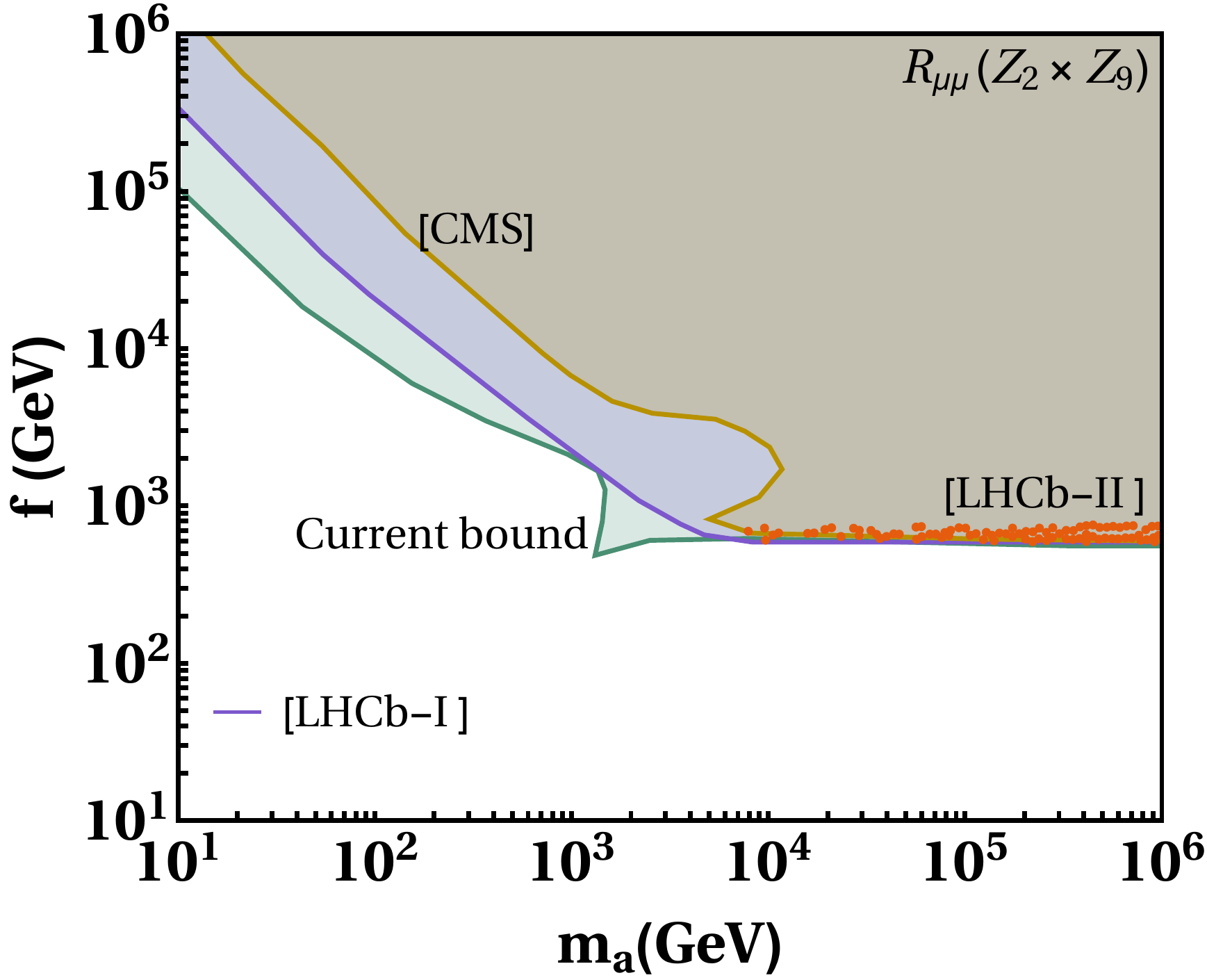}
   \caption{}
         \label{Rmm2}	
 \end{subfigure}
\caption{The left panel represents the allowed parameter space by  $ \mathcal{R}_{\mu\mu} $  for the minimal  ($\mathcal{Z}_2 \times \mathcal{Z}_5$) model with $\lambda_\chi= 2$  with the current measurement  and  for the future projected sensitivity of the LHCb  Phase-\rom{1}  and   LHCb  Phase-\rom{2}.  The same allowed parameter space  for the non-minimal  ($\mathcal{Z}_2 \times \mathcal{Z}_9$) model is shown in the right panel.}
\label{Rmm}
\end{figure}

One of the most important observables for our models is the ratio $ \mathcal{R}_{\mu\mu} $ whose future projected sensitivity will be crucial to constrain our models.  On the left panel in figure  \ref{Rmm1},  we show the bounds arising from the ratio $ \mathcal{R}_{\mu\mu} $ for the minimal model.  These bounds are weaker for the current measurement as well as for the future projected sensitivity of the LHCb  Phase-\rom{1} shown by the red and yellow boundaries,  respectively.   However,  the future projected sensitivity of the LHCb  Phase-\rom{2}  dramatically changes this scenario and provide extremely stringent constraints on the parameter space of the minimal model shown by the purple coloured strip.  Therefore,  the  LHCb  Phase-\rom{2}   will be decisive for the minimal model based on the $\mathcal{Z}_2 \times \mathcal{Z}_5$ flavour symmetry.   We also do not find any improvement over the bounds given by the LHCb  Phase-\rom{1} using the future projected sensitivity of the CMS experiment. Therefore,  we do not show it in figure \ref{Rmm1}.

On the other hand,  for the non-minimal model based on the $\mathcal{Z}_2 \times \mathcal{Z}_9$ symmetry,  the bounds from the ratio $ \mathcal{R}_{\mu\mu} $  are shown in figure  \ref{Rmm2} in the right panel.  The bounds from the current measurement are shown by the green boundary  while the bounds from the  LHCb  Phase-\rom{1}  are surrounded by the  purple coloured curve.  Moreover,  we also have bounds from the future projected sensitivity of the CMS experiment surrounded by olive coloured boundary.  These are more stringent than that of the projected sensitivity of the  LHCb  Phase-\rom{1} .   We note that  similar to  the minimal model,  the bounds for the  projected sensitivity of the  LHCb  Phase-\rom{2} are highly stringent depicted by the orange coloured strip.  This result does not change even  if we deviate from the bench-mark values of the Yukawa couplings used for these bounds.  Therefore,  these bounds are robust,  and crucial to test the parameter space of the non-minimal model based on the $\mathcal{Z}_2 \times \mathcal{Z}_9$ symmetry in the future projected sensitivity of the  LHCb  Phase-\rom{2}.

For the $K_L \rightarrow \mu^+ \mu^-$ decay,   we have only reliable estimate of the so-called short distance (SD) part of  $K_L \rightarrow \mu^+ \mu^-$ decay \cite{Buras:2013rqa}.  We use the SM prediction obtained in reference \cite{Buras:2013rqa} and given by
\begin{align}
C_A^\text{SM}= -V_{ts}^*V_{td}\,   Y\left(\frac{m_t^2}{m_W^2}\right) -  V_{cs}^* V_{cd}         Y_{\rm NNL},
\end{align}
where at NNLO $Y_{\rm NNL} = \lambda^4 P_c(Y)$,  $\lambda = |V_{us}|$ and $P_c(Y) = 0.113 \pm 0.017$\cite{Gorbahn:2006bm}.   The short distance contribution can be extracted from the experimental measurement and has an upper limit\cite{Crivellin:2013wna},
\begin{align}
\br(K_L\rightarrow \mu^+\mu^-)_{\rm SD} & < 2.5   \times 10^{-9}.
\end{align}

For the case of $D \rightarrow \mu^+ \mu^-$ decay,  the SM contribution is plagued by large non-perturbative effects.  Therefore,  we only require that the flavon contribution does not generate more than the experimental upper bound on the branching ratio that is given at $90\%$ C.L.\cite{LHCb:2013jyo},
\begin{align}
\br(D \rightarrow \mu^+\mu^-) & < 6.2   \times 10^{-9}.
\end{align}

\begin{figure}[H]
	\centering
	\includegraphics[width=0.49\linewidth]{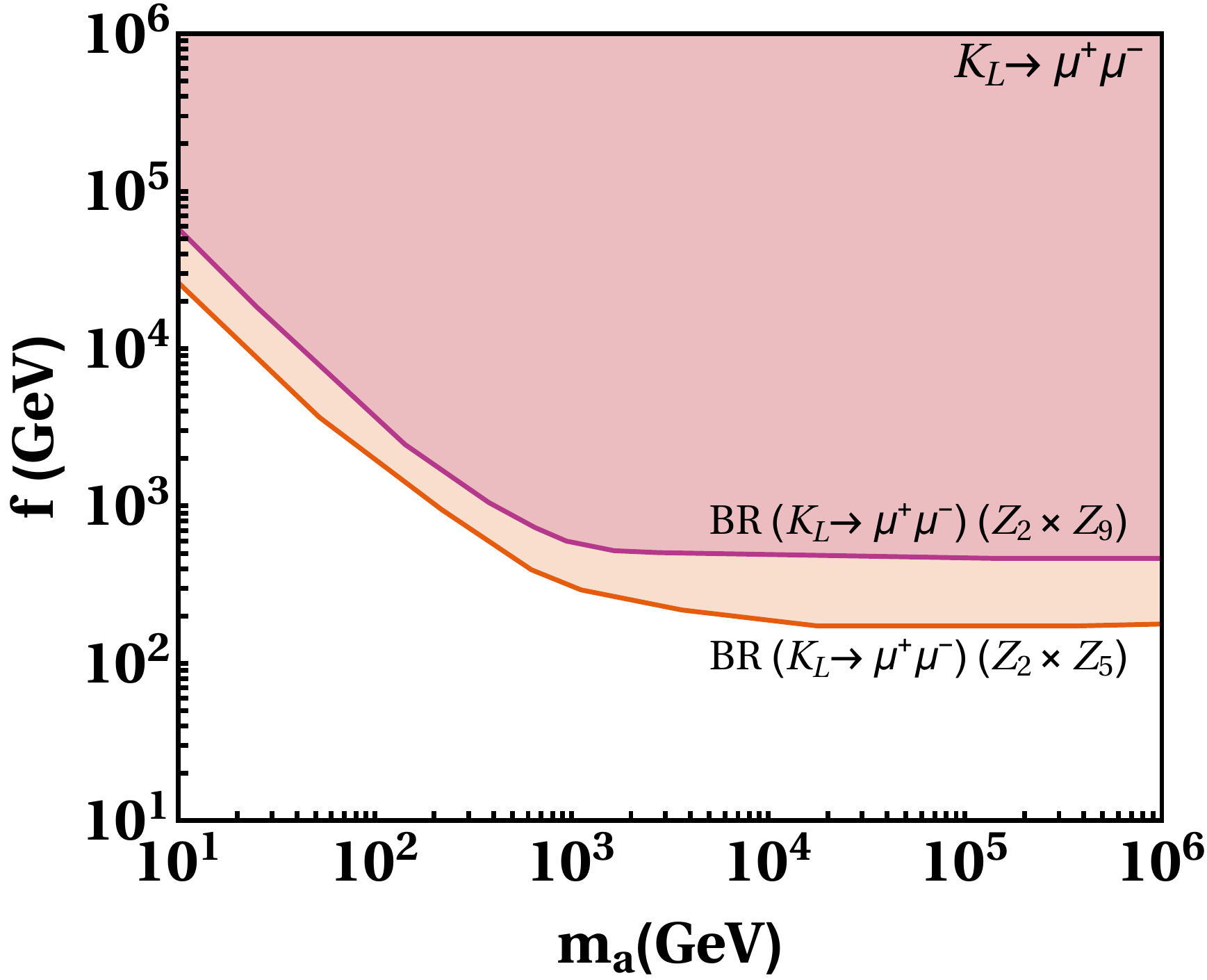}	
	 \caption{The parameter space allowed by  $\br(K_L\rightarrow \mu^+\mu^-)_{\rm SD}$ with $\lambda_\chi= 2$ in the $m_a - f$ plane for the minimal  ($\mathcal{Z}_2 \times \mathcal{Z}_5$) and non-minimal ($\mathcal{Z}_2 \times \mathcal{Z}_9$) model are shown with red and magenta coloured boundaries,  respectively. }
  \label{KDll}
\end{figure}

We show bounds arising from the $\br(K_L\rightarrow \mu^+\mu^-)_{\rm SD}$   for $\lambda_\chi= 2$  for the minimal and the non-minimal models  in the $m_a - f$ plane in figure \ref{KDll}.  It turns out that the constraints on the parameter space of the minimal  model are weaker.  However,  the bounds on the parameter space of the non-minimal model are more stringent.  The constraints from the  $D \rightarrow \mu^+ \mu^-$ decay are much weaker than that from the  $\br(K_L\rightarrow \mu^+\mu^-)_{\rm SD}$.  Therefore,  we do not show them in this work.

\section{Leptonic flavour physics in the minimal and non-minimal $\mathcal{Z}_2 \times \mathcal{Z}_N$ flavour symmetry}
\label{lepton_flavour}
Charged lepton flavour violation (CLFV) has a substantial potential to provide bounds on flavon physics in future upcoming experiments.  The quark flavour constraints currently dominate our flavon model.  However,  it is expected that the future projected sensitivities of CLFV,  which are   shown in table \ref{lfv_exp},   may significantly improve  the  bounds from quark flavour physics.

\begin{table}[H]\centering
\begin{tabular}{l|cccc}
\text{Observables} & \text{ Current sensitivity}  & \text{Ref.}   & \text{Future projection} & \text{Ref.}  \\
\hline
BR($\meg $ )& $ < 4.2 \times 10^{-13}$ & MEG~\cite{MEG:2016leq}
 &$ 6 \times  10^{-14}$   &     MEG\rom{2}~\cite{MEGII:2018kmf} \\
 BR($\tau\to e \gamma$) & $<3.3\times 10^{-8}$&  Babar~\cite{BaBar:2009hkt} & $ \sim 10^{-9} $& Belle \rom{2}~\cite{Belle-II:2018jsg} \\
  BR( $\tau\to \mu \gamma$) & $< 4.4 \times 10^{-8}$&  Babar~\cite{BaBar:2009hkt} & $ \sim 10^{-9} $& Belle \rom{2}~\cite{Belle-II:2018jsg} \\
BR $\text{(}\mu  \to e \text{)}^{\rm Au} $& 
$< 7 \times 10^{-13}$ & SINDRUM \rom{2}~\cite{SINDRUMII:2006dvw}  &
 $ -$  &   $-$    \\
  BR $\text{(} \mu  \to e \text{)}^{\rm Al} $& 
$ -$ & $-$ &
 $ 3 \times 10^{-15}$  &   COMET Phase-\rom{1}~\cite{Wong:2015fzj,Mu2e-2}     \\
  BR $\text{(}\mu  \to e  \text{)}^{\rm Al} $& 
$ -$ & $-$ &
 $ 6 \times 10^{-17}$  &   COMET Phase-\rom{2}~\cite{Wong:2015fzj}     \\
 BR $\text{(}\mu  \to e  \text{)}^{\rm Al} $& 
$ -$ & $-$ &
 $ 6 \times 10^{-17}$  &  Mu2e~\cite{Mu2e:2008sio}     \\
 BR $\text{(}\mu  \to e  \text{)}^{\rm Al} $& 
$ -$ & $-$ &
 $ 3 \times 10^{-18}$  &  Mu2e \rom{2}~\cite{Mu2e-2}     \\
 BR $\text{(}\mu  \to e  \text{)}^{\rm Si} $& 
$ - $ & $-$  &
 $ 2 \times 10^{-14}$   &  DeeMe  \cite{Teshima:2018ugo}     \\
  BR $  \text{(} \mu  \to e  \text{)}^{\rm Ti} $ & & &
  $ \sim 10^{- 20} -  10^{-18}$   &  PRISM/PRIME~\cite{Davidson:2022nnl,Kuno:2012pt} \\
BR(  $\meee $)& $ < 1.0 \times 10^{-12}$
 &  SINDRUM~\cite{SINDRUM:1987nra}   &   $ \sim 10^{-16}$ &   Mu3e~\cite{Blondel:2013ia}  \\
BR($\tau\to 3\mu$ )& $<$ $ 2.1 \times 10^{-8}$  & Belle~\cite{Hayasaka:2010np} & $ \sim 10^{-9} $    &   Belle \rom{2}~\cite{Belle-II:2018jsg}\\
BR($\tau\to 3e$ )& $<$ $ 2.7 \times 10^{-8}$  & Belle~\cite{Hayasaka:2010np}
& $ \sim 10^{-9} $    &   Belle \rom{2}~\cite{Belle-II:2018jsg} \\
\hline
\end{tabular}
\caption{Experimental upper limits on various Leptonic flavour violation (LFV) processes.}
\label{lfv_exp}
\end{table}

\subsection{Radiative leptonic decays}
The effective Lagrangian for the radiative leptonic decays can be written as,
\begin{align}
\lag_{\text{eff}}&=m_{\ell'}\, C_T^L\,\bar \ell \sigma^{\rho\lambda}P_L\,\ell' \,F_{\rho\lambda}+m_{\ell'}\, C_T^R\,\bar \ell \sigma^{\rho\lambda}P_R\,\ell'\,F_{\rho\lambda}.
\label{meg}
\end{align}

The radiative leptonic decays are mediated by  dipole operators and their branching ratio reads,
\begin{align}
\br(\ell'\rightarrow  \ell\gamma)=\frac{m_{\ell'}^5}{4\pi \Gamma_{\ell'}}\left(|C_T^L|^2+|C_T^R|^2\right) \; .
\end{align}

\begin{figure}[h]
	\centering
    \includegraphics[width=0.44\linewidth]{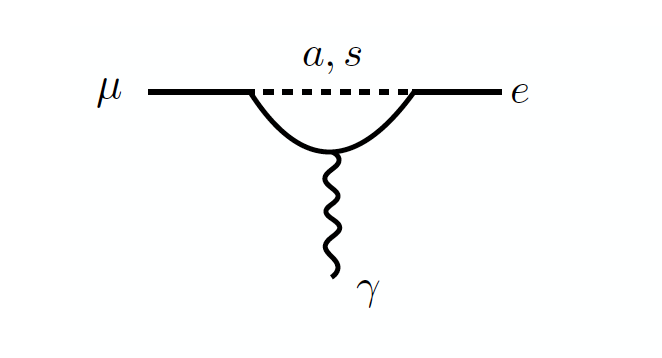}	
 \caption{Feynman diagram representing  $\mu   \rightarrow  e  \gamma$ decay.}
 \label{llg}
	\end{figure}

The one-loop contribution to the radiative leptonic decays  is shown in figure \ref{llg}.  The corresponding  Wilson coefficients are  \cite{Bauer:2015kzy},
\begin{align}
C_T^L = (C_T^R)^*
=\frac{e}{32\pi^2}\sum_{k=e,\mu,\tau} &\bigg\{ \frac{1}{6}\left( \,y^*_{\ell k}y_{\ell' k}+\frac{m_\ell}{m_k}y^*_{k \ell }y_{k\ell' }\right)\left(\frac{1}{m_s^2}-
\frac{1}{m_a^2}\right)\notag\\
&-y_{\ell k}y_{k\ell'}\frac{m_k}{m_{\ell'}}
\left[ \frac{1}{m_s^2}
\left(\frac{3}{2}+\log\frac{m_{\ell'}^2}{m_s^2}\right)-\frac{1}{m_a^2}
\left(\frac{3}{2}+
\log\frac{m_{\ell'}^2}{m_a^2}
\right)\right]
\bigg\} \; .
\label{megw}
\end{align}

\begin{figure}[H]
	\centering
	\begin{subfigure}[]{0.49\linewidth}
	 \includegraphics[width=\linewidth]{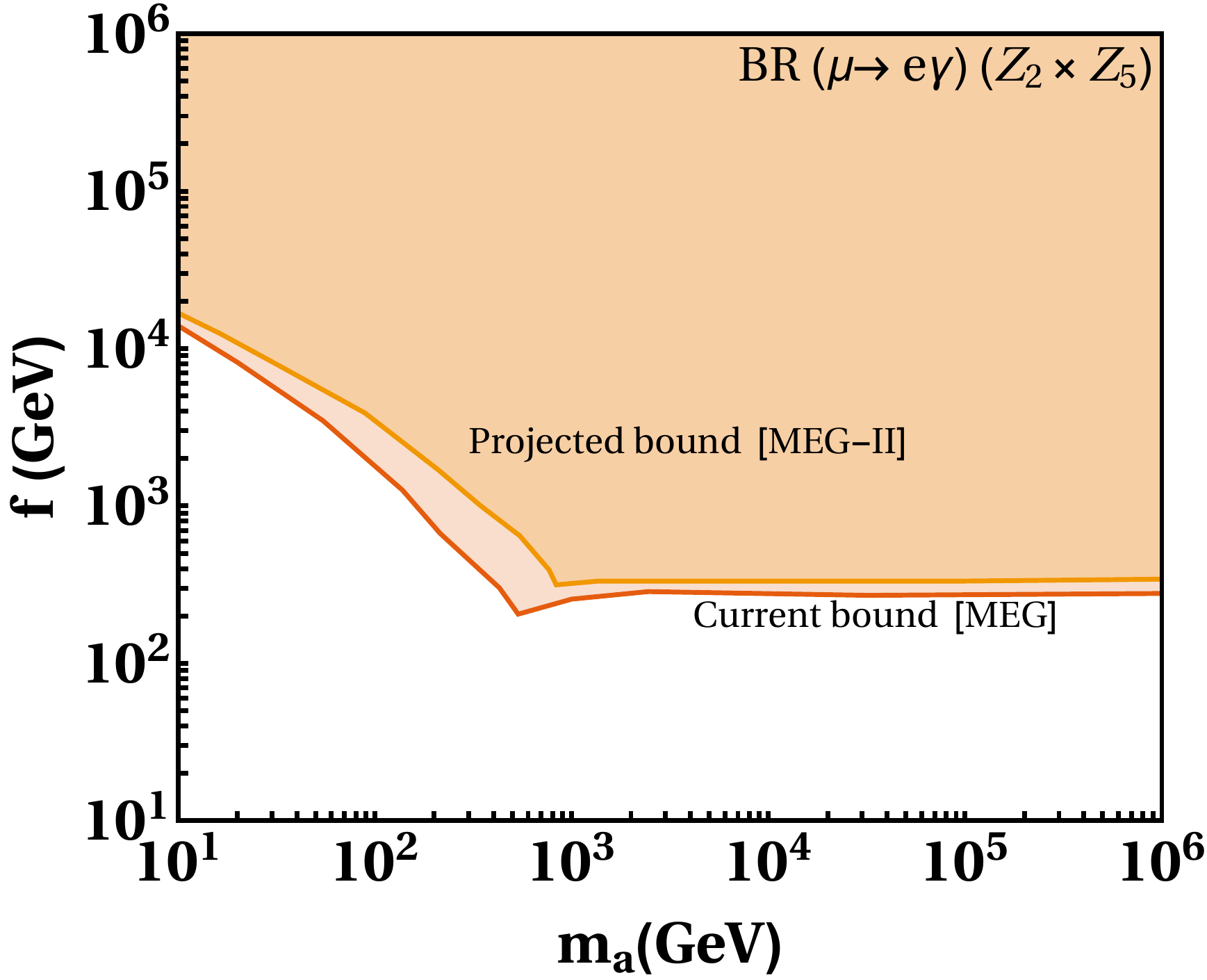}	
	 \caption{}
         \label{rad1}	
\end{subfigure} 
\begin{subfigure}[]{0.49\linewidth}
 \includegraphics[width=\linewidth]{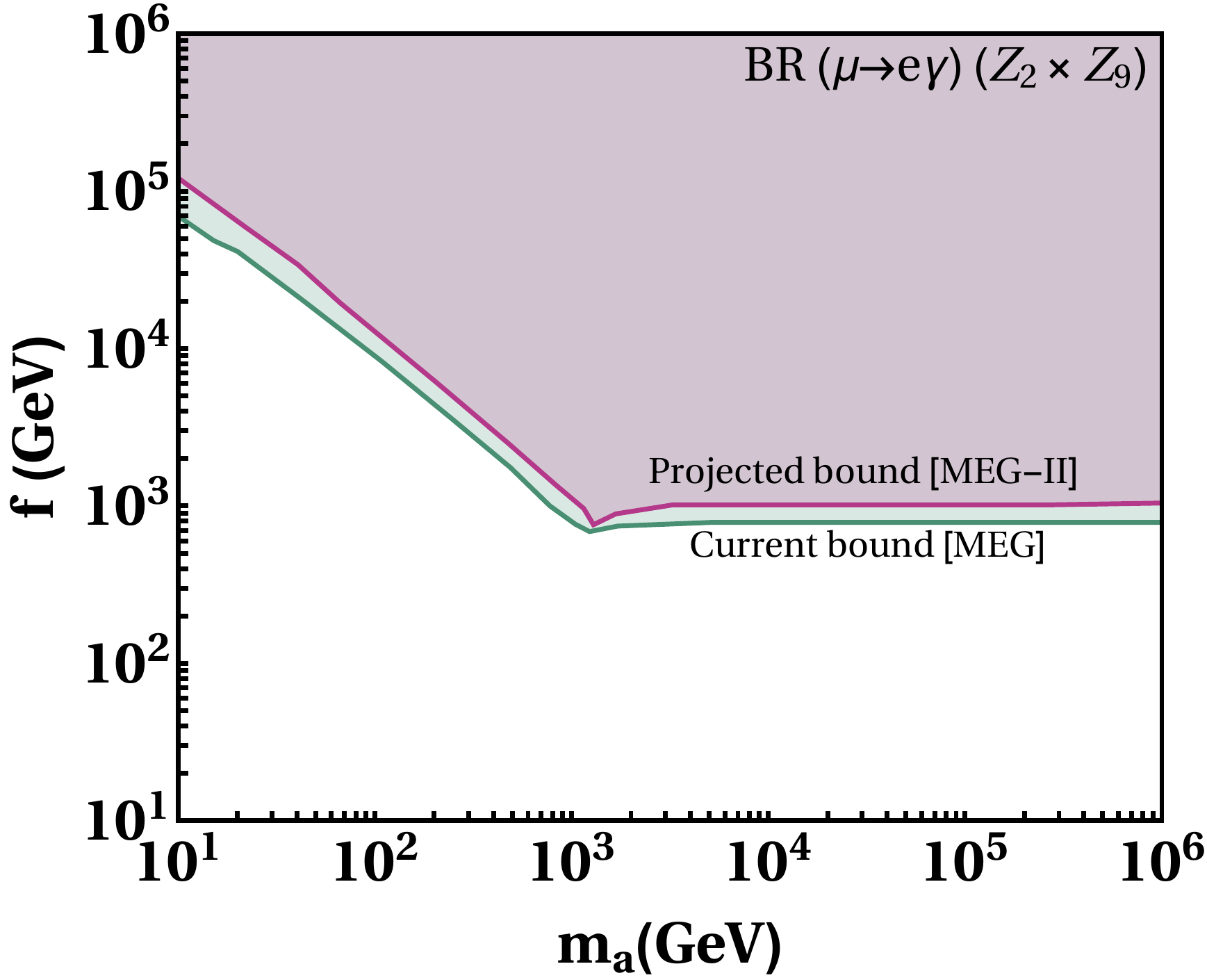}
 \caption{}
         \label{rad2}	
 \end{subfigure}
\caption{In the left panel,  the red and yellow coloured boundaries represent the allowed parameter space by the current and projected sensitivities of the MEG experiment on the $\br(\mu \rightarrow e \gamma)$ for the minimal  ($\mathcal{Z}_2 \times \mathcal{Z}_5$) model with $\lambda_\chi= 2$ in the $m_a - f$ plane.  The same bounds  for the non-minimal  ($\mathcal{Z}_2 \times \mathcal{Z}_9$) model are shown in the right panel with green and magenta coloured curves, respectively .}
\label{rad_lep}
\end{figure}

The radiative leptonic decays $\mu \rightarrow e \gamma$ places weak constraints on the parameter space of the minimal model as shown in figure \ref{rad1}.  This does not change much even for the future projected sensitivities of the MEG-\rom{2} experiment.   For the non-minimal model,  the bounds from the current measurement of the MEG experiment, shown in figure \ref{rad2} by the green boundary,  are quite stringent relative to that of the minimal model.  Moreover,  for the  future projected sensitivities of the MEG\rom{2} experiment,  the constraints are further stronger depicted by the magenta coloured boundary.  The constraints coming from  the decays $ \tau \rightarrow e \gamma$ and $ \tau \rightarrow \mu \gamma$ are  weaker than that of the  decay $\mu \rightarrow e \gamma$.  Therefore,  they are not shown in this work.

\subsection{$A~\mu\rightarrow A~e$ conversion}
The effective Lagrangian describing  $A~ \mu\rightarrow A ~e$ conversion can be written as,
\begin{align}
\lag_{\text{eff}}=
C_{qq}^{VL}\,\bar e \gamma^\nu P_L \mu\, \bar q \gamma_\nu q
+m_\mu m_q\,C_{qq}^{SL}\bar e P_R \mu \,  \bar q q
+m_\mu \alpha_s C_{gg}^L\,\bar e P_R \mu \,G_{\rho\nu}G^{\rho\nu}\,+ (R\leftrightarrow L) \; ,
\label{eq:ConvLag}
\end{align}
Moreover,  there is additional contribution to $A~ \mu\rightarrow A~ e$ conversion from the dipole operators given in equation \eqref{meg}.

\begin{figure}[h]
	\centering
    \includegraphics[width=0.44\linewidth]{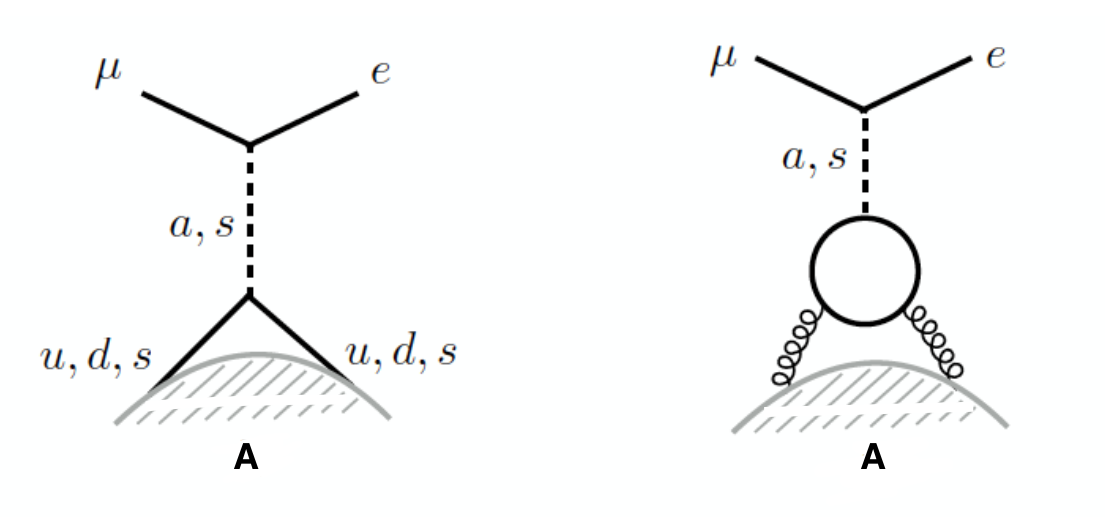}	
 \caption{Feynman diagram showing   $A~ \mu\rightarrow A~ e$ conversion.}
 \label{mec}
	\end{figure}

The Feynman diagram for  $A~ \mu\rightarrow A~ e$ conversion is shown in figure \ref{mec}.  The corresponding Wilson coefficients arise due to the diagram on the left in figure \ref{mec}~\cite{Bauer:2015kzy},
\begin{align}
C^{SL}_{qq}&=\left(\frac{1}{m_s^2}+\frac{1}{m_a^2}\right)y_{\mu e}^*\text{Re}(y_{qq})\,,\notag\\
C^{SR}_{qq}&=\left(\frac{1}{m_s^2}-\frac{1}{m_a^2}\right)y_{ e\mu} \text{Re}(y_{qq})\,.
\end{align}

The nuclear effects which include effects of quarks inside the nucleons as well as the contribution of the Feynman diagram on the right side of figure \ref{mec} are absorbed in the nucleon-level Wilson coefficients defined by,
\begin{align}
\tilde C_p^{VL} &=\sum_{q=u,d} C^{VL}_{qq}\, f_{V_q}^p, \\ \nonumber 
\tilde C_p^{SL} &=\sum_{q=u,d,s} C^{SL}_{qq}\, f_{q}^p-\sum_{Q=c,b,t} C^{SL}_{QQ} \,f_\text{heavy}^p\,,
\end{align}
where  the quark content of the proton is accounted by vector and scalar couplings $f_{V_q}^{p}, f_q^p$, and $f_\text{heavy}^p=2/27\big(1-f_u^p-f_d^p-f_s^p\big)$ \cite{Shifman:1978zn}.   For right-handed operators, analogous expressions are obtained by replacing $L$ with $R$,  and for the neutron $p$ is replaced by $n$.  The vector operators contribute extremely less  than the scalar operators and can be neglected \cite{Bauer:2015kzy}.   The numerical values of vector and scalar couplings are taken from references \cite{Crivellin:2013ipa, Crivellin:2014cta},  which are based on the lattice average given in  reference~\cite{Junnarkar:2013ac},
\begin{align}
f_u^p&=0.0191 \qqquad  f_u^n=0.0171\,,\notag\\
f_d^p&=0.0363 \qqquad f_d^n=0.0404\,,\notag\\
f_s^p&=f_s^n=0.043\,.
\end{align}

The $A~ \mu\rightarrow A~ e$ conversion rate  including nuclear effects can be written as\cite{Bauer:2015kzy},
\begin{align}
\Gamma_{A ~\mu\rightarrow A~ e}=\frac{m_\mu^5}{4}\left|C_T^L D +4\left[m_\mu m_p\tilde C_p^{SL}+\tilde C_p^{VL}V^p+ (p\rightarrow n) \right]\right|^2\, + L \rightarrow R,
\label{eq:Convrate}
\end{align}
where the  dimensionless coefficients $D, S^{p,n}$ and $V^{p,n}$ depend on the overlap integrals of the initial state muon and the final-state electron wave-functions with the target nucleus, and their numerical values are given in table \ref{nucl}~\cite{Kitano:2002mt},
\begin{table}[tb]\centering
\begin{tabular}{p{2.0cm}|p{1.5cm} p{1.5cm} p{1.5cm} p{1.5cm} p{1.5cm} p{2cm}}
\hline
\text{Target}& $D$& $S^p$ & $S^n$ & $V^p$& $V^n$&$\Gamma_\text{capt} [10^{6} \text{s}^{-1}]$\\\hline
\text{Au}&  0.189& 0.0614&0.0918&0.0974&0.146&13.06\\
\text{Al}&  0.0362& 0.0155&0.0167&0.0161&0.0173&0.705\\
\text{Si}&  0.0419& 0.0179&0.0179&0.0187&0.0187&0.871 \\ 
\text{Ti}&  0.0864& 0.0368&0.0435&0.0396&0.0468&2.59 \\ \hline
\end{tabular}
\caption{Numerical values of the  dimensionless coefficients $D, S^{p,n}$,  $V^{p,n}$ and  the muon capture rate for different nuclei.}
\label{nucl}
\end{table}
where  $\Gamma_\text{capt}$ stands for the muon capture rate.

\begin{figure}[H]
	\centering
	\begin{subfigure}[]{0.44\linewidth}
	 \includegraphics[width=\linewidth]{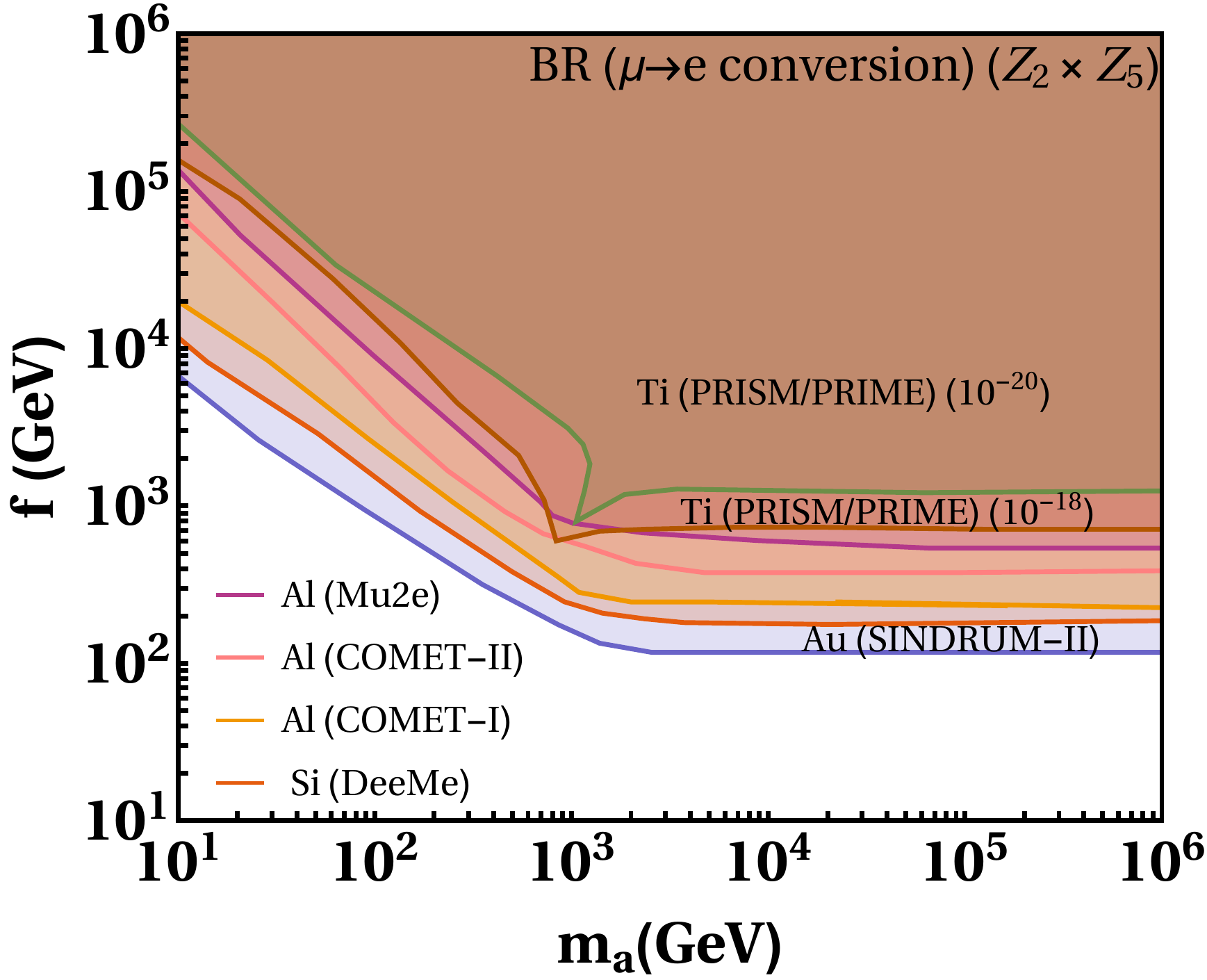}	
	  \caption{}
         \label{mec1}	
\end{subfigure} 
\begin{subfigure}[]{0.44\linewidth}
 \includegraphics[width=\linewidth]{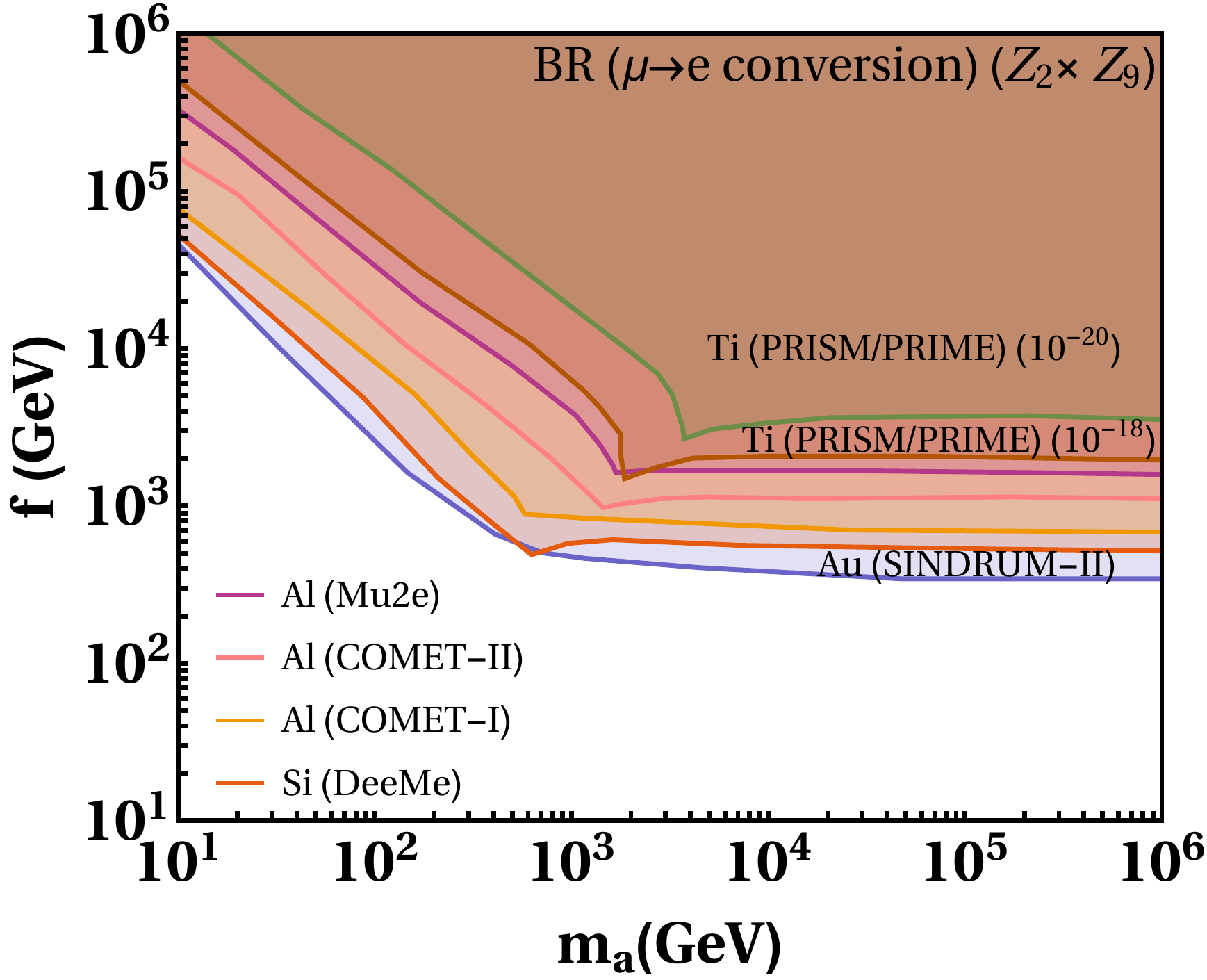}
   \caption{}
         \label{mec2}	
 \end{subfigure}
\caption{The left panel represents the allowed parameter space by $\br (\mu \rightarrow e)$ conversion in the $m_a - f$ plane with different experimental  limits of various target nuclei in table \ref{lfv_exp} with $\lambda_\chi= 2$, for the minimal  ($\mathcal{Z}_2 \times \mathcal{Z}_5$) model.  Similar allowed parameter space  for the  non-minimal  ($\mathcal{Z}_2 \times \mathcal{Z}_9$) model is shown in the right panel.}
\label{mutoe_conv}
\end{figure}

The bounds from  $\br (\mu \rightarrow e)$ conversion for different target nucleus with $\lambda_\chi= 2$ in the $m_a - f$ plane are shown in figure \ref{mutoe_conv}.  The bounds for the minimal model are shown in figure \ref{mec1} for the current measurement and for the different future  projected sensitivities,  and the same for the non-minimal model are shown in figure \ref{mec2}. The strongest bounds among them arise from the projected future sensitivities of  the PRISM/PRIME experiment for the minimal as well as non-minimal models.

\subsection{ $\mu\rightarrow 3e$ and $\tau \rightarrow 3 \mu$ decays}
The three body flavour violating  leptonic decays  $\mu \to 3 e $ and   $\tau \to 3 \ell $ where $\ell= e, \mu$  provide  additional tests of the dipole operators given in equation \eqref{meg}. Their decay width can be written as\cite{Bauer:2015kzy},
\begin{align}
\Gamma(\ell'\rightarrow 3 \ell)=\frac{ \alpha m_\ell'^5}{12 \pi^2}\left|log \frac{{m_\ell'}^2}{{m_\ell}^2} - \frac{11}{4}\right|\left(|C_T^L|^2 + |C_T^R|^2\right).
\end{align}
where the tree-level contribution is ignored due to the strong chiral-suppression which is dominated by the logarithmic enhancement of the dipole operators\cite{Bauer:2015kzy}. Other contributions, such as $Z$-mediated penguin are strongly suppressed and  ignored\cite{Goto:2015iha}.

\begin{figure}[H]
	\centering
	\begin{subfigure}[]{0.49\linewidth}
	 \includegraphics[width=\linewidth]{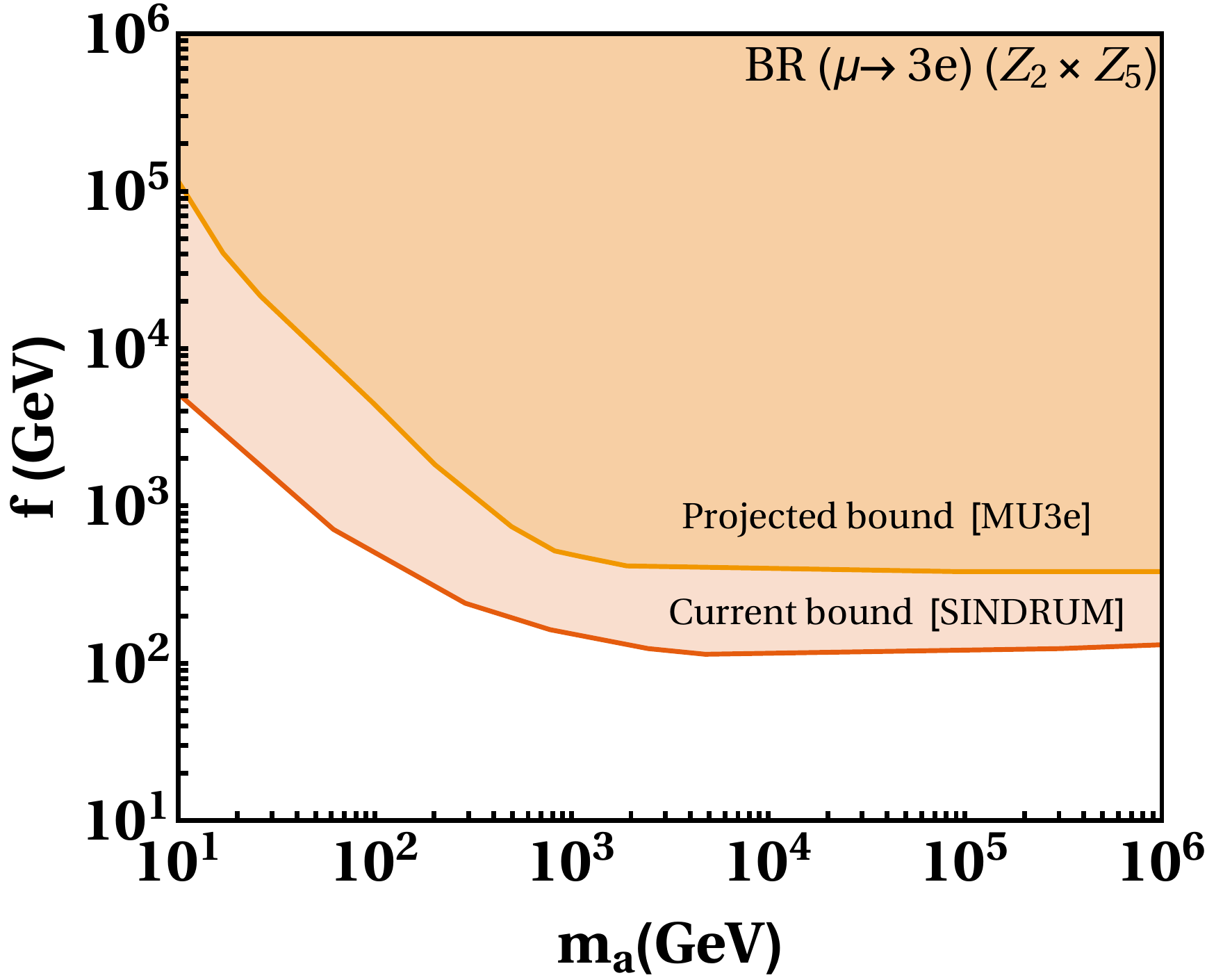}	
	   \caption{}
         \label{tau3e1}	
\end{subfigure} 
\begin{subfigure}[]{0.49\linewidth}
 \includegraphics[width=\linewidth]{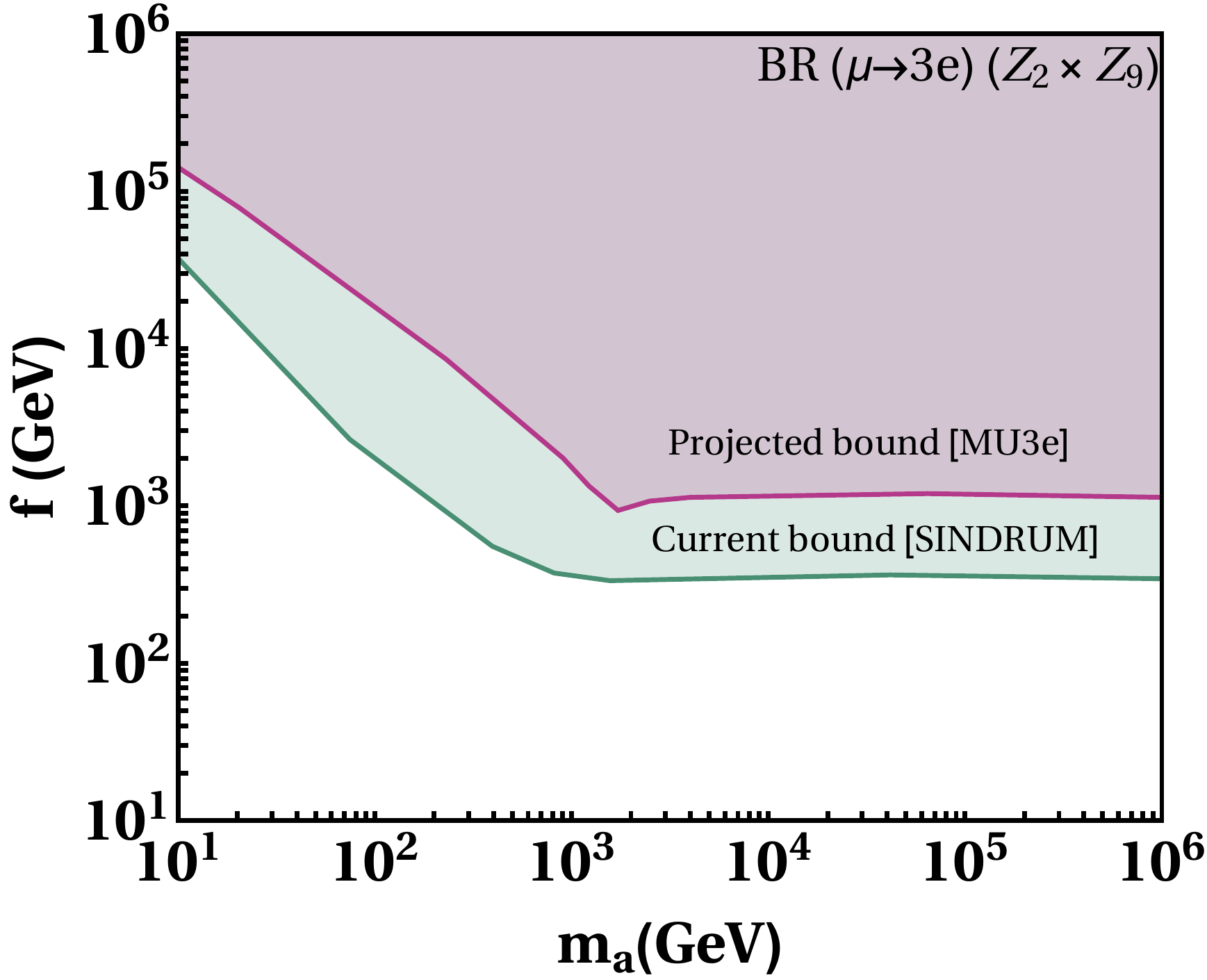}
 \caption{}
         \label{tau3e2}	
 \end{subfigure}
\caption{The allowed parameter space by  $\br( \mu \rightarrow 3e)$ for the minimal  ($\mathcal{Z}_2 \times \mathcal{Z}_5$) model on the left and for the non-minimal ($\mathcal{Z}_2 \times \mathcal{Z}_9$) model on the right with $\lambda_\chi= 2$.   }
\label{tau3l}
\end{figure}

In figure \ref{tau3l},  we show  the allowed parameter space by  $\br( \mu \rightarrow 3e)$    for $\lambda_\chi= 2$ in the $m_a - f$ plane.  For the minimal model,  the bounds are shown in figure \ref{tau3e1}.  They are weaker for the current measurement and for the future projected sensitivity as well.   In figure \ref{tau3e2},  the same bounds on the allowed parameter space are shown for the non-minimal model.  For the current measurement,  the bound is given by the green boundary, and for the future projected sensitivity, it is depicted by the  magenta coloured curve.  As obvious from figure  \ref{tau3e2}, the bounds for the non-minimal model are quite stringent relative to the minimal model.  The bounds from the decay  $\tau \to 3 \ell $ are much weaker and are not shown in this work.

\begin{figure}[H]
	\centering
	\begin{subfigure}[]{0.44\linewidth}
	 \includegraphics[width=\linewidth]{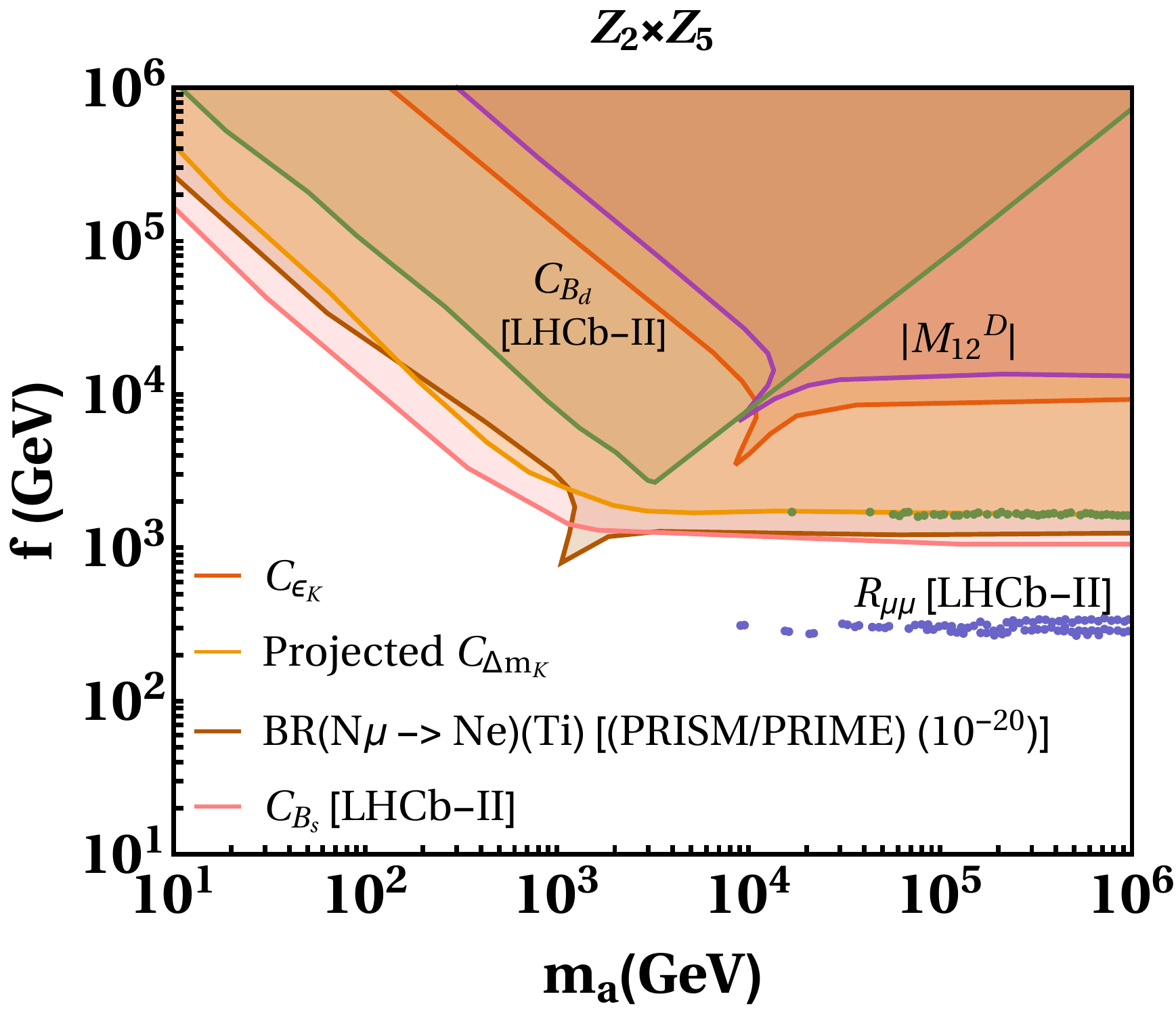}	
	   \caption{}
\end{subfigure} 
\begin{subfigure}[]{0.44\linewidth}
 \includegraphics[width=\linewidth]{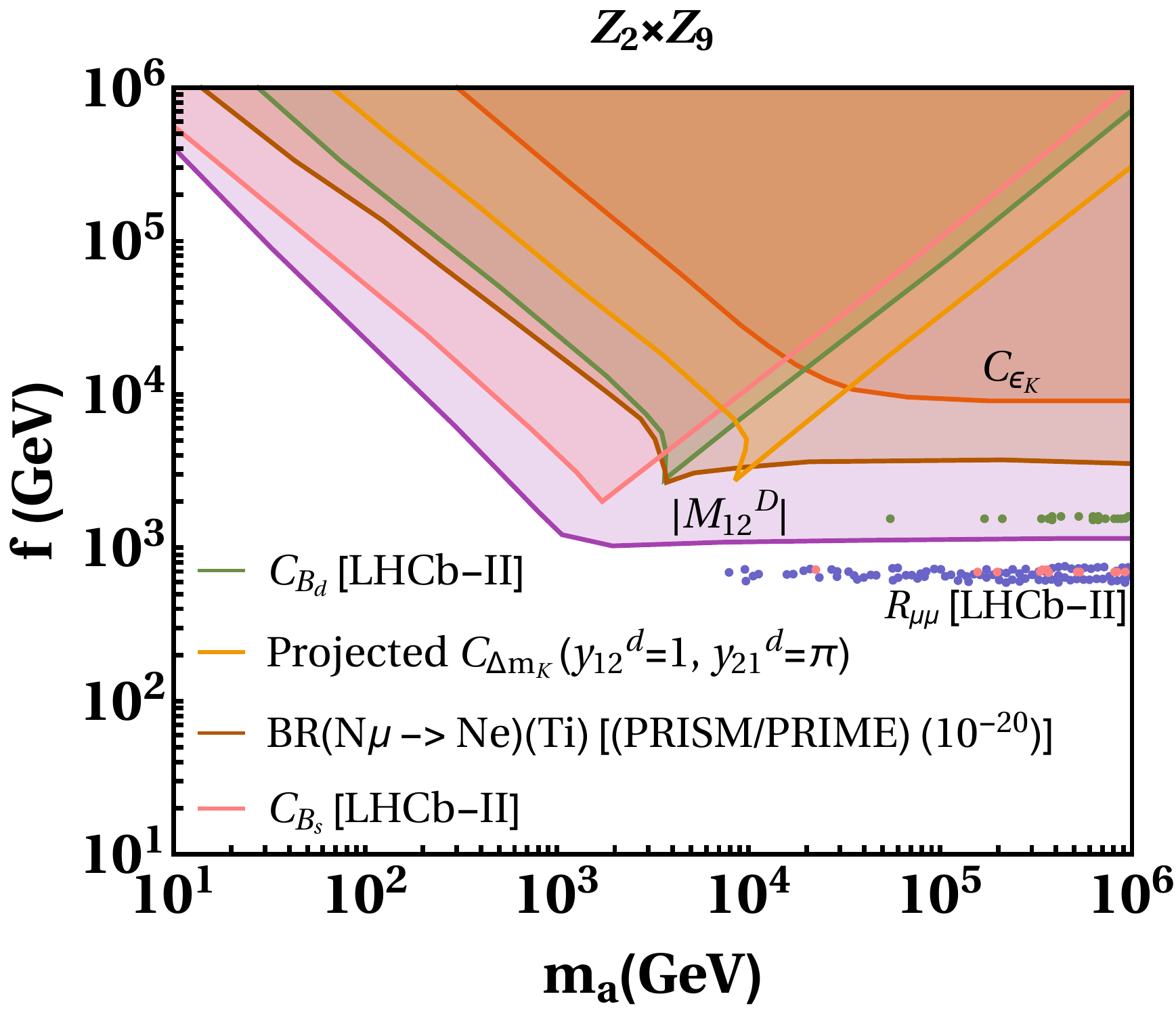}
 \caption{}
 \end{subfigure}
\caption{ Summarized significant bounds on the minimal $\mathcal{Z}_2 \times \mathcal{Z}_5 $  and the non-minimal $\mathcal{Z}_2 \times \mathcal{Z}_9 $ model in the left and the right panel respectively. }
\label{sum_plot}
\end{figure}

Finally,  we show a summary of the most relevant and stringent bounds on the prameter space of the minimal and the non-minimal models in figure \ref{sum_plot}.

\section{Summary}
\label{summ}

We have discussed a simple FN  mechanism in the framework of a new   $\mathcal{Z}_2 \times \mathcal{Z}_N $  flavour symmetry.  This symmetry  is inspired by the extensively explored 2HDM and the MSSM,  where  the $\mathcal{Z}_2$  symmetry is an essential ingredient of the theoretical framework.    We show  that a minimal form of this symmetry,   the $\mathcal{Z}_2 \times \mathcal{Z}_5 $,  is capable of providing an explanation to the charged fermion mass pattern and quark mixing along with a mechanism to predict neutrino masses and mixing angles.    However,  for the  minimal model based on the $\mathcal{Z}_2 \times \mathcal{Z}_5 $ flavour symmetry,  all the Yukawa couplings are not order one,  which is a preferred choice in literature.   This observation leads to a non-minimal model based on the $\mathcal{Z}_2 \times \mathcal{Z}_9 $ flavour symmetry,  where all the couplings are order one.    The  FN mechanism created through the  $\mathcal{Z}_2 \times \mathcal{Z}_N $ flavour symmetry is different from the conventional FN mechanism,  where a continuous $U(1)$ symmetry is employed to  achieve a solution of the flavour problem of the SM.

The leading question  is the scale where the flavour symmetry $\mathcal{Z}_2 \times \mathcal{Z}_N $ is broken.  This is addressed by deriving the bounds on the parameter space of the model using flavour physics data.  Moreover,  future sensitivities of the CMS and the   phase-\rom{1} and \rom{2} of the LHCb for the flavour physics observables   turn out to  be an interesting and fertile ground to investigate the  breaking  scale of  the $\mathcal{Z}_2 \times \mathcal{Z}_N $ flavour symmetry.    Particularly  the reach of the experiments such as MEG \rom{2} and PRISM/PRIME, which are going to test   lepton flavour violating effects,  could play a crucial role in  improving the present limits by orders of magnitude.

For the minimal model based on the flavour symmetry $\mathcal{Z}_2 \times \mathcal{Z}_5 $,  the quark flavour physics plays a crucial role.  For instance,  the most stringent constraints on the parameter space of the minimal model comes from the  $K^0-\bar K^0$  and $D^0 - \bar D^0$ mixing using the present data.    In particular,  the $D^0 - \bar D^0$ provides the very tight bounds since it is highly unsuppressed due to the PMS in the minimal model.  The $B_d -\bar B_d$ mixing provides tighter bounds than  the $B_s -\bar B_s$ mixing.  The future projected sensitivity of the $B_d -\bar B_d$ mixing will further constrain the parameter space of the minimal model in future.  The rare decays $\br(B_{d,s}\rightarrow \mu^+\mu^-) $ branching ratios place relatively weaker bounds on the parameter space of the minimal model for the present measurements as well as for the future projected sensitivities of  phase-\rom{1} and \rom{2} of the LHCb.   However,   the future projected sensitivity of the  rare  $B$ decays observable $ \mathcal{R}_{\mu\mu} $ provides  robust and the most stringent bound on the parameter space of the minimal model.  Among the leptonic observables,  the future projected sensitivity of the PRISM/PRIME for the BR $  \text{(} \mu  \to e  \text{)}^{\rm Ti} $  will substantially be able to constrain the parameter space of the minimal model.

In the case of the non-minimal model based on the flavour symmetry $\mathcal{Z}_2 \times \mathcal{Z}_9 $,  both the quark as well as leptonic flavour physics play decisive role in constraining the parameter space of the non-minimal model.  On the quark flavour side,  the most stringent bounds are coming from the  $K^0-\bar K^0$ mixing while from the leptonic flavour side,  it comes from the radiative leptonic decay  $\mu   \rightarrow  e  \gamma$.    We observe that  every flavour observable including  the branching ratio of the $K_L \rightarrow \mu^+ \mu^-$ decay,  is able to provide important bounds on the parameter space of the non-minimal model.  In particular,  the future projected sensitivity of the  rare  $B$ decays observable $ \mathcal{R}_{\mu\mu} $  for the LHCb and CMS experiments,  will test the non-minimal model rigorously.  Moreover,    the future projected sensitivity of the PRISM/PRIME for the BR $  \text{(} \mu  \to e  \text{)}^{\rm Ti} $  will eliminate a large part of the allowed parameter space.

In short,   quark flavour physics will determine the allowed parameter space of the minimal model,  and leptonic flavour violating observables independently can further probe the parameter space upto a very high scale.   For the non-minimal model based on the $\mathcal{Z}_2 \times \mathcal{Z}_9$  flavour symmetry,   quark and lepton flavour physics play a crucial role in constraining  the allowed parameter space of the model.  The future sensitivities of the experiments such as MEG \rom{2},  Mu3e,  DeeMe, COMET,  Mu2e and PRISM/PRIME will be able to constrain the parameter space of the non-minimal model based on the $\mathcal{Z}_2 \times \mathcal{Z}_9$  flavour symmetry.  On the quark side,   the  $B_d -\bar B_d$ mixing in the   future phase-\rom{1} and \rom{2} of the LHCb will be able to eliminate a sufficient region of the parameter space.  In future phase-\rom{2} of the LHCb,  the ratio  $ \mathcal{R}_{\mu\mu} $  will be crucial in ruling out the major part of the flavon parameter space.  Thus,  future projected sensitivities of the LHCb phase-\rom{1} and \rom{2}   will play a defining role in determining the fate of the flavon models discussed in this work.

\section*{Acknowledgement}
We are extremely thankful to Prof. Srubabati Goswami for very important discussion on our model.  This work is supported by the  Council of Science and Technology,  Govt. of Uttar Pradesh,  India through the  project ``   A new paradigm for flavour problem "  no.   CST/D-1301,  and Science and Engineering Research Board,  Department of Science and Technology, Government of India through the project `` Higgs Physics within and beyond the Standard Model" no. CRG/2022/003237. 
\section*{Appendix}
\begin{appendix}
\section*{Benchmark points for the Yukawa couplings}
\label{benchmark}
We reproduce the fermion masses using the following values of the fermion masses at $ 1$TeV\cite{Xing:2007fb},
\begin{eqnarray}
\{m_t, m_c, m_u\} &\simeq& \{150.7 \pm 3.4,~ 0.532^{+0.074}_{-0.073},~ (1.10^{+0.43}_{-0.37}) \times 10^{-3}\}~{\rm GeV}, \nonumber \\
\{m_b, m_s, m_d\} &\simeq& \{2.43\pm 0.08,~ 4.7^{+1.4}_{-1.3} \times 10^{-2},~ 2.50^{+1.08}_{-1.03} \times 10^{-3}\}~{\rm GeV},
\nonumber \\
\{m_\tau, m_\mu, m_e\} &\simeq& \{1.78\pm 0.2,~ 0.105^{+9.4 \times 10^{-9}}_{-9.3 \times 10^{-9}},~ 4.96\pm 0.00000043 \times 10^{-4}\}~{\rm GeV}.
\end{eqnarray}

The magnitudes and phases  of the CKM mixing elements are \cite{Zyla:2021},
\bea
|V_{ud}| &=& 0.97370 \pm 0.00014,  |V_{cb}| = 0.0410 \pm 0.0014, |V_{ub}| = 0.00382 \pm 0.00024, \\ \nonumber
\sin 2 \beta &=& 0.699 \pm 0.017, ~ \alpha = (84.9^{+5.1}_{-4.5})^\circ,~  \gamma = (72.1^{+4.1}_{-4.5})^\circ, \delta = 1.196^{+0.045}_{-0.043}
\eea

The present scenario of the neutrino physics for the normal hierarchy can be described by the following global fit results\cite{deSalas:2017kay},
\bea
\Delta m_{21}^2 &=& (7.55^{+0.59}_{-0.5}) \times 10^{-5} {\rm eV}^2, |\Delta m_{31}^2| = (2.50\pm 0.09) \times 10^{-3} \rm{eV}^2,  \\ \nonumber
\sin^2 \theta_{12} &=&  (3.20^{+0.59}_{-0.47}) \times 10^{-1},
 \sin^2 \theta_{23} =  (5.47^{+0.52}_{-1.02}) \times 10^{-1},  \sin^2 \theta_{13} =  (2.160^{+0.25}_{-0.20}) \times 10^{-2},
\eea
where range of errors is $3 \sigma$.

We fit quark and charged-lepton masses along with the neutrino oscillation data by defining
\begin{multline*} 
\chi^2 = \dfrac{(m_q - m_q^{\rm{model}} )^2}{\sigma_{m_q}^2}+  \dfrac{(m_\ell - m_\ell^{\rm{model}} )^2}{\sigma_{m_\ell}^2}  + \dfrac{(\sin \theta_{ij} -\sin \theta_{ij}^{\rm{model}} )^2}{\sigma_{\sin \theta_{ij}}^2}     + \dfrac{(\sin 2 \beta  -\sin 2 \beta^{\rm{model}} )^2}{\sigma_{\sin2\beta}^2}   + \dfrac{( \alpha  - \alpha^{\rm{model}} )^2}{(\sigma_{\alpha})^2} \\ + \dfrac{( \gamma   - \gamma^{\rm{model}} )^2}{(\sigma_{\gamma})^2}+ \dfrac{(\Delta m_{21}^2 - \Delta m_{21}^{2 ~ \rm{model}} ) }{\sigma_{\Delta m_{21}^2}^2} + \dfrac{(\Delta m_{31}^2 - \Delta m_{31}^{2 ~ \rm{model}} ) }{\sigma_{\Delta m_{31}^2}^2}  + \dfrac{(\sin \theta_{ij}^\nu -\sin \theta_{ij}^{\nu ~\rm{model}} )^2}{\sigma_{\sin \theta_{ij}^\nu}}
\end{multline*} 
where $q=\{u,d,c,s,t,b\}$, $\ell=\{e,\mu,\tau\}$, $\nu=\{\nu_{e},\nu_{\mu},\nu_{\tau}\}$ and $i,j=1,2,3$.  The phases of the CKM matrix in the standard choice are defined as follows:
\begin{eqnarray}
\beta^{\text{model}} =\text{arg} \left(- \dfrac{V_{cd} V_{cb}^*}{V_{td} V_{tb}^*}\right),~\alpha^{\text{model}} =\text{arg} \left(- \dfrac{V_{td} V_{tb}^*}{V_{ud} V_{ub}^*}\right),~\gamma^{\text{model}} =\text{arg} \left(- \dfrac{V_{ud} V_{ub}^*}{V_{cd} V_{cb}^*}\right).
\end{eqnarray}

\subsection*{The minimal model}
The dimensionless coefficients $y_{ij}^{u,d,\ell,\nu}= |y_{ij}^{u,d,\ell,\nu}| e^{i \phi_{ij}^{q,\ell,\nu}}$  are scanned with $|y_{ij}^{u,d,\ell, \nu}| \in [0.1,4\pi]$ and $ \phi_{ij}^{q,\ell,\nu} \in [0,2\pi]$.  The fit results  are,

\begin{equation*}
Y_u =\begin{pmatrix}
-1.68 - 3.37 i &  -0.09 + 0.03 i  &   -0.1 - 0.02 i    \\
1.53 + 4.95 i    &- 0.57 + 0.55 i  &  0.48 + 0.002 i    \\
0.76 + 0.18 i   &  -1.04 + 0.46 i   &  0.58 - 0.65 i
\end{pmatrix},  
\end{equation*}

\begin{equation*}
Y_d = \begin{pmatrix}
-4.15 + 3.58 i &  2.20 - 0.89 i & 2.62 - 4.20 i   \\
-0.33 - 0.36 i  & 0.07 -0.075 i &  0.17 + 0.47 i  \\
- 0.24 -0.07 i &  -0.06 - 0.084 i   &  -0.07 - 0.12 i
\end{pmatrix},  
\end{equation*}

\begin{equation*}
Y_l = \begin{pmatrix}
-0.07 - 0.06 i &  0.099 - 0.004 i & 0.45 -0.32   i   \\
-0.14 - 0.09 i  & 0.08 - 0.06 i &  -0.63 + 0.24 i  \\
-0.04 + 0.09 i &  -0.09 + 0.06 i   &  0.10 - 0.0003 i
\end{pmatrix},  
\end{equation*}
with $\epsilon = 0.1$ $\epsilon^\prime  =  1.259 \times 10^{-13}$ , $\chi^2 \approx 14$, and the following results for neutrino oscillation data,
\begin{eqnarray*}
\{|y_{11}^\nu|, |y_{22}^\nu|, |y_{33}^\nu|, |c_{11}|, |c_{22}|, |c_{33}|\} & \simeq& \{3.14, 3.14, 1.48, 3.14, 0.90,   2.76 \}, \nonumber \\
\{\phi_{11}^\nu, \phi_{22}^\nu, \phi_{33}^\nu, \phi_{11}^c, \phi_{22}^c, \phi_{33}^c\} & \simeq& \{3.67, 7.73 \times 10^{-7},
1.24, 1.18, 2.10, 5.18 \}.
\end{eqnarray*}

\subsection*{The non-minimal model}
The dimensionless coefficients $y_{ij}^{u,d,\ell,\nu}= |y_{ij}^{u,d,\ell,\nu}| e^{i \phi_{ij}^{q,\ell,\nu}}$  are scanned with $|y_{ij}^{u,d,\ell, \nu}| \in [0.9, 2\pi]$ and $ \phi_{ij}^{q,\ell,\nu} \in [0,2\pi]$.  The fit results are,
\begin{equation*}
Y_u =\begin{pmatrix}
1  &  0.87 - 0.49 i  &   -0.23 + 0.97 i    \\
-0.9 + 1.05 i    &- 0.7 - 0.72 i  &  1    \\
0.94 - 0.33 i   &  0.55 + 0.84 i   &  0.9 
\end{pmatrix}, 
\end{equation*}

\begin{equation*}
Y_d = \begin{pmatrix}
0.99 - 0.09 i &  3.24 - 1.05 i & 1    \\
0.99 - 0.10 i  & 0.92 + 0.39 i &  0.9   \\
1  &  1    &  -1.04 + 0.54 i
\end{pmatrix},  
\end{equation*}

\begin{equation*}
Y_l = \begin{pmatrix}
0.9  &  0.9  & 1.5    \\
0.9   & 1.5  &  1.5  \\
1.5  &  1.5    &  0.9 
\end{pmatrix},
\end{equation*}
with $\epsilon = 0.23$, $\epsilon^\prime  =  1.259 \times 10^{-13}$ , $\chi^2 \approx 9$, and the following results for neutrino oscillation data,
\begin{eqnarray*}
\{|y_{11}^\nu|, |y_{22}^\nu|, |y_{33}^\nu|, |c_{11}|, |c_{22}|, |c_{33}|\} & \simeq& \{3.14, 3.04, 0.9, 2.08, 3.14,   0.9 \}, \nonumber \\
\{\phi_{11}^\nu, \phi_{22}^\nu, \phi_{33}^\nu, \phi_{11}^c, \phi_{22}^c, \phi_{33}^c\} & \simeq& \{2.68, 0.51,
4.92, 0.86, 3.14, 1.99 \}.
\end{eqnarray*}

\section*{Outline of a possible ultraviolet completion of the $\mathcal{Z}_2 \times \mathcal{Z}_N$ model }
We present an outline of the underlying renormalizable theory,  which could be suitable to the models discussed in this work using the idea discussed in  reference \cite{Abbas:2020frs}.   Let us suppose that the underlying theory is a technicolour (TC) theory containing two technicolour symmetries.  The SM Higgs field comes from the conventional TC group and the flavon field $\chi$ is derived from a different dark technicolour symmetry (DTC).  

The TC chiral condensate which play the role of the SM Higgs VEV can be parametrized as,
\begin{equation}
\langle   \bar{\psi}_L^{\rm TC} \psi_R^{TC}     \rangle  =  \left(  \Lambda_{\rm TC} \exp (k_{\rm TC} \Delta \chi^{\rm TC})  \right)^3,
\end{equation}
where $\Delta \chi^{\rm TC}$ is the chirality of the operator on the left of above equation,  $ \Lambda_{\rm TC}$ is the scale of the underlying gauge TC theory,  and $k_{\rm TC}$ is a constant.

In a similar manner,  we can write DTC chiral condensate which  represents the flavon VEV,
\begin{equation}
\langle   \bar{\psi}_L^{\rm DTC} \psi_R^{DTC}     \rangle  =  \left(  \Lambda_{\rm DTC} \exp (k_{\rm DTC} \Delta \chi^{\rm DTC})  \right)^3.
\end{equation}

Now the couplings $y_{ij}$ are given by the following equation:
\begin{equation}
y_{ij} = f ( \Lambda_{\rm TC},   \Lambda_{\rm DTC},  \Lambda_{\rm ETC}),
\end{equation}
where the scale $ \Lambda_{\rm ETC}$ corresponds to the extended TC theory in which the SM,  TC and DTC fermions are embedded.

The  masses of the fermions can be written as,
\begin{equation}
m_f \propto   f ( \Lambda_{\rm TC},   \Lambda_{\rm DTC}, \Lambda_{\rm ETC}).
\end{equation}

We observe that if the function $f ( \Lambda_{\rm TC},   \Lambda_{\rm DTC}, \Lambda_{\rm ETC})$ is  always generated by a tree-level exchange of the underlying theory,  all the couplings $y_{ij}$ will have the same order of magnitude (which could be order one).  However, depending of the structure of the underlying theory,   some of the $y_{ij}$ could be loop-induced and some could come from tree-level contributions.   Therefore,  this will result  in some couplings being suppressed compared to  the tree-level contributions.   This scenario will lead to the numerical couplings  for the minimal model based on the $\mathcal{Z}_2 \times \mathcal{Z}_5$  flavour symmetry.  
\end{appendix}

\end{document}